\documentclass[journal]{IEEEtran}
\usepackage{amsmath,amsfonts}
\usepackage{algorithmic}
\usepackage{array}
\usepackage[caption=false,font=normalsize,labelfont=sf,textfont=sf]{subfig}
\usepackage{textcomp}
\usepackage{stfloats}
\usepackage{url}
\usepackage{verbatim}
\usepackage{graphicx}
\def\BibTeX{{\rm B\kern-.05em{\sc i\kern-.025em b}\kern-.08em
    T\kern-.1667em\lower.7ex\hbox{E}\kern-.125emX}}
\usepackage{balance}

\usepackage{microtype}
\usepackage[numbers,sort&compress]{natbib}
\usepackage{booktabs}
\usepackage{pdflscape}
\usepackage{longtable}

\begin{document}
\title{Telling stories with data - A systematic review}

\author{Kay Schröder, Wiebke Eberhardt, Poornima Belavadi, Batoul Ajdadilish, Nanette van Haften, 
 Ed Overes, Taryn Brouns, André Calero Valdez
\thanks{K. Schröder is with HS Düsseldorf University of Applied Sciences, Düsseldorf, Germany}
\thanks{W. Eberhardt is with University of Twente, Twente, The Netherlands}
\thanks{P. Belavadi is with RWTH Aachen University, Aachen, Germany}
\thanks{B. Ajadilish is with Zuyd University of Applied Sciences, Heerlen, The Netherlands}
\thanks{N. van Haften is with Maastricht University, Maastricht, The Netherlands}
\thanks{E. Overes is with Zuyd University of Applied Sciences, Heerlen, The Netherlands}
\thanks{A. Calero Valdez is with Institute of Multimedia and Interactive Systems, University of Lübeck, Lübeck, Germany}

}

\maketitle

\begin{abstract}
The exponential growth of data has outpaced human ability to process information, necessitating innovative approaches for effective human-data interaction. To transform raw data into meaningful insights, storytelling, and visualization have emerged as powerful techniques for communicating complex information to decision-makers.
This article offers a comprehensive, systematic review of the utilization of storytelling in visualizations. It organizes the existing literature into distinct categories, encompassing frameworks, data and visualization types, application domains, narrative structures, outcome measurements, and design principles. By providing a well-structured overview of this rapidly evolving field, the article serves as a valuable guide for educators, researchers, and practitioners seeking to harness the power of storytelling in data visualization.
\end{abstract}



\section{Introduction}
\IEEEPARstart{S}{torytelling} stands as an age-old tradition, a cornerstone of human culture, ensuring the continuity of knowledge across generations. The power of stories emanates from their ability to engage our entire cognitive and emotional faculties, enabling us to empathize, connect, and understand the narrative's content. In our modern era, this storytelling tradition is being merged with data visualization, thereby becoming a crucial tool in comprehending complex data associated with global phenomena such as climate change, pandemics, and societal changes.

However, the application of traditional storytelling to the rapidly evolving domain of information visualization is not without its challenges. This integration necessitates a deeper understanding of the nuances of classical storytelling and an exploration of its potential applicability to data-driven narratives. With various interpretations of storytelling across different domains, a holistic approach is required, one that merges these diverse perspectives.

This paper addresses these challenges, aiming to provide a comprehensive review of storytelling in the context of data and information visualization. We identify a dual-fold problem: the need to consolidate various perspectives on storytelling and the need to understand the potential application of traditional storytelling techniques in the context of data-driven narratives.

Drawing upon literature studies, information visualization, and psychological and behavioral science, we explore the multifaceted nature of storytelling. Four research questions guide this exploration:

\begin{itemize}
\item What are the theoretical or conceptual foundations of data-driven storytelling, and how is the story creation process structured?
\item How can data-driven storytelling be examined from a narratological perspective? What are the different narrative structures, and what goals do they serve?
\item How can we structure individual data stories within a taxonomy that reflects a user's conceptual understanding of the story?
\item What are the cognitive and non-cognitive effects of storytelling, and how were these effects measured empirically?
\end{itemize}

This study will be of interest to a wide audience, including researchers, practitioners, students, and educators from diverse fields such as narratology, visualization, design, journalism, data science, bioinformatics, communication, and computing. The insights presented here may serve as a primer for those new to visualization and storytelling and as a tool for educators seeking to structure storytelling concepts for their audiences. Moreover, the findings can guide researchers and practitioners in identifying gaps and potential future research directions in their respective fields.

The rest of this paper is structured as follows: We begin with an overview of related work, followed by a description of the method and procedure employed in this review. We then structure the selected literature based on our four research questions, concluding with a discussion on the findings and potential future research directions.
\section{Related Work}

Storytelling, while a long-established concept, has only recently been applied in the context of data visualization.

\citet{203segel2010narrative} provided the first comprehensive review of data storytelling and narrative visualizations in 2010, identifying specific design strategies and relevant interaction paradigms. Their analysis also delineated different categories within genres, a classification we have adopted for our analysis of visualization types and application areas.

Building on this, \citet{279bach2018design} proposed grouping these genres into three categories based on their orientation: spatial, temporal, or a combination of both. \citet{149kosara2013storytelling} further explored the role of the setting and audience within three scenarios: self-running presentations, moderator-guided live presentations, and presentations for individual or small groups, which allow for more specific interactions with the audience~\cite{roslingTed}.

\citet{139tong2018storytelling} offered a two-dimensional classification for the literature, distinguishing between various elements, such as authoring tools, user engagement, narratives, and transitions, as well as between different access methods. Their analysis, however, only briefly touched upon certain aspects, which we address in our review by providing a comprehensive overview of the effects.

More recently, \citet{336zhu2020survey} published a survey on automatic infographic and visualization recommendations detailing how storytelling processes can be automated. Their work spans a range of related tools, models, and frameworks, from data-driven automatic visualizations and annotations to knowledge-based visualizations. We have integrated examples from their survey into our tools overview (see Fig.~\ref{fig:StoryGeneration}). Furthermore, \citet{losev2022embracing} highlighted the need for leveraging diversity within the visualization community to foster new ideas and collaborations.

In the realm of journalism, \citet{freixa2021binomial} provided a comprehensive overview of data-driven storytelling, underlining the importance of integrating interactivity and visualization. They discussed the potential of these techniques for engaging news readers, the tools and resources used in digital newsrooms, and the limitations of visual interactions. Similarly, \citet{lopezosa2022data} argued for further analysis of data storytelling across different journalistic formats and reader perspectives.

Despite these significant contributions, a comprehensive summary of the literature across various application and research domains remains lacking. Our review addresses this gap by offering an interdisciplinary overview of data storytelling, drawing on perspectives from Behavioral science, HCI research, Marketing research, Data-Interaction research, and Communication Science. We discuss theoretical frameworks and foundations, narrative structures and their goals, and data story types across different methodological settings. Moreover, we provide a comprehensive overview of cognitive and non-cognitive effects and their evaluation methods.

\begin{figure}[htb]
\centering
\includegraphics[width=\columnwidth]{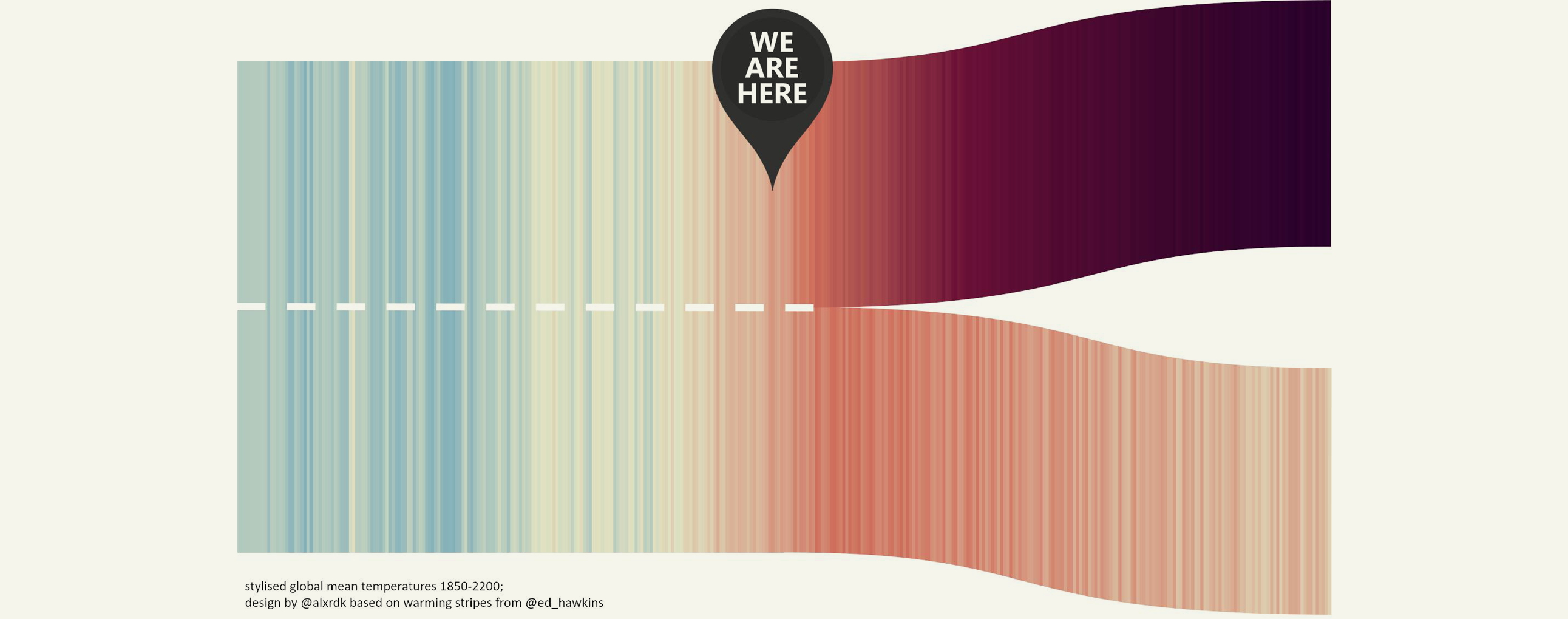}
\caption{The Warming Stripes alternatives by {@}alxrdk based on warming stripes from Ed Hawkins}
\label{fig:example}
\end{figure}

Recent contributions to the field include the work of \citet{aziz2022review}, who explored the link between personality traits and user preferences for visual design styles in data storytelling. \citet{lim2022keeping}
discussed the educational potential of innovative data visualization in news journalism, particularly during the COVID-19 pandemic. \citet{saneiremixing} used data storytelling to foster computational data literacy within the context of socioscientific issues like climate change. \citet{shan2022research} proposed a design strategy for data storytelling in cultural heritage, examining data narrative from the perspectives of data science, visualization, and narratology. Lastly, \citet{dailey2022visualization} conducted a survey on the use of narrative infographics by U.S. municipal governments, emphasizing their role in informing the public about various issues, particularly in the domain of public health and safety.
\section{Method}

Our research methodology follows a mixed-method review approach, aligning with established literature in the field to synthesize related findings comprehensively. We adopt the Preferred Reporting Items for Systematic Reviews and Meta-Analyses (PRISMA) guidelines~\cite{page2021prisma} as our review protocol, which entails four main steps.

\textbf{Firstly}, we initiated the review process with a Scopus database search, targeting the subject areas ``storytelling'' and ``visualization'', restricting to papers published before February 2023. The query string used was \verb|(TITLE-ABS-KEY (storytelling))| \verb|AND (visuali?ation)|. This initial phase yielded 1010 papers.

\textbf{Secondly}, each author received a random sample of 80 papers, with the task of mapping the core elements of their selection into a mind map. Collaboratively, we then identified individual categories and defined relevant inclusion and exclusion criteria. A custom-developed tool facilitated the process, enabling researchers to browse and tag individual papers with details such as source name, title, abstract, keywords, a ``tf-idf'' word cloud, and the publication year. Two authors reviewed each paper, categorizing it for inclusion or exclusion. In case of disagreement, a third author was involved in making a majority decision. Additionally, we manually searched for related work and incorporated additional papers into our sample ($n=9$), leading to a total of 414 papers.

\textbf{Thirdly}, we assigned each category to two authors, who divided all papers within their category for an in-depth read. We corrected any misclassifications and cross-checked them. Publications were rated by relevance and excluded if the full text did not meet our criteria. We held weekly meetings to discuss preliminary conclusions and address papers relevant to multiple or different categories. This stage resulted in 184 remaining publications.

\textbf{Lastly}, in the fourth step, we organized the remaining papers into a coherent context and summarized them within the respective sections.

Three main exclusion criteria guided our review process. Firstly, we considered the relevance to the research question. Since we focus on data-driven stories, we excluded all papers that did not involve storytelling applied in the data context. This includes stories intended for entertainment or those visualizing qualitative data types in narrative visualizations. Secondly, we excluded papers that did not delve deeply into storytelling, such as those where storytelling was used descriptively or as a keyword. Lastly, we applied an incompatibility criterion, excluding papers not written in English or not published in peer-reviewed journals.

\subsection{Definitions}

Our systematic review includes only publications describing data-driven storytelling and related aspects per the following definition. The term ``storytelling'' has been broadly used in the visualization community without a universally accepted definition \cite{245lee2015more, 139tong2018storytelling, 302obie2019study, 305janowski2019mediating, 125lugmayr2017serious}. However, most definitions share a common trait of portraying a process or sequence of events. Therefore, we distill the existing definitions and \textbf{define a data-driven story as a series of related events in a (meaningful) context to facilitate understanding and decision-making concerning data}. To comprehend our perspective on storytelling and its effects, we first define some terms from narratology in relation to visualization and provide an illustrative example (see Fig.~\ref{fig:example}):

\begin{itemize}
\item \textbf{Story subject}: Describes what the data and story are about (e.g., in Fig.~\ref{fig:example}, this is the climate crisis).
\item \textbf{Story object} (actor or carrier): The elements that depict the story (e.g., in our example, the average global temperature and the ``We are here'' sign).
\item \textbf{Story events}: Individual arguments, data representations, or contextual information (e.g., in our example, we have two potential future story events with different outcomes).
\item \textbf{Connections}: The link between the particular story events and how they are structured (e.g., in our example, the positioning of the ``We are here'' sign indicates the current position in time).
\item \textbf{The audience}: The reader or target group of the story (e.g., in our example, people looking at the figure).
\item \textbf{Effects}: Cognitive and non-cognitive effects concerning the story context (e.g., in our example, the storyteller aims to engage the audience emotionally and cognitively).
\end{itemize}

\section{Results}
In this section, we summarize the results of our systematic review, following the four research questions we have outlined above.

\subsection{Frameworks and design principles}

Within this section, we take a closer look at data storytelling from a more holistic point of view and will try to answer questions like: 
What are the underlying mechanisms? 
How can the design process be conceptually/cognitively grounded? 
How can we structure the story creation process?

\subsubsection{Fundamentals of storytelling}

Storytelling is a method that creates a narrative context through a guided combination of explicit (e.g., the evil wolf in Little Red Riding Hood) and implicit knowledge (i.e., the wolf is treacherous, should not be trusted, and let into the house)~\cite{184gershon2001storytelling}. As a result, story perception is an internal construction process whereby an internal model of the story is created. This construction process is influenced by the interaction of sensory impulses, their processing, internal knowledge, and associations. 
As a result, we need to distinguish between different dimensions that influence our understanding of a story: cognitive perception, narrative construction, and persuasive modeling.

\paragraph{Cognitive perception}
\begin{figure}[htb]
    \centering
   \includegraphics[width=\columnwidth]{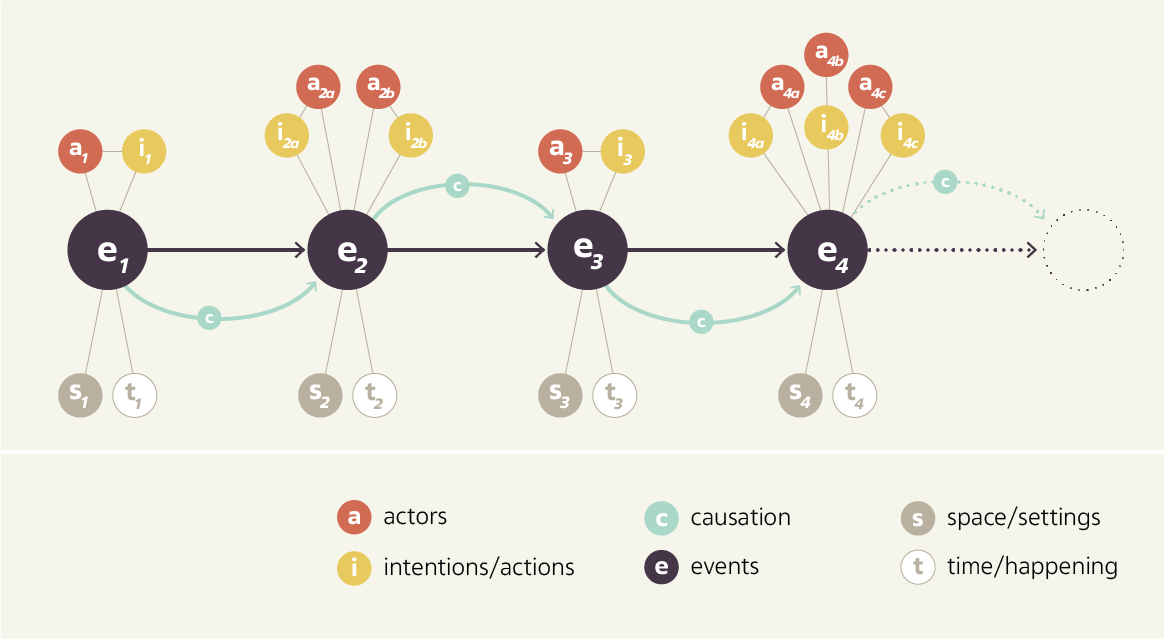}
   \caption{Sequence of events with individual attributes adapted from \citet{277mayr2018once}}
   \label{fig:seqcog}
\end{figure} 

\begin{figure*}[htbp]
    \centering
   \includegraphics[width=1\textwidth]{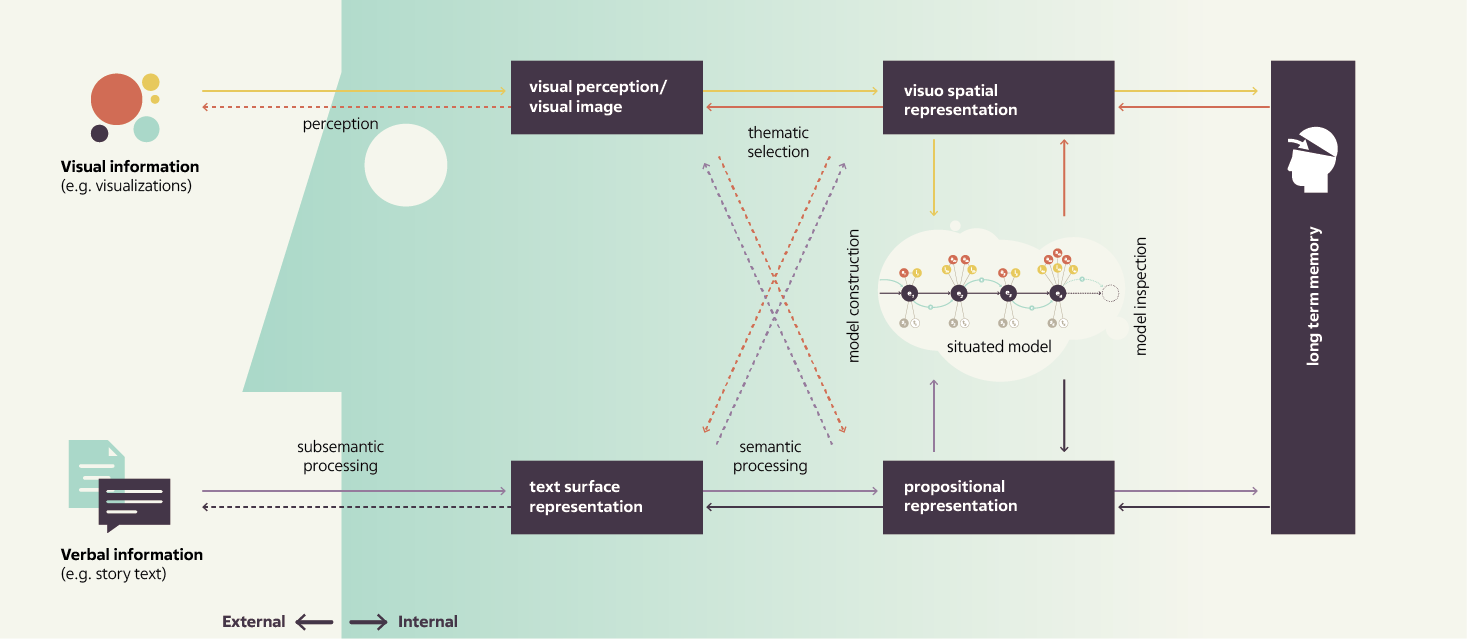}
   \caption{Schematic showing processes involved in multimodal cognition adapted from \citet{277mayr2018once}}
   \label{fig:multimodal}
\end{figure*}

We know from visualization research that the cognitive processing of visual information depends on the type of information and how it is encoded visually.
We look into multimodal information processing to understand how information is processed cognitively ~\cite{058groshans2019digital, 277mayr2018once, 295seyser2018scrollytelling, 016brunet2018actionable} (see Fig.~\ref{fig:multimodal}).

Multimodal cognition is based on the assumption of a dual-layer cognitive architecture. We process verbal (e.g., words or text) and visual information (e.g., figures) in parallel but independent of each other. 
After the stimuli representing an event ``e'' (see Fig.~\ref{fig:seqcog}) is processed within the iconic memory, it gets semantically processed and organized to construct a situated model inside the working memory. This construction process also interacts with prior knowledge stored in the long-term memory, e.g., through supplementing information---whereby it functions as grammar~\cite{184gershon2001storytelling} or contextualization through previous experiences (see Fig.~\ref{fig:multimodal}). 

Some approaches like annotations implicate that we perceive the verbal and visual channel sequentially, simply because we cannot look at both in parallel, \citet{307kong2019understanding} propose to visually encode the data and guide the user's attention with visual cues accompanied by an audio narrative to improve comprehension.
\citet{353smith2020storytelling} propose a framework with a detailed guideline to design an auditory description display for complex interactive environments. This approach especially benefits users with visual impairments, providing access to all relevant information in a highly interactive manner and supporting user engagement. Other examples of using the verbal and visual channels in parallel are classical presentation scenarios, like Rosling's Gapminder talks~\cite{roslingTed} for a general audience. Here, Rosling directs the audience's attention visually to the relevant changes within the visualized data and explains the context verbally. The moderated form of data presentation can also support the communication between domain experts and non-domain experts in a health context~\cite{160hougaard2019telling}.

\paragraph{Narrative construction}\label{narrativeConstruction}
\citet{277mayr2018once} point out that narratives are not just a presentation format but ``a fundamental way of organizing human experience and a tool for constructing models of reality'' as they closely correspond to how we sequentially perceive the world around us. A narrative is considered a chain of related events in cognitive science whereby the individual events consist of several central data dimensions such as time, space, actors, etc. (see Fig.~\ref{fig:seqcog}). 

When we look at the connection between the event ``e'' (see Fig.~\ref{fig:seqcog}) and how we interpret them, a critical component is the story schemata. They form an essential building block as they represent more generic concepts about the relationship of the events, related expectations, and related knowledge, such as when they function as grammar~\cite{184gershon2001storytelling}. \citet{el2020towards} defines this into four layers: factual, intentional, structural, and presentational.

Schemata is an umbrella term for several knowledge structures like frames, plots, scenarios, and references. Describing every detail in an event is often unnecessary, as the audience's schemata can complete the meaning based on prior knowledge. Story schemata refer to the content of the entities and the connection between them, thereby influencing cognitive processing and understanding~\cite{277mayr2018once}. Led by the schemata, the recipient sequentially perceives events and the related dimensions and builds an internal representation of the story called the ``situated model'' ~\cite{277mayr2018once}. The situated model evolves through continuously updated story events that are cognitively connected within the global internal story construct. While perceiving the narrative, the audience might engage with the story described as narrative immersion or transportation~\cite{178Botsis2020,287isenberg2018immersive}. Although immersion is a very elusive term, \citet{287isenberg2018immersive} pointed out that narrative immersion has an emotional as well as a cognitive dimension and distinguishes between(1) immersion in an absorption context, whereby the user engages with the story (emotionally), and (2) immersion in narrative/transportation context, whereby the user is getting connected with the story and the intended message. Narrative immersion is not a binary state; there are multiple levels of involvement and several influencing factors. Therefore, it remains an open question of how we can measure or control this effect to facilitate narrative visualizations. 

\paragraph{Persuasive messaging}
The basic idea of distinguishing between different elements in storytelling is not new. Some authors~\cite{178Botsis2020,243hill2014using} refer to Aristotle, who distinguishes between three relevant elements in storytelling: First, \textbf{Ethos} refers to the author's credibility and demands a clear and accurate presentation of the findings and transparency regarding the underlying data and analytical process that led to the conclusion. The relevance of a clear message contributing to the author's credibility was pointed to in different contexts~\cite{058groshans2019digital,296burkhard20184d,352martinez2020data}. Second, \textbf{Pathos} refers to the (emotional) relationship between the reader and the story. Due to the assumed asymmetry in literacy, prior knowledge, and perspective, it is proposed to consider the emotional connection with the topic to influence the accessibility and remove emotional barriers~\cite{330jacob2020visualising, 348fish2020storytelling, 178Botsis2020}. Pathos can also refer to the concept of narrative immersion~\cite{287isenberg2018immersive} or the general principle of effective storytelling (``lure people in'') mentioned by \citet{058groshans2019digital}. \citet{lan2021kineticharts} propose an animated design scheme and measured that it can convey positive affective emotions like amusement, surprise, tenderness, and excitement.
Furthermore, the domain and visualization literacy of the audience needs to be considered within the story creation process. And third, \textbf{Logos} is about constructing the arguments, which demands considering the reader's perspective and carefully constructing the logical arguments. 
Other authors distinguish between the multiple aspects of storytelling: \citet{235figuerias2014narrative} explored three dimensions that can facilitate storytelling: (1) context of the data, (2) emotional or empathetic dimension (the personal relationship with the topic, which can facilitate motivation and memorability), and (3) temporal structure of the event (which can be linear and non-linear). 

\citet{zhang2022visual} present a framework for information unit-based data storytelling that combines multiple disciplines. The framework utilizes game development and machine learning methods to assist in the composition of data and story elements.

Several visualizations were analyzed to investigate subjectivity and identify specific techniques~\cite{267carpendale2017subjectivity}. The authors point out that subjectivity is a controversial term within the visualization community, with objectivity as a core value. However, communication without subjectivity is hardly possible. A better understanding of subjective aspects can help us develop visual encodings that more closely reflect individual experiences and facilitate a broader view of personal perspectives from individuals. \citet{331lyu2020visual} combine the Data-Insight-Knowledge-Wisdom method (a common method to explain human understanding in the perceptual and cognitive space) with six constitutive factors in Jakobson’s communication model to create a framework for turning data into a story. Their research focuses on a theoretically grounded story creation process by splitting it into two parts: the creation of wisdom (analysis, translation, implementation) and the creation of a story (investigation, representation, development, implementation).

\subsubsection{Structuring the storytelling process}
The early work on defining storytelling within information visualization focused on categorizing formal aspects and definitions (e.g., genres) to create a comprehensive overview of the design space and related strategies~\cite{203segel2010narrative,071jern2010explore}. Based on this inclusive perspective, almost all visualizations would be considered narrative: we could even describe a line chart with a headline, sub-line, and annotations as a narrative visualization (annotated chart). Over time, the definition was narrowed down by emphasizing the sequential dimension (such as a set of story pieces). This delineated exploration from intention and the message of a visualization~\cite{125lugmayr2017serious,149kosara2013storytelling,205chen2010photomagnets,157RodriguezTellingStories,178Botsis2020}. As a result, the storytelling process and its underlying mechanisms moved more into the foreground, calling for a reflection of the process, related tools, and algorithms, as can be seen in the work of \citet{chotisarn2021deep} and \citet{shi2020calliope}. 

\paragraph{Focus on the Data Exploration}
In many cases, the structure of the working process was initially derived from the context itself (mostly in data journalism or visual analytics), whereby exploration was an integral part of the process~\cite{015van2012web, 014ho2011web, 224lundblad2013geovisual, stalph2021exploring}. For example, one of the \textbf{critical elements of exploration in storytelling is identifying and selecting key data features for presentation} (e.g., through snapshots) and emphasizing their contextual meaning (e.g., through captions or annotations)~\cite{014ho2011web, 015van2012web, 041jern2009collaborative, 071jern2010explore,224lundblad2013geovisual} or re-ordering ~\cite{tyagi2022pc}. As a result of this, the core statements are easier to understand in the resulting presentation scenario. The viewer does not have to go through all the necessary steps to identify the key aspects. To facilitate the reproducibility of the discovery process, \citet{080gratzl2016visual} propose the ``CLUE'' approach---an interactive provenance graph that indexes and displays the relevant steps of the exploration process. The resulting transparency of the process directly contributes to the author's credibility. However, a comprehensive understanding of the exploration process usually requires specific domain knowledge, and therefore, this approach might not be applicable within scenarios with a general audience. 
The authors found this particularly the case in business analytics, where users have a task to bridge the gap between raw data and business insights to become a data-driven organization.  To create a data-driven organization, \citet{147BoldosovaStorytellingBusinessAnalytics} propose a conceptual framework with propositions about the relationship between business analytics, data-driven storytelling, and the intention to use business analytics. They state that the intention to use business analytics regularly depends on users' positive or negative experiences while interacting with it and that storytelling contributes to a positive experience. The difference between traditional data interpretation and data interpretation supported by data-driven storytelling is thus reflected in the user's experience while interpreting data and making decisions. 
\citet{258MarjanovicEmpoweringBusiness}, on the other hand, suggests a more pragmatic approach to increase visual analytics skills within the business. A good narrative for the data interpretation journey can help to accomplish that task. While \citet{147BoldosovaStorytellingBusinessAnalytics} recommend that the key to implementing storytelling within business analytics is behavioral change.

\citet{258MarjanovicEmpoweringBusiness} focuses more on empowering business users to develop visual data exploration skills. Using visual stories as boundary objects among primary (developers) and secondary designers (users of visual analytics). Boundary objects need to be co-created rather than exchanged.  

In line with this, \citet{179minelli2014visual} proposed visualizing development sessions to support understanding developer behavior. All authors require users to be involved in the process of creating stories and require them to be inspired by those stories. \citet{ya2021analytical} take another perspective and propose a framework to facilitate business decisions based on analytical reasoning features from three parts of visual analytics representation: higher-level structure, interconnection, and lower-level structure.

Considering multiple provenance paths in parallel can also enhance understanding in various contexts. \citet{diamond2021canadian} developed a dashboard for data storytelling for the cultural sector where provenance was used to identify data sources. The framework used for this dashboard was the four editorial layers defined by \citet{000Hullman2011Framing}. \citet{372yousuf2014constructing} proposed a framework to increase student engagement where a visualization system uses storytelling to present complex data. The main focus of the framework is to construct narratives with multiple exploration paths that are personalized for individual end-users, primarily to support weak students. The authors found that personalized visual narratives facilitate understanding and engagement. Similarly, \citet{park2022storyfacets} propose a system called ``Storyfacets''. The system provides different views of the same analysis session to support provenance exploration to a different audience---experts, managers, and laypersons.

\begin{figure}[htb]
    \centering
   \includegraphics[width=\columnwidth]{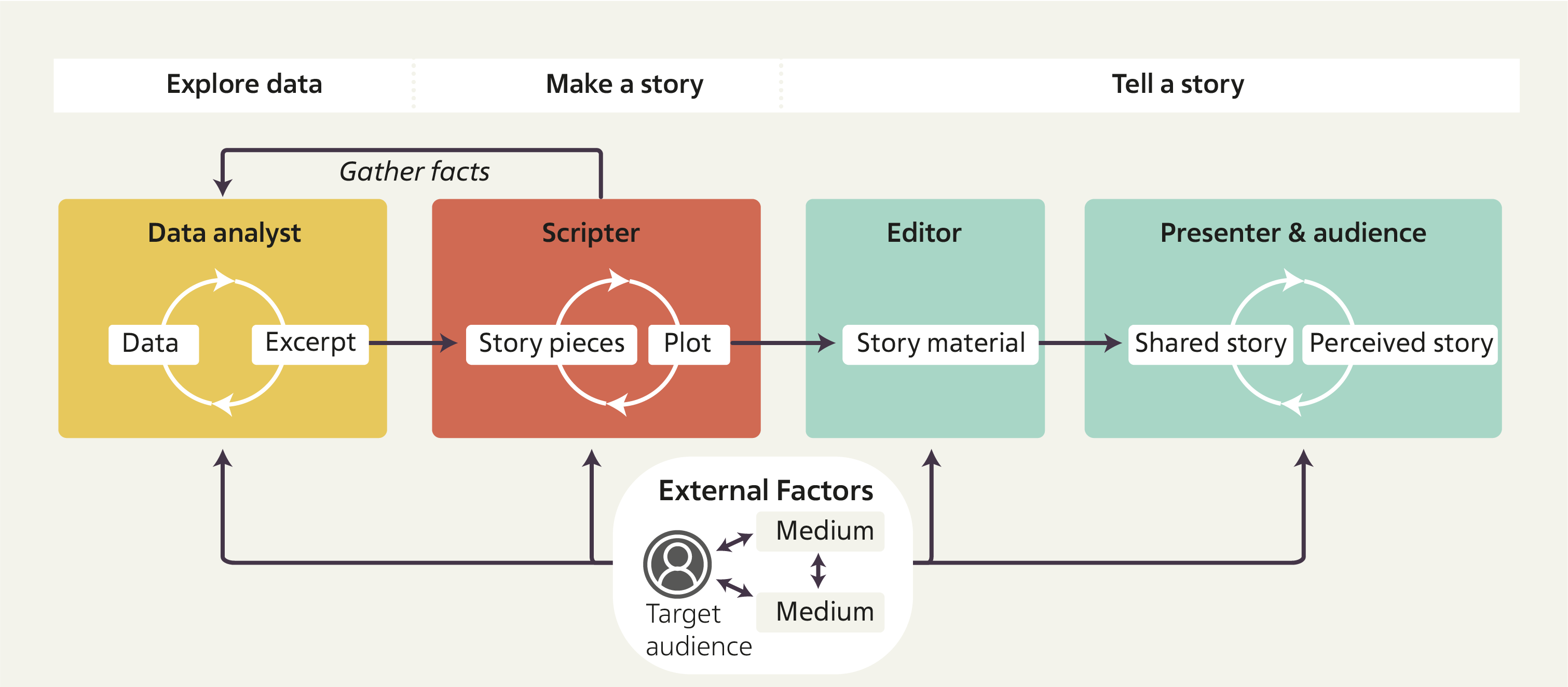}
   \caption{Sequence of events with individual attributes}
   \label{fig:vdsp}
\end{figure}

\paragraph{Focus on mass communication}
\citet{lopezosa2022data} reviewed the current developments, evolving technologies, and challenges in the field of data journalism.
Building on the foundations laid by the previous models within the field, \citet{245lee2015more} proposed a high-level overview: the visual data storytelling process, which integrated all relevant steps (see Fig.~\ref{fig:vdsp}) to communicate data using storytelling effectively. 
In this context, the individual stages of the process, especially exploration and story creation, were considered iterative and not necessarily linear processes where external factors like audience, setting, and medium influence all steps. 
The authors raise an essential concern that the transformation of data to an understandable format might also result in a (un)intended misuse, which leads to relevant ethical questions, particularly in connection with transparency concerning the exploration process, related choices, and the underlying data.

Later, this perspective was extended by identifying seven key features of data-driven storytelling within data journalism, thereby offering a more nuanced perspective on the multifaceted role of data in both analytical and transparency/ethical context ~\cite{050weber2018data} as well as the discussion of narrative techniques and interaction \cite{figueiras2022information}. 

Examining the use of narrative info-graphics by the United States municipal government for public information dissemination, \citet{dailey2022visualization} summarize the design elements of the narrative info-graphics instrumental in fostering transparent communication. These elements included: \textit{use of clear and concise language, the use of visuals to support the narrative, the use of color to highlight key information, the use of annotations to emphasize key observations, and the use of subheadings to organize information.}
\citet{shi2022breaking} overviews six design patterns related to user input and interaction to facilitate engagement, information recall, and subjective connection to the data story. 
Another illustrative example of innovative public data presentation is described by \citet{225de2013new}. Grounded in the principles of accessibility and clarity in the context of European statistics, the authors used information and communication technologies to provide new opportunities for disseminating statistical data. The information is described using interactive graphics, sector-specific glossaries, references to publications, and other auxiliary resources. Another upcoming domain is the field of legal design, aiming to utilize data storytelling to facilitate interaction and understanding with legal information~\cite{polania2022designing}.

Current developments in open data lead to new challenges. \citet{305janowski2019mediating} proposed several data story patterns that can enhance the discovery process of open data. 
\citet{074BrolchainOpenData} investigated the data communication process in an open data context. The authors stated that \textbf{open data platforms should provide storytelling features} to enable users to find and present insights within the data. The authors propose a framework (YourDataStories-YDS platform) that consists of five stages to create a data story that should facilitate users to understand and get to Open Data. First, \textbf{discovery} during which the data set is explored and traversed, including information on metadata, completeness, and highlighting outliers and anomalies. Second, \textbf{assistance}, during which the data is cleaned and processed. Third, \textbf{insight}, which is considered the most important phase, consisting of two main features: explanatory features (such as availability of visualizations and automatic narrative generation) and social features to gain more information to read the dataset (such as engaging with data-owners and discussing with other dataset users). Fourth, \textbf{leverage}, during which conclusions from the analysis are shared and discussed also to estimate validity. Fifth, \textbf{trust} including transparency that privacy and other rights are respected.

\begin{figure}[htb]
    \centering
   \includegraphics[width=\columnwidth]{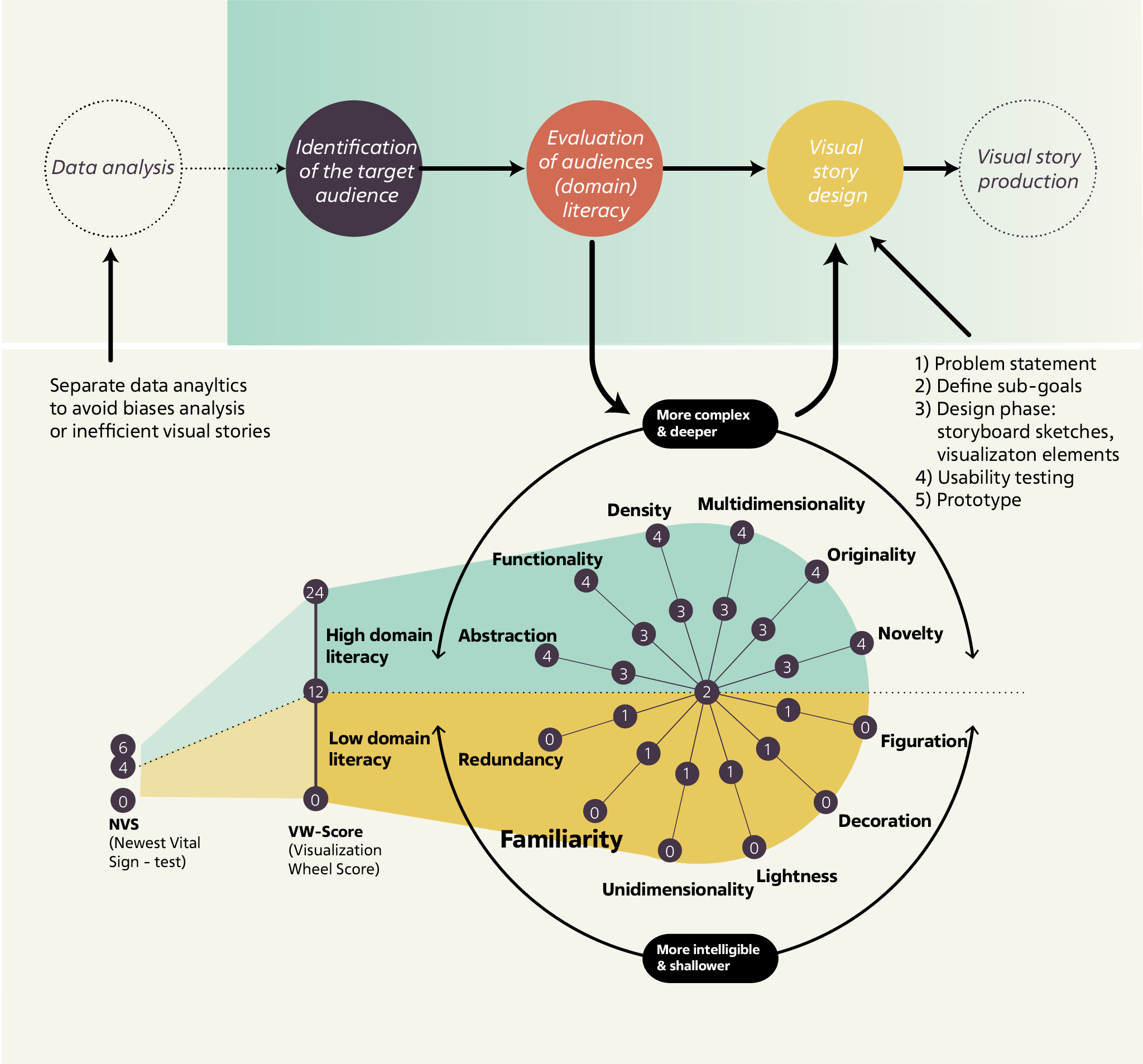}
   \caption{Story creation process within knowledge dissemination according to \citet{178Botsis2020}}
   \label{fig:storygenoverview}
\end{figure}

\paragraph{Focus on Scientific communication.}
Typically, knowledge dissemination in science is a different role of the data exploration process and involves a specific perspective of the (target) audience~\cite{178Botsis2020,296burkhard20184d} as well as design strategies~\cite{rickhaus2022visual}. Analytics and exploration are part of the research that led to the knowledge that needs to be disseminated. Therefore, \citet{178Botsis2020} highlighted the importance of separating exploration and story creation to avoid the risk of biases. These can be influenced by the needs of the narrative or the other way around, resulting in inefficient stories. The major challenge in this application domain is a general asymmetry of knowledge and (domain) literacy between author and reader. Inspired by the general concept of Aristotle's ethos, pathos, and logos, the authors proposed to start with the target audience and understand the user by analyzing (domain) literacy and interviews, defining goals and sub-goals to support the clarity of the created story. To facilitate the implications of the audience towards design choices, they adapted Cairo's visualization wheel and combined it with the Newest Vital Sign (NVS) score to help story creators with design choices (see Fig.~\ref{fig:storygenoverview}).

To provide an illustrative example for this concept, \citet{fernandez2022beyond} point out that traditional learning analytics dashboards pose a significant challenge in interpretation due to the often limited data literacy of their primary users, which includes teachers and students. To address this issue, they propose using alternative ways to communicate data insights using visual narrative interfaces.
\citet{rickhaus2022visual} emphasized how visualizations enhance scientific storytelling and lateral thinking, helping the readers understand complex information. He also recognizes the effective use of visual elements, such as color, shape, and layout, as strategic tools for emphasizing critical data points and directing the reader's focus toward key information. Consequently improving the reader's engagement.
\citet{296burkhard20184d} introduced a visual storytelling design board to align design choices with literacy considerations. They integrated the external factors as a dedicated stage within their process. Data and conclusions already existed in their scenario as the pre-existing research findings. They created a user scenario narrative to understand better the individual steps, narrative structure, audience's perspective, and related expectations. Subsequently, the resulting findings were transformed into mock-ups for further iterations together with users through visual and interactive prototyping~\cite{17saini2019aesop,296burkhard20184d}.

\paragraph{Focus on Collaboration}
The story creation process can also directly involve users and their individual views on the data. Collaboration can involve human-machine interaction supporting the collaborative editing process ~\cite{sun2022erato}.
Several authors proposed to design the exploration and story creation phase as a collaborative process of multiple users~\cite{340jiang2020data,52jiang2019data,309axelrod2019intergenerational}, while modeling personal family migration narratives with data. In user studies, they analyzed the behavior of several family groups and identified ten different socio-technical practices of data wrangling. They developed a conceptual model of the data wrangling process, differentiating the individual and global/data perspective.

\paragraph{Focus on the user's conceptual model of time and space.}
Geospatial visualization has a long tradition of developing visualizations in a spatiotemporal context, focusing on the idea of incorporating the space and time dimensions into narrative visualizations. The commonly used phrase in storytelling---``Once upon a time in a land far away''---elucidates the relationship between the dimensions of time and space and storytelling~\cite{114rodrigues2019once}. \citet{114rodrigues2019once} emphasize the aspect of employing spatio-temporal dimensions when creating interactive visualizations by providing a set of guidelines. Agreeing with the concept of the connection of narratives and spatial dimension resulting in spatial identities that, in turn, contribute to the shaping of the story, \citet{087caquard2014can} designed an application to explore the geographic structure of the story and to perceive the impact of stories in the creation of places. They achieved this by developing a map of contemporary Canadian films by characterizing the spatio-temporal dimensions of narratives. The dimension of spatio-temporal narratives leverages strongly in geographic visualization systems. In describing geo-data, the spatial dimension can represent the spatial relationship between the data entities, and the temporal dimension would illustrate the changes in the data over time. The additional information about the data could then be represented in a thematic dimension~\cite{144thony2018storytelling}. \citet{144thony2018storytelling} explore storytelling techniques in 3D geographic visualizations to help conceptualize geospatial data over time. Visualizing geodata in a three-dimensional manner provides the opportunity to make it interactive and present the information in a captivating and intuitive way. 
This way of presenting geo data motivates the otherwise overwhelmed user to explore the relevant information more effectively. They discuss using various components in the 3D maps that help tell data stories. Story maps are built by considering whether the data belongs to spatial, temporal, or thematic dimensions. For data storytelling, storylines follow a theme or a person from a  data-related perspective. Geographical 3D maps have exploited the use of storylines to provide a better immersive experience of the story to the users. Storyboards and scene components are the remaining essential elements that \citet{144thony2018storytelling} discuss, which can be used to create interactive stories. Another example where location mapping has been used for creating effective visualizations is the work done by \citet{141chaudhary2019storytelling}, where they use geographical visualizations to present the citizen complaints in India to the responsible authorities to help in decision making. 
Along the same lines, \citet{lan2022data} introduce a new web mapping platform where users can tailor their own story maps and effectively identify geographic patterns of social and economic phenomena through narrative mapping. This would prove to be a great tool for social scientists and policymakers

\subsubsection{Automated systems to structure the storytelling process}
\paragraph{Automated visualization recommendations and reasoning.}
The evolution of data-driven storytelling has generated a demand for automated tools to assist story creation. Upcoming genres have led to the development of different storytelling forms, including DataToon~\cite{053kim2019datatoon} and Timeline Storyteller~\cite{273brehmer2016timelines} and others designed to effectively communicate data or visualization and videos~\cite{29amini2016authoring,kim2017visualizing}. 

In the field of automated systems specifically for visualizing data in quantitative research, we found one paper in our survey by \citet{choudhry2020once} who presented CAUSEWORKS, which is a textual narrative system that uses visualization and text to describe causal data. The system provided narratives summarizing data changes and projecting trends. Although we found many works focusing on specific types of data based on the domain it was collected from, for example, \citet{342chotisarn2021bubble} developed a prototype to support bubble chart animations based on Twitter data. The tool automatically creates animations and facilitates authoring through captions and filtering functionalities. Another example of this type of data presentation is Rosling's presentation of economic data~\cite{roslingTed}. This inspired \citet{shin2022roslingifier} to design a semi-automatic storytelling system for data presentation called Roslingifier. The system provides three views supporting data presentation with auto-detected events based on storytelling techniques such as natural language narratives, visual effects highlighting events, and temporal branching. To summarize time-varying data in a comprehensive narrative, \citet{161bryan2016temporal} propose ``Temporal Summary Images'' (TSI), that automatically identify points of interest through computing noteworthy changes (e.g., strong increase/decrease) to recommend potentially relevant data features. 

We also reviewed several works on data videos. For example, \citet{shi2021autoclips} propose Autoclips for automatically generating data videos when a sequence of data facts is given as input. They construct an algorithm that creates videos that have comparable quality to human-made videos. The work of \citet{346shu2021dancingwords} adds Word Clouds as an authoring tool that interactively crafts word clouds and animations to generate storytelling videos.

\paragraph{Automated sequencing.}
Along with efficient visual encoding and identifying key aspects, ordering events is crucial in developing data-driven stories. The section narrative structures (see section~\ref{sec:narrative}) elaborates extensively on theoretical aspects of sequences and narrative structures in general. In comparison, this part gives an overview of sequencing support tools. \citet{008hullman2013deeper} have significantly contributed to the field by researching automatic sequencing in narrative visualization. The underlying idea is to look at a story as a series of views connected by transitions. The authors incorporated the transition costs and a global weighting to calculate the most efficient transition type. As a result, the GraphScape method was created to illustrate the theory~\cite{kim2017graphscape}. Later, this approach was combined with visual encoding support. \citet{009obie2019framework, obie2022gravity++} proposed a framework for logically ordered data stories and the related tool ``gravity''---built on Vega lite for visualization support and recommendation for effective narrative ordering based on Graphscape supporting collaboration and presentation~\cite{030obie2020authoring}. The DataToon tool~\cite{053kim2019datatoon} also suggests automatic transitions and panel recommendations for dynamic network data comics, which support narrative ideation and storyboarding. Another approach by \citet{321Walsh2019Modelling} used an algorithmic method to find stories in large data lines. This technique is dynamic by allowing for displaying and collapsing the timeline. Focusing on the provenance of the exploration process, some authors~\cite{343Barczewski2020Storyline, 345Cao2020Sequence} created a machine-learning model to develop storylines for data visualization users. The model learns from past user exploration of the data set to build a preferable storyline. For example, the algorithms create a drill-down sequence based on the user's exploration habits~\cite{345Cao2020Sequence}. An archival library study took a similar approach by adapting the narrative sequence to the user input~\cite{075Battad2019Library}. Looking at the individual story nodes,~\citet{313zanda2019technological} used the Story Network Principle to generate a tree of story paths for data visualization. Overall, these techniques have been shown useful as they allow for customization to individual user preferences.

\subsubsection{Tools to create data stories}
\begin{figure}[h!tb]
    \centering
   \includegraphics[width=\columnwidth]{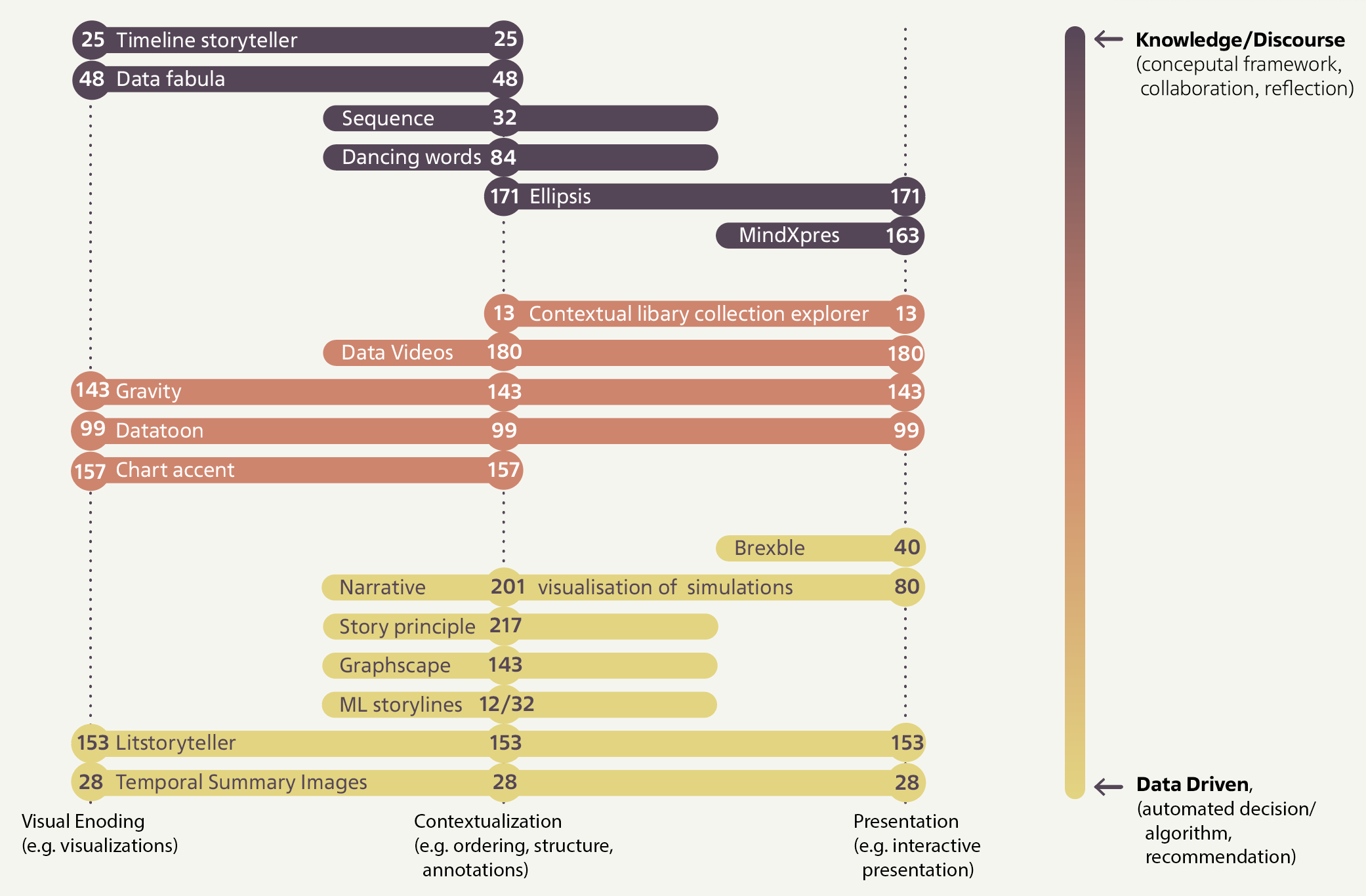}
   \caption{Story Generation Tools: the number refers to the reference number}
   \label{fig:StoryGeneration}
\end{figure}

Tools may help the user by choosing efficient visual encodings, by choosing accurate contextualization, or by simplifying the process of presentation.
In the following, we present the three categories of tools that we found in the articles we reviewed (see Fig.~\ref{fig:StoryGeneration}). 

The first set of tools generally tries to help users \textbf{explore and present} their data. As one example, \citet{271roels2016interactive} proposed a set of general requirements and introduced MindXpres, a presentation platform that combines interactive data exploration with storytelling. To help the general audience gain insights and have a better understanding of time-series data, \citet{089lu2020illustrating}
proposed an interactive authoring tool that converts 2-D time series to data videos. To detect changes in time-series data and illustrate them in video sequences, the authors employ several algorithms to complete tasks that otherwise would have required much effort if done manually. 
Another category includes tools that help users annotate existing data visualizations and present them as visual data stories on the web. 
Typical tools from this category allow importing existing visualizations and help the user to design scenes and annotations for parts of the story. 
Toolkits come in a variety of feature sets. Tools like ChartAccent 
provide simple tools to annotate figures from data~\cite{39ren2017chartaccent}. 
Tools like Ellipsis~\cite{31satyanarayan2014authoring} additionally provide multiple tools to generate different narrative structures, and tools like Flourish provide fully-fledged data exploration and visualization software with a focus on data-driven storytelling. 
The last category contains tools designed to automate the full process of data-driven storytelling, typically with the help of visualization. Tools in this category use various algorithms to find meaning in the data (e.g., unsupervised machine learning) and automatically pick or recommend ideal narrative structures for the given data. As an example \citet{206cruz2011generative} have developed a conceptual framework for telling a story from data, introducing Data Fabulas, ``the set of events, agents, actions and chronology extracted from a dataset'' with the idea of enhancing cognition about information.
\citet{105ping2018litstoryteller} proposed an interactive system that helps researchers understand a field of research from the scientific literature. Their system is designed to be used as supplementary information for systematic reviews to not only grasp the overall research trends in a scientific domain but also get down to research details embedded in a collection of core papers. Their tool supports interactive visual storytelling at multiple levels. It allows for answering various research questions, covering macro-scale and micro-level questions. 
As entities of investigation, the authors use concepts or terminologies gathered by employing various text-mining methods and then map them to basic visual elements to form visual storylines. We have visualized tool examples in Fig.~\ref{fig:StoryGeneration}.

\subsection{Narrative structures}
\label{sec:narrative}
To further understand how we can utilize storytelling to enhance our understanding of quantitative information, we first need to understand the building blocks of a story. The core components of storytelling are the collection of events and their temporal and causal relationship that we call narrative structure. Relevant questions are: How is the information organized? What different narrative structures are there, and what goals are pursued by them? 

\begin{figure}[htbp]
   \includegraphics[width=\columnwidth]{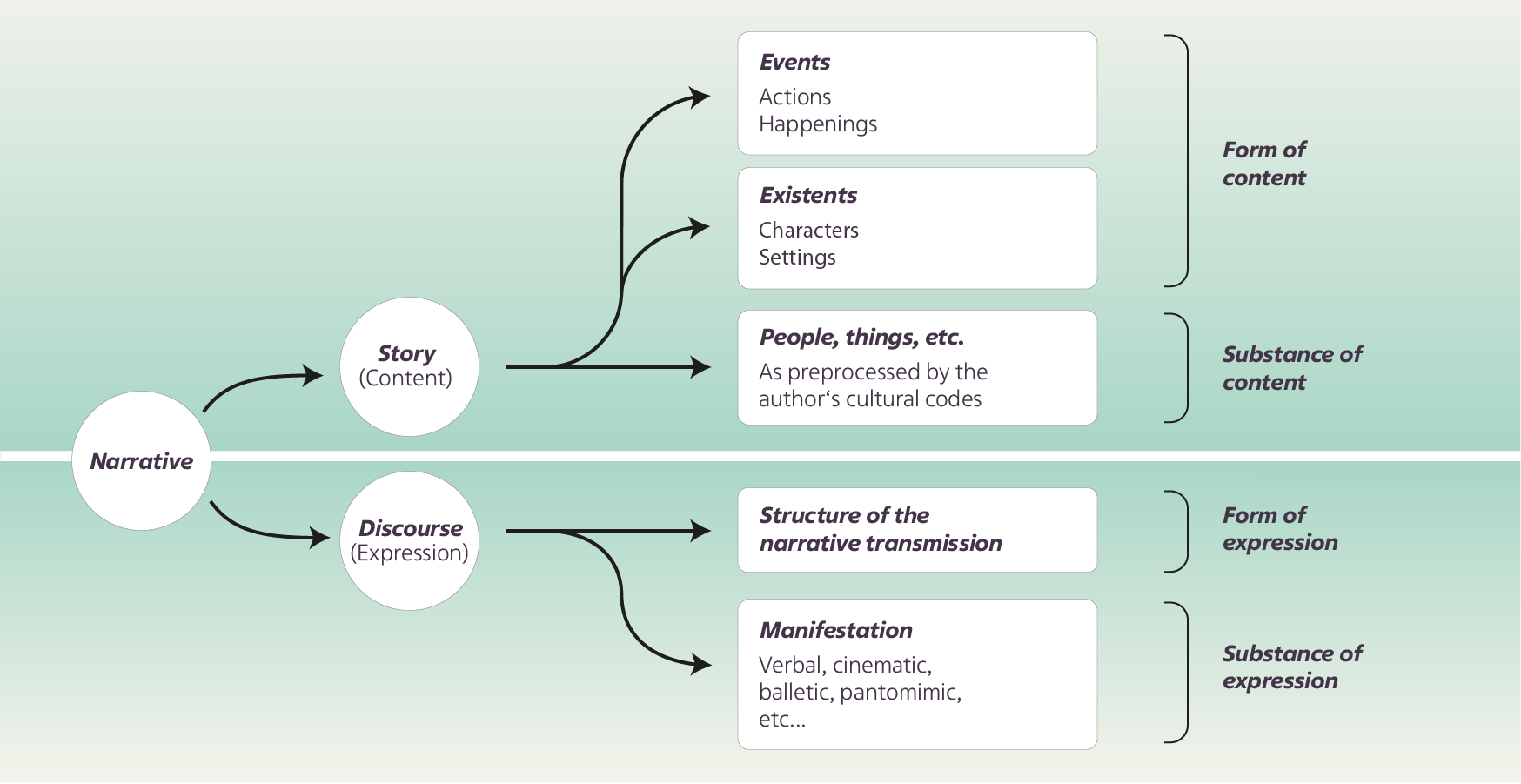}
   \caption{Narrative structures according to Chatman ~\cite{000Amerian2015KeyConceptsNarratology}}
   \label{fig:structure}
\end{figure}

In traditional narratology, \textbf{a narrative has two components} (see Fig.~\ref{fig:structure}): \textbf{the story} (relating to the content, what is told?) \textbf{and the discourse} (relating to the expression, how is it told?)~\cite{000Chatman1980StoryAndDiscourse}.
While these refer to texts or films, visualizations are based on a different narrative structure, namely the space in which the story can be presented~\cite{000Cohn2013VisualNarrativeStructure}.
A fundamental difference lies in the additional dimension created by the data representation and has implications for layout and the temporal structure~\cite{273brehmer2016timelines}. 

\subsubsection{Existents}
If we look at the formal content, we can consider a story a collection of existents and events. Existents refer to actors/characters and settings, while events include multiple dynamic entities such as actions and happenings that shape the causal relationships~\cite{000Amerian2015KeyConceptsNarratology}.

\paragraph{Characters} Characters are a fundamental building block of classical stories, but how do they behave when data plays a central role? Multiple articles have mentioned the need for a sequence with a clear protagonist and other characters for data stories via data visualization~\cite{135arevalo2020storylines, 170zhao2019understanding, 246nardi2016form, 121ma2011scientific, 328roth2020cartographic} or data videos~\cite{076hook2018facts}. The data in data visualization can be presented as characters in a story~\cite{328roth2020cartographic}. Additionally, a character can be separated from the data by presenting an animated agent with audio~\cite{108hilviu2015narrating} or without audio~\cite{319blackmore2019evaluating}. When the data stories include personal history data, the user him-/herself can also be seen as the character or player~\cite{256jung2017serendipity}. The user can also become an actor in the story through interactivity, allowing the user to steer the story~\cite{121ma2011scientific}. \citet{328roth2020cartographic} indicates that data stories in geography can have a protagonist and an antagonist. According to the characters, different structures for the story can be shaped. Similarly, \citet{257bounegru2017narrating} illustrated five narrative patterns in network visualizations which were created around characters, either single actors, multiple actors, entire organizations, or the association/opposition of multiple characters. The two articles illustrate a rare example of how traditional storytelling with the use of characters is applied to data visualization. While articles mention the use of characters, there is no consistent notion of what is and what is not a character, 
nor how they can be used. The knowledge is due to a limited academic review of used characters in data stories by practitioners.

\paragraph{Settings}
Data-driven storytelling settings can be used to place the visualized data into a particular spatial and temporal context. The context can assist in characterizing a story to create a purpose~\cite{108hilviu2015narrating}.  It can be set by the story itself or by the environment of the user~\cite{256jung2017serendipity}. This can be realized in multiple ways, e.g., for timeline data visualization, the scale can be used and manipulated to place the data visualization in a context~\cite{273brehmer2016timelines}. The path of the timeline can also be formed and shaped to represent the story's setting. The use of maps is an example of spatial settings~\cite{328roth2020cartographic}.
   
\subsubsection{Events and sequence of events} A storyline consists of a sequence of events. The basis for this is establishing a logical connection that enables the user to interpret the individual events as a coherent overall plot. During a story, multiple storylines can emerge in parallel and work up to a conclusion, where all the storylines of the plot are drawn together. This can also relate to the message~\cite{245lee2015more}. 

\paragraph{Structure} In data storytelling, the terms exploratory and communicative (or explanatory) are taken as the general way to structure a story~\cite{157RodriguezTellingStories}. However, it can be argued whether the two types are purposes instead of narrative structures. A clearer insight into narrative data story structures is given by \citet{203segel2010narrative} who argue that there are three different structures: linear, exploratory, and a combination thereof~\cite{203segel2010narrative}. The latter refers to hybrid linear-nonlinear patterns, for example, a Martini Glass structure where the story starts explanatory and afterwards opens into an exploratory panel. The three structures have been the basis of many data storytelling studies, such as those by \citet{050weber2018data} and \citet{157RodriguezTellingStories}. Other studies have shared similar findings stating that data stories can be structured as guided, guided exploratory and exploratory-viewing~\cite{264wang2016guided} or information-seeking, comparative, and iterative for scientific storytelling~\cite{121ma2011scientific}. 
Parallelism can be seen as a fourth structure~\cite{008hullman2013deeper,139tong2018storytelling}. The structure tells the story with repetitive sequences. The parallel structure is explored for timeline visualizations by \citet{273brehmer2016timelines} who state that the stories can be unified, faceted, segmented, as well as faceted and segmented. 

Although the general structures give little insight into the sequence of the story, commonly data stories use temporal or chronological sequences~\cite{108hilviu2015narrating}. The plot of a story with a (partial) explanatory structure can be told in a different sequence.  Besides, the three-act structure is referred to create data stories~\cite{356concannon2020brooke,328roth2020cartographic}. The three acts include various narrative elements such as setting, hook, and plot twist~\cite{108hilviu2015narrating,328roth2020cartographic}. \citet{341lidskog2020cold} frame the three acts in the historical account (the what, the why) and what to do about it. The narrative moves beyond a `story with a morale' by inciting action on the audience's side. Often, these narratives can be taken apart further, as illustrated by \citet{121ma2011scientific}, who states that traditional stories develop as follows: know the audience and their knowledge level, set the stage, character introduction, develop plot, show relevant story, and implication for the reader. Laurel refined Freytag's triangle, which is referred to by \citet{144thony2018storytelling}. The paper states that tension is created by the phases of exposition, (trigger) incident, critical action, and crisis to a climax. Then, the (re)solution occurs, and the plot ends in a relaxation phase.

\paragraph{Elements} The central element of the story-narrative should be the problem from which the setting, purpose, agent, acts/events, and means/helpers depart. \citet{135arevalo2020storylines} follow the theory of \citet{000Jones2017ScienceStories} and \citet{000Murray2014Narrativeresearch}. Their papers adapted the central elements from \citet{000ElShafi2018}: connecting with your audience, raising problem awareness, relating to a practitioner's world, acknowledging remaining challenges for practice, and giving a take-home message. 
\citet{169amini2015understanding, yang2021design, steinert2022mobility}, expands further on the theory by Freytag~\cite{144thony2018storytelling, 169amini2015understanding,000freytag1894freytag} as adjusted for visual narratives by \citet{000Cohn2013VisualNarrativeStructure} by examining dramatic structures for data videos and proposing guidelines ~\cite{yang2021design}
\citet{000Cohn2013VisualNarrativeStructure}'s definitions for \textbf{the four story blocks are Establisher, Initial, Peak, and Release} (E, I, P, R). 
Patterns can be examined in data stories to create narrative categories by the following method [Element+] where Element is one of {E, I, P, R} and the ``+'' sign indicates repetition of the preceding element. 
According to \citet{169amini2015understanding} ``E+I+PR+'' pattern was the most common category, as well as ``E+I+P'' and its subset pattern of ``EIP.'' 
So far, the study by \citet{328roth2020cartographic} is the only study that has mapped several sequences of events that reflect traditional story plots. \citet{328roth2020cartographic} states that for geographical visualization, eight different structures can be used: destruction, genesis, emergence, metamorphosis, cause \& effect, convergence, divergence, and oscillation. These are closely related to the development of the different characters in the stories. \citet{246nardi2016form} describes the approach of re-storying through which a story is constructed from original pieces of data involving the creation of characters, settings, and events. Here, it is not solely recounting a sequence of events, but re-storying is about composing a new story through evaluating and interpreting.

\subsubsection{Representation of a story}
\citet{036behera2019big} briefly touch on the potential for data stories to differentiate themselves in their stories' purpose and the targeted audiences. Other articles agree that narrowing down the audience for the story is important~\cite{135arevalo2020storylines}. However, they fall short of explaining different audience groups for data stories. Regarding the purpose of the data story, it is often the story's topic or main message, which can be explained in a few sentences~\cite{256jung2017serendipity}. \citet{050weber2018data} states that three purposes for data stories exist: to tell, explain, or argue visually. The article bases the definition on theoretical considerations of traditional narratives.

\citet{116ojo2018patterns} took a different approach and defined seven types of data stories after examining 44 winning projects, namely reveal information of personal interest, enable a deeper understanding of a phenomenon, reveal anomalies and deficiencies in systems, track changes in systems, refute claims, reveal information about an entity in increasing levels of details, and reveal unintended consequences. Similarly, \citet{257bounegru2017narrating} states five narrative view types: Exploring associations around single actors, detecting key players, mapping alliances and oppositions, exploring the evolution of associations over time, and revealing hidden ties.

\paragraph{Narrator} The narrator of the data storytelling influences the narrative structure. Three types of narrators have currently been reviewed in data storytelling: the designer, the user, and the visualized character. A few papers have studied the narrators~\cite{328roth2020cartographic,000zhao2015data}. The user can guide the story in interactive visualizations~\cite{121ma2011scientific}.
\citet{000Heyer2020} took it further and allowed the user to act as the narrator by stating prior beliefs. Designers are narrators by framing the story~\cite{320Cunningham2019ProvenanceNarratives,000Hullman2011Framing} such as using annotations~\cite{320Cunningham2019ProvenanceNarratives}. 
Lastly, the data visualization itself can be the narrator by using voice narrations~\cite{319blackmore2019evaluating} or animated agents that narrate through voice~\cite{108hilviu2015narrating}. \citet{000Hullman2011Framing} illustrate various design rhetoric and viewing codes that the designer can use to emphasize a specific message or guide the structure of the story. Moreover, the presentation structure of data videos is another influence the narrator can have~\cite{076hook2018facts}.

\paragraph{Genres}
In chapter~\ref{visualization-types-and-areas-of-application}, the data visualization types (or genres) for storytelling have been discussed. As aforementioned, the prominent research of \citet{203segel2010narrative} has mentioned key genres that have been saturated by other authors, pointing out new narrative visualization genres. Similarly, \citet{328roth2020cartographic} points out seven genres (static visual, long-form infographic, dynamic slideshow, narrated animations, multimedia, personalized story map, and compilations) specifically for geographic maps. However, it is non-conclusive if these time and geographic layouts can be used for other visualization types. \citet{244michel2015snow} discussed a multi-media long format through which the narrative can continuously expand by adding new exhibits to the structures. At the same time, the genre uses specific codes that allow customizing the narrative to different formats and media. Similarly, \citet{295seyser2018scrollytelling} coined the term \textit{Scrollytelling} adapted from the long-form articles used in journalism. Again, the authors highlighted the vertical aspect along with multi-modality. As opposed to \citet{244michel2015snow}, the \textit{Scrollytelling} genre does have a definite endpoint. \citet{289majooni2018eye} took a different perspective and analyzed the performance of different layout designs. The natural left-right (up-down) eye movement could improve comprehension of infographics and data visualization. Other research has applied these types of genres, such as \citet{318metoyer2020supporting} who took the drill-down genre of \citet{203segel2010narrative}. In addition, the medium can also be seen as the structure of the story in the spatial layout. \citet{170zhao2019understanding} illustrated that data visualization stories could be partitioned and sequenced into data comics to create a meaningful order in the visual space. Common data stories such as long formats or slideshows are designed vertically or horizontally in the visual space. On the other hand, \citet{273brehmer2016timelines} constructs four genres for timeline visualizations: linear, radial, grid, spiral, and arbitrary. These visual flows can help outline the story structures for data visualization.

The genre of a narrative gives structures to the overall type of story. The term genre would refer to romance, action, comedy, or documentary in classic movies or story books. Data storytelling uses similar genres such as scientific~\cite{341lidskog2020cold,121ma2011scientific}, geographical~\cite{328roth2020cartographic}, and biography data storytelling~\cite{273brehmer2016timelines}. \citet{076hook2018facts} found that nine genres were used for data videos: advertisement, comedy, cookery, documentary, drama, horror, music, science fiction, and unclassified.

\subsubsection{Combination of the discourse and story}
A few papers highlight the combination of narrative structure techniques used for the discourse and story for the following genres: timeline visualization~\cite{273brehmer2016timelines}, interactive data videos~\cite{076hook2018facts}, spatio-temporal visualizations~\cite{114rodrigues2019once}, and geographical visualizations~\cite{328roth2020cartographic}. 
Other research has looked into visualizing story plots~\cite{133Qiang2016StoryCake,358Willemnsen2020Storyworlds}. For example, \citet{133Qiang2016StoryCake} visualized the story structure of the movie Inception, which has a multitude of narrative layers.
In Human-Computer Interaction, the generation of visual storylines is closely related to computer games and educational applications. This type of research has focused on creating story paths for interactive environments. Accordingly, algorithms generated relationships between entities and visualized the different storylines for one narrative in a tree-like graph~\cite{320Cunningham2019ProvenanceNarratives,190riedl2006linear}. The underlying story path structure is often based on established theories such as Rhetorical Structure Theory~\cite{320Cunningham2019ProvenanceNarratives}.

\subsection{Data story types}
 In this section, we propose a classification scheme for individual framework types to help future researchers within their respective research domains:

\begin{figure}[htb]
   \includegraphics[width=\columnwidth]{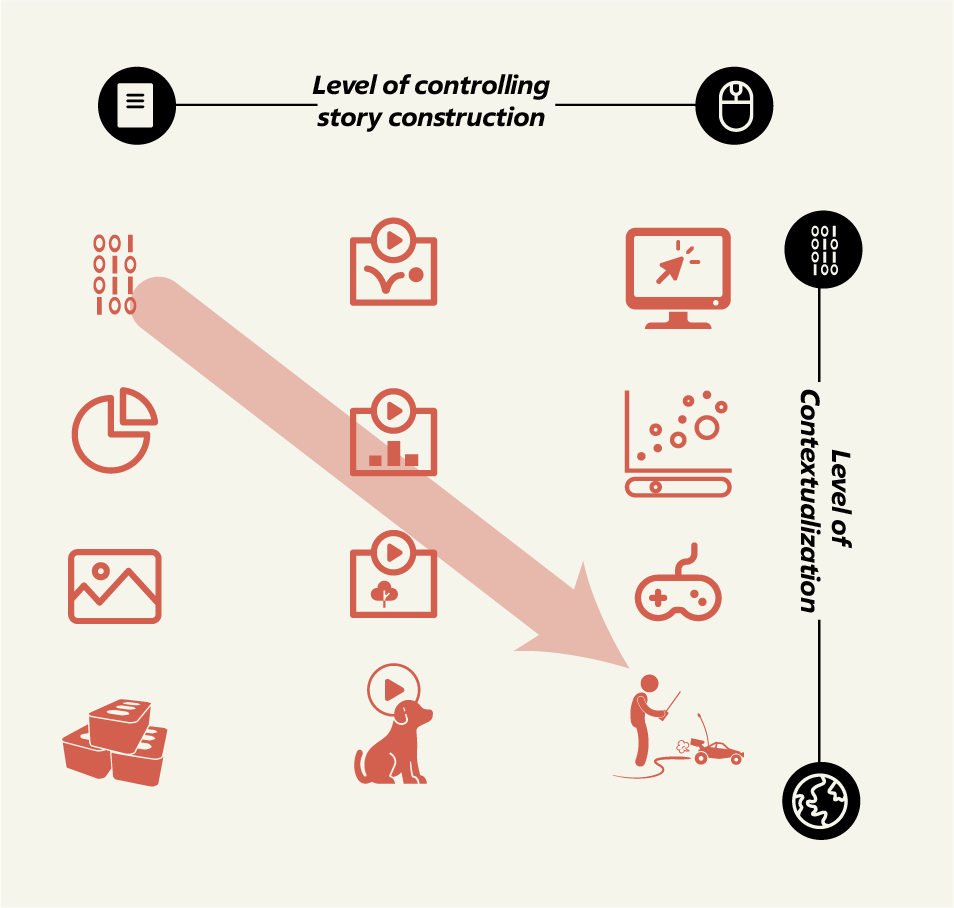}
   \caption{Overview of the individual level of contextualization from raw data on the top left to the real or virtual world on the bottom right. }
   \label{fig:visualizationtype}
\end{figure}

Within the conceptualization process of a story, we can distinguish between two dimensions: the level of contextualization and the level of control within the story construction.
The first one refers to the semiotic distance of the visual language utilized to represent the data through perceivable means, while the latter one refers to how the temporal and causal structure of the story are constructed by the user.

Contextualization helps to understand what the data stands for (the meaning), while story control can facilitate understanding the underlying concepts and relationships (the related concept).
While the author-driven and reader-driven approach introduced by \citet{203segel2010narrative} describes data storytelling from the perspective of information visualization,  we describe data storytelling from an interdisciplinary perspective of human data interaction, narratology, and psychology. 
Here, we look at the amount of contextual information and the level of control and how they influence the internal construction process of the final data story.
We view the information presented in a data story as a combination of implicit and explicit knowledge aiming to explain the meaning of the data, where the level of contextualization plays a major role. 
Depending on the amount of contextual information in the data story, we categorize the data stories into four categories: stories presenting the literal information in the data, stories that use visualizations that represent the data along with textual explanations, stories that use metaphors representing the contextual information along with the information about the data and finally physically or virtually representing the data in a real/virtual world, where the contextual information is highlighted even more. 

As described in the previous section \ref{narrativeConstruction}, a story is constructed as a combination of events with several attributes that are temporally and causally connected. Controlling the order of events and other attributes can impact the story. Hence the level of control can be determined based on how many attributes are controlled directly and indirectly and how they are presented temporally.
Looking at the users' perception, how story construction is controlled is another dimension that can generally be designed for any level of contextualization. 
While perceiving a narrative construct, we distinguish between three individual levels: 
\textbf{a static presentation}, where the user can experience and explore the information solely with his eyes (e.g., an infographic, a data comic, or an annotated chart)
\textbf{a dynamic presentation} where the temporal structure of how the individual story blocks are perceived is controlled (e.g., a data video or animation)
\textbf{an interactive presentation}
There, the temporal structure and possibly also the content can be manipulated by the user 
(e.g., an interactive data story, see Fig. \ref{fig:visualizationtype})
This section gives an overview of the used visualization and data types within specific categorization schemes.

\subsubsection{Low Context - Verbatim}
\paragraph{Annotated charts} Annotated charts combine visualizations by additional forms of communication to emphasize and explain specific parts of the dataset. Textual and graphical annotations are often used in visualizations to highlight points of interest and draw readers' attention to them. Examples of these additional forms have been given by \citet{149kosara2013storytelling}, such as written text, audio, video, or links to more information. They also point out that guidance of the reader can be done via visual cues like arrows or other methods. \citet{31satyanarayan2014authoring} indicate this as being base-level annotations. 
Although annotations are often positioned statically, annotations can be made dynamic by binding them to data points~\cite{31satyanarayan2014authoring, latif2021kori}. Annotated charts seem to be well connected by creating empathy, derived from \citet{146peng2017storytelling} and \citet{235figuerias2014narrative}. We also found that annotated charts are used above average for enhancing time series data.

Interactivity, in general, can enhance the contextual dimension of data~\cite{041jern2009collaborative,234figueiras2014tell}. By allowing the users to interact with the data and visualizations, interactions provide a sense of control to the users to explore the data and thereby understand the context better. At a basic level, interactions like slide transitions, click and zoom, or context menus are used to provide additional information on the data. 

\subsubsection{Medium Context - Narrative Visualization}
\paragraph{Infographics} Infographics, or information graphics, combine graphics, images, and text in one visualization, are seen as an efficient way to communicate complex data, information, or knowledge~\cite{295seyser2018scrollytelling}. Since Charles Minard's work in 1858 and 1869, infographics have been widely used. \citet{330jacob2020visualising} report an increase in the use of infographics in journalism since the 1980s, as they often have replaced photographs. In the analyzed literature, there have been several categorizations of infographics. \citet{otten2015infographics} have defined three main categories (data graphics, maps, and diagrams). In comparison, \citet{albers2015infographics} recognizes four categories (simple infographic, snapshot with graphic needs, complex infographic, and information flow/process). \citet{295seyser2018scrollytelling}  argue that because of the strength in visualizing complex data, infographics also form the basis of Scrollytelling, which we discuss in this article as interactive visualization. 
\paragraph{Posters.} In the articles we reviewed, posters were hardly used. \citet{161bryan2016temporal} investigated creating posters using time-varying data to identify and visualize points of interest and create data stories from them. He combined a time-based visualization extended with comic strip-style data snapshots of relevant steps and annotations in his result.

\paragraph{Slideshows}
\citet{149kosara2013storytelling} describe slideshows as an efficient way of storytelling to visualize and explain data, not to analyze, whereby the data is presented in a linear, controlled way. Their primary focus is on slideshows for large audiences. They argue that this form of storytelling is missing interaction, considered one of visualization's most important aspects. They also stress that interaction facilitates faster and more practical data analysis because of the reader's ability to change the view and potentially add different data quickly. This would define slideshows (and data-videos) as being author-driven rather than reader-driven~\cite{118alexandre2016promoting}.  Schroeder et al. proposed automated slideshows to explain financial data~\cite{schroeder2023show}  utilizing abstract~\cite{schroder2022pension} and metaphorical~\cite{schroder2022rethinking} representations. 
However, \citet{203segel2010narrative} emphasize that slideshows work well to make complex datasets and narratives accessible. 



\paragraph{Interactive visualizations} Interactive visualizations recently became increasingly popular, especially in journalism~\cite{218george2013storytelling}. The concept of interaction (as applied in computer games, for example) differs significantly from what we associate with a classical narrative, which is traditionally told without interaction, which helps to keep focus in a storyline~\cite{149kosara2013storytelling}. 
A non-linear interactive data story, on the other hand, is characterized by a stronger focus on free exploration and has been enabled by open and accessible large-scale datasets and interactive data visualization tools. 
Interactive non-linear narratives seem to be supported by the increasingly growing role of data and the growing preference for using data visualizations in the public news media as these tools are powerful for enriching narratives about data-related topics~\cite{203segel2010narrative}.
Terms often used in interactive non-linear contexts are data wrangling and collaborative storytelling. \citet{340jiang2020data} examined the learning opportunities related to data wrangling practices, such as when social data scientists wrangled data and created models to explain social changes. \citet{214Wong2011Collaborative} make a connection with collaborative storytelling to emphasize that storytelling is being used for fostering creativity, using common techniques like brainstorming, elaboration, and associative thinking. In alignment with these principles, \citet{hasan2022playing} employ interactive data comics as a means to enhance the co-design process.
Then, dynamic interactive narratives change the perspective of the user from being a passive (reader) into an active (meaning-maker and actor). By adding user experiences into the process of transforming author-driven narratives to reader-driven, a story can also be structured by dynamic factors such as the data or individual characteristics of the user. \citet{302obie2019study} conducted a confirmatory user study to compare author-driven narratives based on presentation videos in relation to interactive visualizations in terms of comprehension and memorability. They concluded that users prefer interactive visualizations based on accuracy but found the data in the presentation videos is easier to understand. In this context,~\citet{shi2022breaking} identify six Breaking The Fourth Wall (BTFW) design patterns, a technique of integrating interaction with narrative to increase reader engagement. Upon conducting a user study to assess their benefits and concerns, the authors concluded that BTFW interaction improved self-story connection, user engagement and information recall, while raising concerns about balance, privacy, and learning.

\subsubsection{Medium/High Context - Metaphorical Visualization}
\paragraph{Data comics.} Data comics are an upcoming genre within data-driven storytelling. They combine the freedom of 2D spatial layout presented in infographics and annotated charts with the linearity of narration found in videos and live presentations, enabling readers to consume the story at their own pace~\cite{279bach2018design} and providing high interactivity and interoperability~\cite{236chu2014optimized, wang2021interactive}. 
Although the term comic strip seems to exclude scientific applications, this genre has been used for uncovering dynamic networks~\cite{155bach2016telling}, trying to model business processes by key stakeholders
~\cite{065simoes2018eliciting}, statistics~\cite{155bach2016telling}, physics~\cite{000bahr2016CartoonGuide}, and data clips~\cite{155bach2016telling}.  
\citet{065simoes2018eliciting} found that comic strips are handy for capturing tacit process knowledge but less effective in getting a complete process overview. On the other hand,\citet{155bach2016telling} found that graph comics can improve understanding of complex temporal changes. 
\citet{170zhao2019understanding} declare data comics to be more engaging, space-efficient, faster, and easier compared to infographics. They concluded that the comic layout helps readers view the set of visualizations as a whole story without explicit instruction. Although we found that data comics are also used for exploratory purposes, the comic genre was largely used for explanatory purposes. \citet{154wang2019teaching} presented a successful and reproducible case study of how to teach data comics workshops to interdisciplinary students.  

\paragraph{Storyline visualizations} \citet{136watson2019storyprint} mention that visualization techniques on storyline visualization platforms are often used to attract new ways of human interaction. Our starting point to look into this interaction was the exploratory empirical study about users’ reception and usage behavior with interactive information graphics done by \citet{88burmester2010users}. They found that graphic usage duration differed between users and that story-based approaches, although motivating readers, might lead to less intensive reception of information as interactive information graphics tend to expose readers to too much information~\cite{88burmester2010users}. \citet{165endert2014human} tried to merge human intuitive capabilities of interactive visualization with the big data processing capabilities of analytics. They argue that the involvement of human analysts in the task of creating storylines from large amounts of data is too explorative, questions are ill-defined or unknown a priori, and training data is not available. They conclude that purely visual methods are insufficient, but using visualization as a medium for human-data interaction is recommended. They suggest design principles where both user input and visual feedback are placed in the context of the process of an analyst. \citet{131danner2020story} studied the dynamics between writers of stories and organizations that influence the shape and effectiveness of stories. Story writers must ensure that storylines are statistically true, actionable, and humanizing~\cite{131danner2020story}. The work of \citet{136watson2019storyprint} and \citet{105ping2018litstoryteller} give a good overview of various storyline visualization tools and applications.

\paragraph{Video.} Videos have long served as tools to tell stories and to inform people, as in documentaries.
Videos when combined with data visualizations referred as data videos~\cite{317tang2020design} can have a positive effect on learning~\cite{tversky2002animation}, engagement~\cite{weng2018srvis}, and viewers' focus~\cite{heer2007animated}. 
This would explain the popularity of data videos in recent years as found by \citet{29amini2016authoring} ~\cite{169amini2015understanding}. 
However, creating effective visual data stories can be challenging due to the need for interdisciplinary skills. 
\citet{xu2022wow} propose multiple opening styles for data videos. 
\citet{sun2022erato}developed Erato as a human-machine cooperative data story editing system that uses an interpolation algorithm to help users create smoother transitions between frames. 
Additionally, adaptive narratives and personalized data stories are promising presentation approaches for enhancing engagement, as noted by \citet{067bonacini2019Engaging}.

Interactivity, in general, can enhance the contextual dimension of data~\cite{041jern2009collaborative,234figueiras2014tell}
Interactivity is sometimes seen as a way to present visual information in interactive infographics. Interactive narratives can be linear, non-linear, or dynamic. Linear narratives are often associated with Scrollytelling, which combines storytelling and scrolling, as it allows the reader to explore the topic in depth by scrolling through the visual. The central element is a mostly full-screen animation with embedded elements like visualizations, video, textual, audible, or photographic content~\cite{295seyser2018scrollytelling, morth2022scrollyvis}.
\citet{295seyser2018scrollytelling} noted that scrollytelling articles are often text-centric and recognize image-centric articles where images/photos and videos order the elements. An elastic narrative allows following a predetermined order. In that respect, it should be considered that the story will branch off on specific points to allow a deeper understanding of the story. \citet{190riedl2006linear} demonstrated that branching is effective as ``they let users perform actions concurrently with system-controlled character actions.''

\subsubsection{High Context - Multimodal Experience}

\label{visualization-types-and-areas-of-application}

\paragraph{VR Data stories.} With Virtual reality(VR) technologies acquiring greater immersive capabilities and increased levels of interaction, the possibilities of research for interactive visualizations and data research have increased. In fact, interactive visualizations made up nearly 42\% of the analyzed literature in this review. Data representation in VR is no longer confined to static representation of the image. Data can now be visualized in 3D along with narration, sounds, and sensations giving the user a full immersive experience while interacting with the data. Although the field is relatively new, a significant amount of research has been done in understanding the data by visualizing it in a 3D environment~\cite{VRrubio2018digital}.
When analyzing the work on visualizing data in an immersive environment, we looked into what are the different contexts in which representing data in VR would be helpful. How would VR technologies be helpful for data visualization? What are the different interactions studied by different research groups?

\citet{303lugmayr2018financial} found several advantages for using 3D, VR, and immersive technologies for visualizing financial information. Among these advantages are presenting large amounts of data in a limited space and overcoming limitations due to a restricted physical space. By combining quantitative and qualitative data, information can be presented holistically and viewed from different perspectives. It was also concluded that using VR to visualize data supports data exploration, can increase user engagement, and allows the separation of different sets of data intuitively. A rare example in this field is the work
by \citet{187RenXRcreator}, who developed a prototype system to create immersive data-driven stories, supporting collaborative authoring and the use of different VR and AR devices. This tool was based on the work of \citet{31satyanarayan2014authoring}.

According to \citet{marques2022narrative}, augmented reality (AR) can enhance narrative visualization by improving view, focus, and sequence through simulation. Although research in this field is still in its early stages, future studies could investigate no-code AR authoring tools for designers. For instance, \citet{187RenXRcreator} created a prototype system that allows for collaborative authoring and supports using various VR and AR devices to produce immersive data-driven stories. Their work builds upon the research of \citet{31satyanarayan2014authoring}, who also explored authoring tools for AR.
In another example, ~\citet{186VRsoler2019workflow} proposed a set of scripted tools to accommodate storytelling in a 3D environment by collecting and visualizing the user's navigational data. \citet{chen2021augmenting} present a prototyping tool for creating augmented table tennis videos along with providing a design space for characterizing augmented sports videos based on their constituents and how to organize them. 
\citet{VRmadni2015expanding} employ storytelling techniques in VR to promote collaborative decision-making among different stakeholders in the process of systems engineering.
Data storytelling is also getting increased attention in the VR research community. The review article by \citet{VRrubio2018digital} provides a detailed summary of the state of the art of research and discusses ways in which collaboration in VR could benefit from data storytelling. Along the same line, \citet{VRliestol2018story} explore the possibility of how to keep the balance and divide the control between the explanatory and the exploratory aspects of storytelling in VR. \citet{VRxu2019effects} provide an overview of the most used methods to preserve narrative control in VR. To address the problem of attention guidance, they propose using a virtual character to direct users' attention towards the target in a VR environment. Another example of using virtual characters for storytelling is to deliver complex geospatial information in the form of narratives. They compare the narratives by virtual characters with audio/text-only modalities~\cite{VRblackmoreevaluating}.

\subsection{Storytelling effects and evaluation}

\begin{figure*}[htbp]
    \centering
    \includegraphics[width=\textwidth]{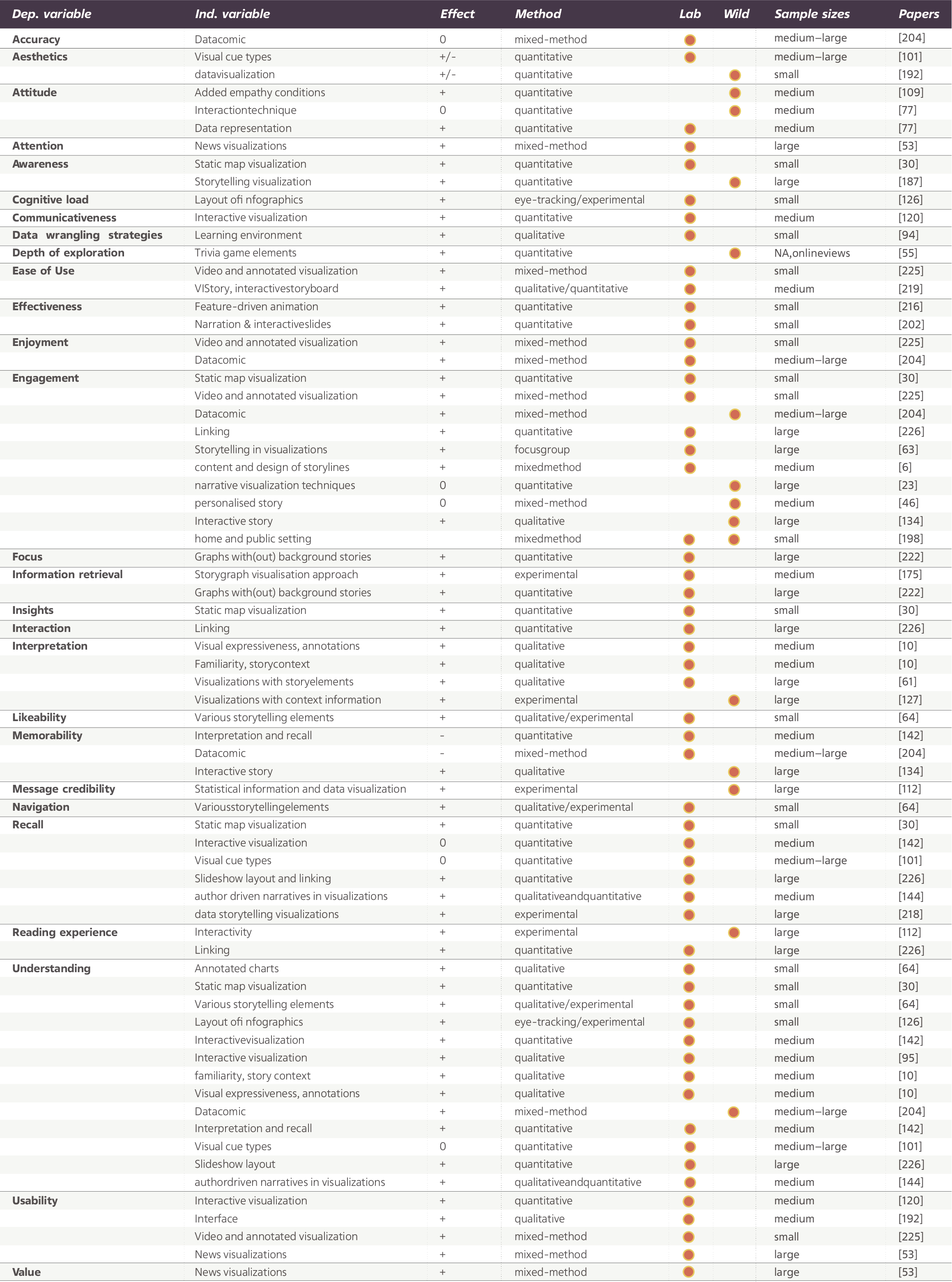}
    \caption{Overview of all dependent and independent variables in papers that contained empirical studies on the effects of storytelling}
    \label{fig:table}
\end{figure*}

While the narrative building blocks shape how the story unfolds, the influence of the story, however, builds upon the user's interaction with the resulting artifact. When we listen to stories, our brains immediately react and both cognitive and non-cognitive areas of the brain are stimulated~\cite{111abdulsabur2014neural,000Virtue2008Inferences}. To intentionally design data-driven stories that support as intended, it is crucial to understand the effects of storytelling and the mechanisms behind them.
For this purpose, we summarize how different effects have been studied in storytelling visualization research.
To provide an overview of our findings, we analyze the papers with regard to five categories---affective, cognitive, interactivity, indirect, and behavioral effects.
For all studies, we identify the independent and dependent variables and related effects.
We identify what methods were used to study the effects and whether studies were conducted in the lab or in the field. Lastly, we look at the individual sample sizes to conclude the expected replicability of the effects (see Fig.~\ref{fig:table}).


\subsubsection{Variables of interest}
Using storytelling in visualization has been claimed to improve several different aspects of communication~\cite{203segel2010narrative}. It is critical to first look at the difference between information and knowledge. Information, in many cases, refers to numerical facts directly retrievable from the visualization, while knowledge refers to the change in cognition in a user that enables them to act or think differently about the subject. Knowledge requires interpretation~\cite{000chen2008data}.  
Therefore, it is critical to separate these concepts into possible effects and variables. Before we identify how dependent and independent variables are interconnected, we identify which variables have been studied. The papers we selected for our review that used storytelling in data visualization utilized the following dependent variables:
\begin{itemize}
    \item \textbf{Accuracy} refers to readers having a more correct understanding after viewing the visualization \cite{43wang2019comparing}
    \item \textbf{Aesthetics} refers to stronger perceptions regarding aesthetics \cite{307kong2019understanding,317tang2020design} 
    \item \textbf{Attention} refers to the brain's process of selecting (visual) information for processing, measured using eye-tracking~\cite{185de2018does}. Yet, attention in the psychological sense would require a more extensive experimental paradigm. 
    \item \textbf{Attitude} refers to beliefs and convictions a person has in the real world. Attitudes can be expressed in cognitive, affective, and behavioral manners. Studies measure changes in the willingness to judge the real world referent~\cite{326liem2020structure,so2020humane}.
    \item \textbf{Awareness} refers to the ability to improve situational awareness (e.g., of changes in a project)~\cite{359burkhard2005tube,fernandez2021storytelling,so2020humane}
    \item \textbf{Cognitive Load} refers to the perceived mental exertion required to extract knowledge from a visualization.~\cite{289majooni2018eye}
    \item \textbf{Communicativeness} refers the quality of the communication induced by the visualization~\cite{70lunterova2019explorative}
    \item \textbf{Data wrangling strategies} refers to activating users to perform a more wide variety of data analysis strategies\cite{52jiang2019data} 
    \item \textbf{Depth of exploration} refers to activating users to perform deeper analyses\cite{081diakopoulos2010game} 
    \item \textbf{Ease of Use} refers to how easy the visualization is to use \cite{82zhi2019gameviews,174zeng2020vistory}
    \item \textbf{Effectiveness} refers to how effective the visualization is to convey data \cite{264wang2016guided, 63yu2016effectiveness} 
    \item \textbf{Enjoyment} defines the subject's joy from interacting with a data visualization~\cite{43wang2019comparing}.
    \item \textbf{Engagement} refers to the depth of the emotional connection between the topic and the users' emotions.~\cite{43wang2019comparing,359burkhard2005tube}
    \item \textbf{Focus} refers to an improvement in filtering out irrelevant information which is not part of the visualization~\cite{43wang2019comparing,zhao2021evaluating}.
    \item \textbf{Information retrieval} refers to the quality of how factual information is retrieved from the visualization~\cite{019schumann2013approach, zhao2021evaluating}.
    visualization.~\cite{234figueiras2014tell,link2021credibility}
    \item \textbf{Insights} refers to the number of knowledge units new to the reader extracted from a visualization~\cite{359burkhard2005tube}
    \item \textbf{Interaction} refers to the number of interactions users do with an interactive visualization \cite{82zhi2019gameviews} 
    \item \textbf{Insights} refers to the number of individual novel insights gathered from the visualization \cite{359burkhard2005tube} 
    \item \textbf{Interpretation} refers to how data or information is converted to knowledge. Visualizations can aid interpretation by helping the viewer to attain the ``correct'' or ``intended'' knowledge from the provided information~\cite{155bach2016telling, fernandez2021storytelling, link2021credibility,mantri2022viewers}.
    \item \textbf{Likeability} refers to the overall positive affective reaction of the user towards the 
    \item \textbf{Memorability} refers to the quality of being memorable, without necessarily measuring recall~\cite{139tong2018storytelling, obie2020effect}.
    \item \textbf{Understanding} refers to the amount of correctly extracted knowledge.~\cite{43wang2019comparing,302obie2019study,235figuerias2014narrative,320Cunningham2019ProvenanceNarratives,359burkhard2005tube,234figueiras2014tell,289majooni2018eye, obie2020effect,302obie2019study,307kong2019understanding,144thony2018storytelling,336zhu2020survey} 
    \item \textbf{Message credibility} refers to the credibility of a message~\cite{link2021credibility} 
    \item \textbf{Navigation} refers to the ease of explorability of a visualization.~\cite{234figueiras2014tell}
    \item \textbf{Recall} refers to the amount of correctly recalled information from a visualization.~\cite{43wang2019comparing,302obie2019study,359burkhard2005tube, zdanovic2022influence}
    \item \textbf{Reading Experience} refers to the reported quality of experience while reading data \cite{link2021credibility,103zhi2019linking} 
    \item \textbf{Usability} refers the overall improvement in usability when using storytelling \cite{70lunterova2019explorative,317tang2020design} 
    \item \textbf{Value} refers to the perceived value of the reported data in journalistic contexts~\cite{185de2018does} 

\end{itemize}

\subsubsection{Evaluation methods used}
Among all the eligible studies 
we found a variety of different approaches being used to understand storytelling in empirical research. Mixed-method designs are frequently used to better understand the impact of storytelling~\cite{43wang2019comparing, 185de2018does}. 
To investigate the impact of storytelling on the user, most studies use lab or field experiments~\cite{43wang2019comparing,340jiang2020data,235figuerias2014narrative}. From all eligible papers, we found that only a few papers 
used crowd-sourced samples as a data-gathering technique.
We expect to find only a few papers using crowd-sourcing because methods such as eye-tracking were often used to understand storytelling but are impossible to apply in crowd-sourcing~\cite{289majooni2018eye}.
Moreover, this is possibly due to the novel interest in storytelling in visualization research. When a field is new, internal validity may be valued more highly than external validity. As a consequence, field experiments are rare~\cite{43wang2019comparing,359burkhard2005tube}.

\subsubsection{Sample sizes}
To understand the potential for replicability, we also \textbf{look at the sample sizes} in the studies that we review. We assume that larger samples should yield effects that are more likely to replicate. By using the definition of effect sizes from \citet{000cohen2013statistical} ($\delta=0.8$ as large, $\delta=0.5$ as medium, $\delta=0.2$ as small), assuming a between-subject difference in means design, and using power analysis (error rates: $\alpha < .05$, $\beta < 0.2$) we derive approximate sample sizes for experiments (i.e., small effects need large samples). For survey data---which are more prone to noise---we use the stricter error rate of $\alpha < .01$ and round our sample size to the nearest hundred. For qualitative data, we use sample sizes commonly used in research practice~\cite{000nielsen2000you,000mason2010sample}. We picked the lowest recommendations we could find for the categories.
We categorize sample sizes as follows. For experimental data, we chose small (less than 25), medium (25--50), and large (over 50). For survey data, we chose small (less than 100), medium (100--500), and large (over 500). For qualitative data (Interviews, Focus Groups, etc.), we chose: small (less than 5~\cite{000nielsen2000you}), medium (5--15~\cite{000mason2010sample}), and large (over 15).

\subsubsection{The importance of engagement}
In this section, we want to summarize the findings distilled from empirical studies on storytelling in data visualization. 
While in the first place, narrative visualizations are designed to get the individual’s attention and facilitate understanding, if a visualization is designed effectively, the individual will stop scanning and will engage with the content~\cite{203segel2010narrative}. This note aligns with the two-systems approach to \textit{judgment and choice} as presented by \citet{kahneman2011thinking}: the fast and instinctive thinking---system 1---generates first impressions, intentions, and feelings. These are used as suggestions for the slow, deliberate system 2, which may then turn these into conscious beliefs and actions. Effective communication needs to connect with system 1. Otherwise, system 2 is less likely to engage, and people might subsequently not integrate the information ~\cite{shleifer2012psychologists}. Naturally, most effects of data-driven storytelling that are being researched, therefore, fall under the system 1 thinking mode. Both cognitive and affective involvement with the story is crucial for engagement: appealing features not only call for attention but also increase memorability and, thereby comprehension~\cite{135arevalo2020storylines}.
Whether communication was effective in achieving engagement can be determined by evaluating content, process, and outcome~\cite{rohrmann1992evaluation}. Content-wise, a story should feel relevant and be perceived as useful. Ideally, the process of designing the story includes defining key actors. Moreover, outcome measures should include measures of engagement~\cite{135arevalo2020storylines}. 
\textbf{Engagement can be defined as an ``affective, cognitive, and behavioral connection''}~\cite{o2008user}. We investigate the state of research and discuss these three different measures of engagement. 
\paragraph{Affective}
The first reaction to a story is often an emotional one, and authors have explored different elements that contribute to this emotional response, such as attractiveness, interest~\cite{135arevalo2020storylines},
and likeability~\cite{234figueiras2014tell, 235figuerias2014narrative}.
These effects have been studied particularly well in journalistic contexts. For example, the work of~\citet{185de2018does}, 
shows that news consumers notice, read, and appreciate news visualizations more when they are seamlessly integrated with a news story. 
However, the impact of data storytelling extends beyond journalism. \citet{234figueiras2014tell} found that storytelling visualizations improve the likability, comprehensibility, and navigation of data. In an effort to gain insight into the spontaneous behavior of groups of people, \citet{van2022more} presented a personalized weather forecast system that used interactive data videos to trigger social reflection that, in turn, induced emotional and narrative engagement among the family members. 
Additionally, data comics have been shown to improve enjoyment and engagement with visualized content~\cite{43wang2019comparing}. They also appear to strengthen the focus of the reader by reducing the text-picture distance, thereby, facilitating understanding and reducing cognitive load. It is important to note, however, that some studies have found no impact on engagement when combining storytelling visualizations with exploratory data analysis~\cite{143boy2015storytelling}.

An additional example of tube-map-like visualizations includes participants reporting higher levels of inspiration, activation, and energy, and showed more ``eureka-like experiences'' when using the visualization~\cite{359burkhard2005tube}.
According to \citet{341lidskog2020cold}, a compelling story should include both an emotional appeal and a normative orientation. As happens with simple storytelling, data stories can also make people feel like they are losing track of time and space~\cite{207lugrin2010exploring}, also described as \textit{narrative immersion} or \textit{flow}.
When individuals are confronted with their prior beliefs, emotions such as surprise or doubt can develop and prejudices are exposed~\cite{327heyer2020pushing}. As visualizations also carry a risk of oversimplification, annoyance or irritation is an emotion that can arise~\cite{356concannon2020brooke, 341lidskog2020cold}, especially when people have more knowledge than what is being displayed in the story.
Researchers also look at the change in attitude---or the persuasiveness of data-driven narratives---which is related to the cognitive as well as the behavioral layer of engagement~\cite{so2020humane}. \citet{326liem2020structure} cautioned that while it is a common assumption that visuals change attitude, there is little empirical evidence for it. They also do not find support for this assumption in their studies. \citet{327heyer2020pushing} found that eliciting individuals’ prior beliefs together with the display of the visualization does not increase attitude change---which is in line with the limited impact of visuals on attitude that \citet{326liem2020structure} mentioned---but does result in people feeling surprised. 
Viewers of the visualizations also expressed doubt and prejudice. When given a text, people gave more prejudicial comments compared to a visualization. Summarizing their results, \citet{327heyer2020pushing} also call for investigating different design strategies for informing people on the one hand and changing their beliefs or attitudes on the other. 
In addition to attitude, \citet{095perez2018interaction} have investigated attributions that individuals make towards a brand that has been featured in a data story, along with identification with that brand and trust.
Stories can enhance perceptions and appeal~\cite{317tang2020design} and raise awareness for complex topics~\cite{356concannon2020brooke,fernandez2021storytelling}, yet the goal is to also connect with people on a cognitive and behavioral level. 

\paragraph{Cognitive}
Besides affective reactions to storytelling, there are several cognitive effects as well.
Compared to reading a text with the same information, people seem to learn faster and more when looking at visualizations~\cite{327heyer2020pushing}. Visual cues (such as transparency or introducing additional elements) that guide the individuals’ attention towards certain story parts have also been shown to help people comprehend the story better~\cite{307kong2019understanding}. This and the overall user experience can be evaluated by testing factors such as salience and relevance using simple Likert scales~\cite{hullman2013contextifier}. 
Comprehension of stories evaluated with the QUEST model (e.g.,  \cite{graesser1991question}) has shown that visualization strategies matter for both understanding and perceived coherence of the story~\cite{098jhala2009comparing}. 
Telling stories with data can reduce the cognitive load of processing the information~\cite{232liao2014storytelling, 317tang2020design,meuschke2022narrative}, enhance decision-making quality~\cite{147BoldosovaStorytellingBusinessAnalytics}, 
improve understanding of data~\cite{356concannon2020brooke,235figuerias2014narrative,zhao2021evaluating}, 
ease understanding and perceived usefulness~\cite{135arevalo2020storylines}, and improve comprehensibility~\cite{234figueiras2014tell}.
The effect of visual cues in the presence of verbal (audio) narrations, for example, has been studied by \citet{307kong2019understanding}. Using data-driven storytelling, their results show that integrated and separate visual cues such as glow, desaturation, depth of field, etc., help to direct attention to relevant visualizations faster and maintain the reader's focus. The authors also provide suggestions based on the role of cues and their effectiveness. They found that brightness-based cueing was perceived as both most effective and aesthetically pleasing. Yet, their study did not show an actual influence on the participants' comprehension and recall. 
These findings tie into the results by~\cite{359burkhard2005tube}, who also found that using a tube-map-like visualization helps to gain attention and improves focus. Similarly,~\citet{zhao2021evaluating} found that background story influences participants’ focus areas during interactive graph explorations.
\citet{borkin2015beyond} offer a clear overview using a visualization taxonomy and experimental evidence on which techniques work best for recognition and recall. Using a tube-map-like visualization to show a project plan helped participants understand project goals better. The visualization improved the quality of discussions about the plan, improved individuals' attention to details, improved memorability of information, as well as recall~\cite{359burkhard2005tube}.
While the evidence for improvement of understanding and recall seems to increase, it is still unclear whether this also leads to changes in opinion or attitudes.
Shedding light on cognitive mechanisms in information synthesis,  \citet{mantri2022viewers} conducted four experiments where they presented a pair of line charts to the users sequentially. They found that when the two charts depicted relationships in opposite directions, participants tended to weigh the positive slope more. 
In a medium-sized study,~\citet{326liem2020structure} investigated the effect of using visual data storytelling on changes in attitude. While changes were detectable, the results showed smaller effects on participants' attitudes than expected~\cite{326liem2020structure}.
After having attained the reader's attention it is necessary to ensure understanding or comprehension by the viewer of a visualization. The effect of storytelling on understanding has been investigated in multiple studies. Most studies that address understanding investigate the effect of different visualization types or narrative structures on the amount of correctly extracted information from the visualization. 
By simply connecting data visually in a way that fosters imagining a story behind the data---as proposed by~\citet{019schumann2013approach}---users are asynchronously supported in developing an awareness of the underlying data. \citet{019schumann2013approach} used graphs to visually connect audio recordings in a repository to help collaborators understand the state of the project using a story graph. Their evaluation finds that this approach improved understanding of important information about the repository.
For data comics, findings even provide instructive insights. \citet{155bach2016telling} identified eight design factors for creating data comics that helped participants interpret the visualizations without training and only minimal annotations. They used data comics to communicate changes in dynamic networks.
Similarly, \citet{43wang2019comparing} found that data storytelling fosters a more enjoyable and engaging experience if combined with a comic-style presentation, and the use of comic-style improved understanding and recall.
Using a slide show layout, \citet{103zhi2019linking} found that comprehension tasks were significantly performed better; moreover, recall by participants was improved using a slide show layout where additional interaction was provided. Using an interactive slide show was preferred by the participants. \citet{correa2022diving} conducted a focus group with data visualization professionals to understand how designers can use narrative concepts to allow readers to personalize their story experience. The results revealed that using narrative genres while designing helps maintain the narrative intent.
Similarly, participants preferred an interactive storyboard visualization---VIStory---over a timeline-based visualization. \citet{174zeng2020vistory} demonstrated the efficiency of the technique with a qualitative and quantitative study, which showed better understanding and faster time on task. Their tool was designed to help understand the body of literature in visualization research and participants were able to answer more questions more quickly.

When participants perceived information through featured-driven animation, they were quicker than common interactive visualizations---albeit with equally accurate results. Therefore, \citet{63yu2016effectiveness} recommend their use in scientific simulations.
Adding a video of a human narrator and using an author-driven narrative structure significantly 
improved memorability. Yet, it had no significant effect on the long-term recall of information 
when compared against a reader-driven narrative structure~\cite{302obie2019study}.
Overall, we see evidence that the use of storytelling both improves the amount of correct information derived from the visualization and the time on task, improving the effectiveness and efficiency of the visualizations. However, most sample sizes of the studies shown here were either small or medium. Larger studies and even replications may be necessary to put the found effects on a more solid basis. 
By adding additional features like interactivity, the effectiveness of storytelling visualizations can be further improved.

\paragraph{Effects of Interactivity}
When switching from an author-driven approach (what the author wants to tell) to a reader-driven approach, storytelling visualization becomes a tool of exploration. This is achieved by letting the user interact with the visualization. This changes the process from a perception-focused to a more interrogative procedure. The reader understands some information that helps form new questions, which the visualization may then answer. The use of interactivity impacts both the overall message of the visualization, understanding, and recall by changing the engagement of the user. Interactivity allows the reader to focus on aspects of the visualization that they connect with~\cite{203segel2010narrative}.

Investigating the effects of interaction on message credibility and reading experience \citet{link2021credibility} found that interactions do not affect them. However, the perceived interactivity was seen to improve the reading experience.
In an exploratory empirical study about user reception and usage behavior,~\cite{88burmester2010users} investigated engagement and motivation using interactive information graphics.
The results showed that the usage durations were heterogeneously distributed for different users and different interactive information graphics. 
Users spent more time for initial orientation without interacting when introductions were shown.
They also found that story-based approaches motivated users but led to less intensive reception of information. Thus, interactivity is not without risk, as users need to be able to identify when and where to interact with the visualization. Otherwise, large parts of the visualization could remain unexplored.
This effect can be addressed using elements of games. A user study comparing two information graphics---either using or not using trivia questions---\citet{081diakopoulos2010game} found that users showed increased exploration of the data space when the trivia game elements were included. These findings were collected in a field setting, further strengthening the evidence of these results. Many forms of interactivity have been used in digital journalism, as found by~\citet{118alexandre2016promoting}. In their review of interactive storytelling visualizations from journalism, they find different patterns of how interactivity is achieved. Most frequently, the details-on-demand technique was used in the field. More importantly, the authors conclude that more research is needed to understand the effects on and the experience of readers when using interactivity in storytelling visualizations in journalism.

\paragraph{Indirect effects}
Next to the several positive effects of data-driven storytelling, there are certainly also indirect effects, some of which may be undesirable.
The current practice of using storytelling visualizations in journalism, for example, has become focused on the application of technology rather than providing the public with knowledge and insights (for a better understanding of current events)~\cite{180zhang2018visualization}. This bears the risk of letting readers deduct ``facts'' from these visualizations that could be inaccurate from the journalist's point of view. While reporting the data visually and seemingly neutral, visualizations may provide more transparency in journalism; they also bear the risk of further blurring the line between factual reporting, opinionated reporting, and manipulation and leave more leeway for misinterpretation. Using storytelling visualization requires a deep understanding of the domain, a strong scientific skill-set, and a thorough ethical reflection, as their use can have real-life consequences. 
According to a study by \citet{holder2022dispersion}, data visualization design can shape readers' perceptions of the people represented in the data and perpetuate harmful stereotypes. This suggests that design choices play a critical role in interpreting data visualizations related to social inequity and can contribute to stereotyping.

On the other hand, an improvement in data literacy was observed as a side effect of storytelling visualizations by \citet{340jiang2020data} when students were asked to create visualizations themselves. By 
``getting personal with big data'', an entry point to visualization is created, making information personally relatable, the interaction with big data for the youth can be enhanced~\cite{52jiang2019data,340jiang2020data}. 
In alignment with this approach, \citet{ploehn2020tsuga} proposes 'using data visualizations as tools of control' by showing the actual 'bodies' that the data represents within the visualizations. This strategy can create a sense of control and provide a transparent platform for discussions surrounding the data. 

\citet{hullman2011impact} have called for more research on the impact of psychological biases (such as anchoring and social proof) on the effectiveness of visualizations. Additionally, for designing a story sequence most effectively, it is important to consider user experience measures~\cite{kim2017graphscape}. Drawing insights from other scientific disciplines such as behavioral economics---where biases and framing effects have been researched extensively~\cite{tversky1974judgment,ariely2008predictably}---and include those in the design process of stories can help prevent undesired effects~\cite{000Hullman2011Framing}. 

\paragraph{Behavioral}
According to \citet{061echeverria2018driving}, 
a data story should always include a clear call to action. 
Behavioral effects can be divided into intentional and actual behavioral measures. \citet{095perez2018interaction} 
looked at purchase and recommendation intention.
In a study examining the intention of the viewers to alter their health-related behavior after watching a data video, \citet{sallam2022towards} found the influence of personality traits in the behavior change. They also found that providing solutions to the health problem presented in the video increased the video's actionability and induced higher behavioral change intentions. 
\citet{boy2015can} investigated low-level interaction with the data story itself.  
They started by analyzing Google Analytics clicking data and then took a closer look at the users' behavioral measures. Their findings revealed no significant differences between a visualization-savvy population and those less acquainted with visual stories. 

While \citet{203segel2010narrative} among others 
have called for more research on the engagement of the reader with the story and the data. Of the 35 papers we have included in this section, only 19 have looked at effects somehow. While many do not study but recognize the importance of researching effects, the remaining papers do not mention it at all. Definitions and degree scales of engagement such as the one put forward by \citet{mahyar2015towards} are certainly very helpful in this process---in line with our argument, they put decision-making based on the visualization as the highest level of engagement.

In a study examining viewers' willingness to change their health-related behavior after watching a data video, \citet{sallam2022towards} discovered that personality traits played a role in behavior change. They also found that providing solutions to the health problem presented in the video increased its actionability and induced higher behavioral change intentions.

\section{Discussion}

In this article, we have examined the literature on storytelling in data visualization, focusing on the use of frameworks, data story types, application areas, narrative structures, and their effects and measurement. This analysis was conducted to help researchers identify pertinent research gaps in storytelling in data visualization. We present an overview of the domain development and future research directions for each topic and address the questions stated in the introduction.

\textbf{\textit{What are the theoretical or conceptual foundations of the story creation process, and how is it structured?}}
Several frameworks for developing narrative visualizations exist, each tailored to specific storytelling contexts, resulting in heterogeneity among frameworks. These frameworks aim to provide narrative recommendations based on the data and context at hand. Researchers have summarized methods for deriving recommendations from existing narrative visualizations and their creation processes. However, the relationship between underlying cognitive processes and data storytelling is still under investigation. To achieve consistency in framework design, it is essential to comprehend the cognitive aspects of information processing, narrative structures, and their influence on our perception of information.

Our review highlights recent studies that specifically focus on underlying cognitive and behavioral processes. The internal construction of stories and the attributes involved can be explained using multimodal processing theory, psychology, and basic narratological concepts. While recent work has increasingly concentrated on storytelling automation, primarily using low-level perception metrics, future work will incorporate ideas from recommendations based on more individualized metrics and narrative structures. An improved understanding of underlying cognitive mechanisms will enable visualization researchers to develop frameworks for creating compelling and effective visual data stories more automatically or semi-automatically.

\textbf{\textit{How can individual data stories be structured within a taxonomy that reflects story construction?}}
Based on our survey, we propose a classification scheme centered on the level of contextualization and the level of control within story construction. We focus on the presented data combined with the audience's explicit and implicit knowledge. Contextualization assists users in understanding the data, with different levels ranging from presenting raw data to constructing physical metaphors that represent the data in context. The level of control, on the other hand, enables users to explore and interact with the data through a data story, with static, dynamic, and interactive levels depending on the user's control over the story's event order. This classification scheme offers an overview of data storytelling's evolution and provides insights into the research direction in this field.

\textbf{\textit{What data types were used in the analyzed stories, and in which areas were they applied?}}
Recent publications on storytelling in data visualization have shifted their focus from explanatory to interactive and exploratory visualization. This is understandable, as designers of storytelling visualizations often reflect on concrete messages when designing. Storytelling is gaining a more prominent role in general communication scenarios, particularly in the public sector, including science, open data, education, journalism, and healthcare. Since the categorization by Segel and Heer, computer-driven visualization has become more dominant in fields such as interactivity and virtual reality. This shift has moved storytelling from primarily author-driven to more reader-driven, enabling increased exploration. No strong correlation between the purpose and application areas was found, although interactive storytelling seems more frequently used for exploratory purposes in education and science. We advocate for more research to elevate storytelling in the private sector.

\textbf{\textit{How is the information organized? What are the different narrative structures, and which goals are pursued?}}
While Segel and Heer focused mainly on visualization "genre," other aspects of storytelling have played an increasingly important role in understanding how narrative structures can enhance visualizations. Recent research has incorporated narratological aspects into the features investigated. Understanding the role of events, existents, content, and different dimensions of expression in a narrative visualization context

can facilitate story creation and understanding of effects. Narratology, with its long history of reflecting elements and sequencing within narratives, is gaining a more prominent role in recent and future research directions.

Surprisingly, elements central to classical literature or film studies, such as characters, the role of the protagonist, or the antagonist, have been minimally studied. However, these elements significantly contribute to the user forming an emotional connection to the story, motivating them to follow it and contextualize the content.

\textbf{\textit{What are the cognitive and non-cognitive effects of storytelling, and how were they measured empirically?}}
Studies on storytelling in data visualization have considered various outcome variables using different methods and approaches across various frameworks and application areas. Many papers utilize qualitative methods to comprehend the effects of storytelling visualization on the reader, offering deep insights into the reception process during storytelling visualization design. These qualitative approaches reveal unexpected effects and provide insights for developing theories about storytelling effects in visualizations. However, a quantitative perspective could strengthen the evidence for many of these findings.

Improvement in understanding and recall is the most frequently studied quantitative outcome, as evidenced in several studies. Nevertheless, sample sizes were not always large, and measurements were inconsistent across studies. Notably, we found few negative results.

The use of interactivity in storytelling visualizations presents conflicting evidence. Although it fosters a deeper connection between the reader and data, it risks making the visualization less transparent and navigable, potentially even reducing the breadth and depth of information retrieval. Therefore, interactivity must be planned with care.

Many papers that report techniques and tools resort to ad-hoc evaluation methods. While the community should value the engineering approach of designing a technique or tool on its own, adding subpar post-hoc evaluation into a paper—just to meet scientific progress expectations—negates much of the effort in developing a technique and might introduce problematic findings. There is growing evidence in psychological research that inconsistent and ad-hoc operationalization may induce false positives in the research body, especially if researchers can manipulate the measurements. Combined with a publication bias towards papers showing significant results, there is a risk of more papers with instrumentation biased towards false positives.

This does not mean that these findings are non-existent. However, research on storytelling and outcomes should consider consolidating research methods and instrumentation to ensure the replicability of findings. This can be accomplished by providing web repositories of measurement methods and scales, conducting specific studies on measurement validity, and conducting more field studies. While laboratory and survey experiments show high internal validity and might replicate well in other lab studies, it is crucial to study how well these findings translate to real-world settings. To establish a theory of storytelling effectiveness in visualization, we recommend engaging more real end-users in real-life settings. Using causal modeling approaches could ensure the applicability of theory to real-world problems.
\section{Conclusion and Future Work}

This paper has provided an extensive overview of the role of storytelling in visualization. While storytelling is a historical phenomenon, predating even the concept of visualization, it is becoming increasingly critical in the realm of data dissemination and communication. Given the interdisciplinary nature of storytelling research, it is unsurprising that various fields focus on different aspects. The challenge for future research is to bridge these gaps and illuminate the field from a comprehensive perspective. This opens up several new lines of inquiry: 

\begin{itemize}
\item \textbf{Integration of existents:} Traditional narratology can enhance data-driven storytelling. For instance, how can we integrate characters, protagonists, or the relationship between users and data into our narratives?
\item \textbf{Understanding and utilizing the setting:} With emerging contexts like virtual reality or physicalization, we need to comprehend how these unique settings impact storytelling.
\item \textbf{Evaluation approach:} Future evaluations need to consider not just cognitive effects but also emotional and contextual impacts.
\item \textbf{Participatory story creation:} Storytelling thrives on the construction of an internal model. Therefore, we must explore how to create narratives across multiple disciplines.
\item \textbf{Cross-disciplinary research:} There is a strong need for multi-disciplinary research approaches to incorporate relevant perspectives.
\item \textbf{Understanding temporal structures:} We need to extend our understanding of temporal structures to accommodate non-linear, interactive approaches.
\end{itemize}

We observed that storytelling has been primarily utilized in the public and journalism sectors for explanatory purposes, often based on small sample sizes. Increasingly, non-linear interactive visualizations are being employed for exploratory purposes. There seems to be a scarcity of articles analyzing storytelling in the private domain, except for journalism and sports sectors, with some exceptions in the business domain.

Data-driven stories have shown immense potential in enhancing understanding and informing decision-making. In a world that generates an ever-growing volume of data, we can think of data not only as the ``new coal," but also as a raw material that requires refined skills to distill into valuable ``data diamonds." Storytelling in visualization lies at the heart of these skills, and its impact on visualization research has been steadily growing. We expect this trend to continue and even accelerate in the future. Data without context is meaningless; for data to have an impact, it requires a story.

{\footnotesize
\bibliography{00_bibliography001_100,00_bibliography201_300,00_bibliography301_400,00_bibliography101_200,00_bibliography}}

\begin{thebibliography}{227}
\providecommand{\natexlab}[1]{#1}
\providecommand{\url}[1]{\texttt{#1}}
\expandafter\ifx\csname urlstyle\endcsname\relax
  \providecommand{\doi}[1]{doi: #1}\else
  \providecommand{\doi}{doi: \begingroup \urlstyle{rm}\Url}\fi

\bibitem[AbdulSabur et~al.(2014)AbdulSabur, Xu, Liu, Chow, Baxter, Carson, and Braun]{111abdulsabur2014neural}
N.~Y. AbdulSabur, Y.~Xu, S.~Liu, H.~M. Chow, M.~Baxter, J.~Carson, and A.~R. Braun.
\newblock Neural correlates and network connectivity underlying narrative production and comprehension: A combined fmri and pet study.
\newblock \emph{Cortex}, 57:\penalty0 107--127, 2014.

\bibitem[Albers(2015)]{albers2015infographics}
M.~J. Albers.
\newblock Infographics and communicating complex information.
\newblock In \emph{International conference of design, user experience, and usability}, pages 267--276. Springer, 2015.

\bibitem[Alexandre(2016)]{118alexandre2016promoting}
I.~Alexandre.
\newblock Promoting insight: A case study of how to incorporate interaction in existing data visualizations.
\newblock In \emph{2016 20th International Conference Information Visualisation (IV)}, pages 203--208. IEEE, IEEE, 2016.

\bibitem[Amini et~al.(2015)Amini, Henry~Riche, Lee, Hurter, and Irani]{169amini2015understanding}
F.~Amini, N.~Henry~Riche, B.~Lee, C.~Hurter, and P.~Irani.
\newblock Understanding data videos: Looking at narrative visualization through the cinematography lens.
\newblock In \emph{Proceedings of the 33rd Annual ACM Conference on Human Factors in Computing Systems}, pages 1459--1468, 2015.

\bibitem[Amini et~al.(2016)Amini, Riche, Lee, Monroy-Hernandez, and Irani]{29amini2016authoring}
F.~Amini, N.~H. Riche, B.~Lee, A.~Monroy-Hernandez, and P.~Irani.
\newblock Authoring data-driven videos with dataclips.
\newblock \emph{IEEE transactions on visualization and computer graphics}, 23\penalty0 (1):\penalty0 501--510, 2016.

\bibitem[Arevalo et~al.(2020)Arevalo, Verbrugge, Sools, Brugnach, Wolterink, van Denderen, Candel, and Hulscher]{135arevalo2020storylines}
V.~J.~C. Arevalo, L.~N. Verbrugge, A.~Sools, M.~Brugnach, R.~Wolterink, R.~P. van Denderen, J.~H. Candel, and S.~J. Hulscher.
\newblock Storylines for practice: A visual storytelling approach to strengthen the science-practice interface.
\newblock \emph{Sustainability Science}, pages 1--20, 2020.

\bibitem[Ariely(2008)]{ariely2008predictably}
D.~Ariely.
\newblock \emph{Predictably irrational - the hidden forces that shape our decisions.}
\newblock Harper Collins New York, NY, 2008.

\bibitem[Axelrod and Kahn(2019)]{309axelrod2019intergenerational}
D.~B. Axelrod and J.~Kahn.
\newblock Intergenerational family storytelling and modeling with large-scale data sets.
\newblock In \emph{Proceedings of the 18th ACM International Conference on Interaction Design and Children}, pages 352--360, 2019.

\bibitem[Aziz et~al.(2022)]{aziz2022review}
M.~S.~A. Aziz et~al.
\newblock A review on the visual design styles in data storytelling based on user preferences and personality differences.
\newblock In \emph{2022 IEEE 7th International Conference on Information Technology and Digital Applications (ICITDA)}, pages 1--7. IEEE, 2022.

\bibitem[Bach et~al.(2016)Bach, Kerracher, Hall, Carpendale, Kennedy, and Henry~Riche]{155bach2016telling}
B.~Bach, N.~Kerracher, K.~W. Hall, S.~Carpendale, J.~Kennedy, and N.~Henry~Riche.
\newblock Telling stories about dynamic networks with graph comics.
\newblock In \emph{Proceedings of the 2016 CHI Conference on Human Factors in Computing Systems}, pages 3670--3682, 2016.

\bibitem[Bach et~al.(2018)Bach, Wang, Farinella, Murray-Rust, and Henry~Riche]{279bach2018design}
B.~Bach, Z.~Wang, M.~Farinella, D.~Murray-Rust, and N.~Henry~Riche.
\newblock Design patterns for data comics.
\newblock In \emph{Proceedings of the 2018 chi conference on human factors in computing systems}, pages 1--12, 2018.

\bibitem[Barczewski et~al.(2020)Barczewski, Bezerianos, and Boukhelifa]{343Barczewski2020Storyline}
A.~Barczewski, A.~Bezerianos, and N.~Boukhelifa.
\newblock How domain experts structure their exploratory data analysis: Towards a machine-learned storyline.
\newblock In \emph{Extended Abstracts of the 2020 CHI Conference on Human Factors in Computing Systems}, CHI EA '20, page 1–8, New York, NY, USA, 2020. Association for Computing Machinery.
\newblock ISBN 9781450368193.
\newblock \doi{10.1145/3334480.3382845}.
\newblock URL \url{https://doi.org/10.1145/3334480.3382845}.

\bibitem[Battad et~al.(2019)Battad, White, and Si]{075Battad2019Library}
Z.~Battad, A.~White, and M.~Si.
\newblock Facilitating information exploration of archival library materials through multi-modal storytelling.
\newblock In R.~E. Cardona-Rivera, A.~Sullivan, and R.~M. Young, editors, \emph{Interactive Storytelling}, pages 120--127, Cham, 2019. Springer International Publishing.
\newblock ISBN 978-3-030-33894-7.

\bibitem[Behera and Swain(2019)]{036behera2019big}
R.~K. Behera and A.~K. Swain.
\newblock Big data real-time storytelling with self-service visualization.
\newblock In \emph{Emerging Technologies in Data Mining and Information Security}, pages 405--415. Springer, 2019.

\bibitem[Benjamin~Bahr(2016)]{000bahr2016CartoonGuide}
R.~P. Benjamin~Bahr, Boris~Lemmer.
\newblock \emph{Quirky Quarks}, pages XVIII, 319.
\newblock Springer-Verlag Berlin Heidelberg, 2016.
\newblock ISBN 978-3-662-49509-4.
\newblock \doi{10.1007/978-3-662-49509-4}.
\newblock URL \url{https://doi.org/10.1007/978-3-662-49509-4}.

\bibitem[Blackmore et~al.()Blackmore, Smith, Nesbitt, North, Wark, and Nowina-Krowicki]{VRblackmoreevaluating}
K.~Blackmore, S.~Smith, K.~Nesbitt, L.~North, S.~Wark, and M.~Nowina-Krowicki.
\newblock Evaluating a virtual human storyteller for improved decision support.

\bibitem[Blackmore et~al.(2019)Blackmore, Smith, Nesbitt, North, Wark, and Nowina-Krowicki]{319blackmore2019evaluating}
K.~Blackmore, S.~Smith, K.~Nesbitt, L.~North, S.~Wark, and M.~Nowina-Krowicki.
\newblock Evaluating a virtual human storyteller for improved decision support.
\newblock 2019.

\bibitem[Boldosova and Luoto(2019)]{147BoldosovaStorytellingBusinessAnalytics}
V.~Boldosova and S.~Luoto.
\newblock Storytelling, business analytics and big data interpretation: Literature review and theoretical propositions.
\newblock \emph{Management Research Review}, 43 No. 2:\penalty0 204--222, 2019.
\newblock \doi{10.1108/MRR-03-2019-0106}.

\bibitem[Bonacini(2019)]{067bonacini2019Engaging}
E.~Bonacini.
\newblock Engaging participative communities in cultural heritage: Using digital storytelling in sicily (italy).
\newblock \emph{International Information \& Library Review}, 51\penalty0 (1):\penalty0 42--50, 2019.
\newblock \doi{10.1080/10572317.2019.1568786}.
\newblock URL \url{https://doi.org/10.1080/10572317.2019.1568786}.

\bibitem[Borkin et~al.(2015)Borkin, Bylinskii, Kim, Bainbridge, Yeh, Borkin, Pfister, and Oliva]{borkin2015beyond}
M.~A. Borkin, Z.~Bylinskii, N.~W. Kim, C.~M. Bainbridge, C.~S. Yeh, D.~Borkin, H.~Pfister, and A.~Oliva.
\newblock Beyond memorability: Visualization recognition and recall.
\newblock \emph{IEEE transactions on visualization and computer graphics}, 22\penalty0 (1):\penalty0 519--528, 2015.

\bibitem[Botsis et~al.(2020)Botsis, Fairman, Moran, and Anagnostou]{178Botsis2020}
T.~Botsis, J.~E. Fairman, M.~B. Moran, and V.~Anagnostou.
\newblock Visual storytelling enhances knowledge dissemination in biomedical science.
\newblock \emph{Journal of Biomedical Informatics}, 107:\penalty0 103458, jul 2020.
\newblock \doi{10.1016/j.jbi.2020.103458}.
\newblock URL \url{https://doi.org/10.1016/j.jbi.2020.103458}.

\bibitem[Bounegru et~al.(2017)Bounegru, Venturini, Gray, and Jacomy]{257bounegru2017narrating}
L.~Bounegru, T.~Venturini, J.~Gray, and M.~Jacomy.
\newblock Narrating networks: Exploring the affordances of networks as storytelling devices in journalism.
\newblock \emph{Digital Journalism}, 5\penalty0 (6):\penalty0 699--730, 2017.

\bibitem[Boy et~al.(2015{\natexlab{a}})Boy, Detienne, and Fekete]{143boy2015storytelling}
J.~Boy, F.~Detienne, and J.-D. Fekete.
\newblock Storytelling in information visualizations: Does it engage users to explore data?
\newblock In \emph{Proceedings of the 33rd Annual ACM Conference on Human Factors in Computing Systems}, pages 1449--1458, 2015{\natexlab{a}}.

\bibitem[Boy et~al.(2015{\natexlab{b}})Boy, Detienne, and Fekete]{boy2015can}
J.~Boy, F.~Detienne, and J.-D. Fekete.
\newblock Can initial narrative visualizationtechniques and storytelling help engage online-users with exploratory information visualizations.
\newblock \emph{Proceedings of the CHI, Seoul, Korea}, pages 18--23, 2015{\natexlab{b}}.

\bibitem[Brehmer et~al.(2016)Brehmer, Lee, Bach, Riche, and Munzner]{273brehmer2016timelines}
M.~Brehmer, B.~Lee, B.~Bach, N.~H. Riche, and T.~Munzner.
\newblock Timelines revisited: A design space and considerations for expressive storytelling.
\newblock \emph{IEEE transactions on visualization and computer graphics}, 23\penalty0 (9):\penalty0 2151--2164, 2016.

\bibitem[Brolch\'{a}in et~al.(2017)Brolch\'{a}in, Porwol, Ojo, Wagner, Lopez, and Karstens]{074BrolchainOpenData}
N.~O. Brolch\'{a}in, L.~Porwol, A.~Ojo, T.~Wagner, E.~T. Lopez, and E.~Karstens.
\newblock Extending open data platforms with storytelling features.
\newblock In \emph{Proceedings of the 18th Annual International Conference on Digital Government Research}, dg.o '17, page 48–53, New York, NY, USA, 2017. Association for Computing Machinery.
\newblock ISBN 9781450353175.
\newblock \doi{10.1145/3085228.3085283}.
\newblock URL \url{https://doi.org/10.1145/3085228.3085283}.

\bibitem[Brunet et~al.(2018)Brunet, Tuomisaari, Lavorel, Crouzat, Bierry, Peltola, and Arpin]{016brunet2018actionable}
L.~Brunet, J.~Tuomisaari, S.~Lavorel, E.~Crouzat, A.~Bierry, T.~Peltola, and I.~Arpin.
\newblock Actionable knowledge for land use planning: Making ecosystem services operational.
\newblock \emph{Land Use Policy}, 72:\penalty0 27--34, 2018.

\bibitem[Bryan et~al.(2016)Bryan, Ma, and Woodring]{161bryan2016temporal}
C.~Bryan, K.-L. Ma, and J.~Woodring.
\newblock Temporal summary images: An approach to narrative visualization via interactive annotation generation and placement.
\newblock \emph{IEEE transactions on visualization and computer graphics}, 23\penalty0 (1):\penalty0 511--520, 2016.

\bibitem[Burkhard et~al.(2018)Burkhard, Perhac, Asada, Troyanov, Zhong, Jiang, and Schubiger]{296burkhard20184d}
R.~Burkhard, J.~Perhac, S.~Asada, A.~Troyanov, S.~Zhong, Y.~Jiang, and S.~Schubiger.
\newblock 4d-ux: User experience design principles for coupling multidimensional visual representations in presentations.
\newblock In \emph{2018 22nd International Conference Information Visualisation (IV)}, pages 379--385. IEEE, 2018.

\bibitem[Burkhard and Meier(2005)]{359burkhard2005tube}
R.~A. Burkhard and M.~Meier.
\newblock Tube map visualization: Evaluation of a novel knowledge visualization application for the transfer of knowledge in long-term projects.
\newblock \emph{J. UCS}, 11\penalty0 (4):\penalty0 473--494, 2005.

\bibitem[Burmester et~al.(2010)Burmester, Mast, Tille, and Weber]{88burmester2010users}
M.~Burmester, M.~Mast, R.~Tille, and W.~Weber.
\newblock How users perceive and use interactive information graphics: An exploratory study.
\newblock In \emph{2010 14th International Conference Information Visualisation}, pages 361--368. IEEE, 2010.

\bibitem[Cao et~al.(2020)Cao, Pan, and Lin]{345Cao2020Sequence}
Y.-R. Cao, J.-Y. Pan, and W.-C. Lin.
\newblock User-oriented generation of contextual visualization sequences.
\newblock In \emph{Extended Abstracts of the 2020 CHI Conference on Human Factors in Computing Systems}, CHI EA '20, page 1–8, New York, NY, USA, 2020. Association for Computing Machinery.
\newblock ISBN 9781450368193.
\newblock \doi{10.1145/3334480.3383057}.
\newblock URL \url{https://doi.org/10.1145/3334480.3383057}.

\bibitem[Caquard and Fiset(2014)]{087caquard2014can}
S.~Caquard and J.-P. Fiset.
\newblock How can we map stories? a cybercartographic application for narrative cartography.
\newblock \emph{Journal of Maps}, 10\penalty0 (1):\penalty0 18--25, 2014.

\bibitem[Carpendale et~al.(2017)Carpendale, Thudt, Perin, and Willett]{267carpendale2017subjectivity}
S.~Carpendale, A.~Thudt, C.~Perin, and W.~Willett.
\newblock Subjectivity in personal storytelling with visualization.
\newblock \emph{Information Design Journal}, 23\penalty0 (1):\penalty0 48--64, 2017.

\bibitem[Chatman(1980)]{000Chatman1980StoryAndDiscourse}
S.~Chatman.
\newblock \emph{Story and Discourse}, page 288.
\newblock Cornell University Press, 1980.
\newblock ISBN 9781501741616.
\newblock \doi{10.1515/9781501741616}.
\newblock URL \url{https://doi.org/10.1515/9781501741616}.

\bibitem[Chaudhary and Arora(2019)]{141chaudhary2019storytelling}
A.~S. Chaudhary and A.~Arora.
\newblock Storytelling data visualization for grievances management system.
\newblock In \emph{International Conference on Futuristic Trends in Networks and Computing Technologies}, pages 395--405. Springer, 2019.

\bibitem[Chen et~al.(2008)Chen, Ebert, Hagen, Laramee, Van~Liere, Ma, Ribarsky, Scheuermann, and Silver]{000chen2008data}
M.~Chen, D.~Ebert, H.~Hagen, R.~S. Laramee, R.~Van~Liere, K.-L. Ma, W.~Ribarsky, G.~Scheuermann, and D.~Silver.
\newblock Data, information, and knowledge in visualization.
\newblock \emph{IEEE computer graphics and applications}, 29\penalty0 (1):\penalty0 12--19, 2008.

\bibitem[Chen et~al.(2010)Chen, Reiter, and Butz]{205chen2010photomagnets}
Y.-X. Chen, M.~Reiter, and A.~Butz.
\newblock Photomagnets: supporting flexible browsing and searching in photo collections.
\newblock In \emph{International Conference on Multimodal Interfaces and the Workshop on Machine Learning for Multimodal Interaction}, pages 1--8, 2010.

\bibitem[Chen et~al.(2021)Chen, Ye, Chu, Xia, Zhang, Qu, and Wu]{chen2021augmenting}
Z.~Chen, S.~Ye, X.~Chu, H.~Xia, H.~Zhang, H.~Qu, and Y.~Wu.
\newblock Augmenting sports videos with viscommentator.
\newblock \emph{IEEE Transactions on Visualization and Computer Graphics}, 28\penalty0 (1):\penalty0 824--834, 2021.

\bibitem[Chotisarn et~al.(2021{\natexlab{a}})Chotisarn, Lu, Ma, Xu, Meng, Lin, Xu, Luo, and Chen]{342chotisarn2021bubble}
N.~Chotisarn, J.~Lu, L.~Ma, J.~Xu, L.~Meng, B.~Lin, Y.~Xu, X.~Luo, and W.~Chen.
\newblock Bubble storytelling with automated animation: a brexit hashtag activism case study.
\newblock \emph{Journal of Visualization}, 24\penalty0 (1):\penalty0 101--115, 2021{\natexlab{a}}.

\bibitem[Chotisarn et~al.(2021{\natexlab{b}})Chotisarn, Pimanmassuriya, and Gulyanon]{chotisarn2021deep}
N.~Chotisarn, W.~Pimanmassuriya, and S.~Gulyanon.
\newblock Deep learning visualization for underspecification analysis in product design matching model development.
\newblock \emph{IEEE Access}, 9:\penalty0 108049--108061, 2021{\natexlab{b}}.

\bibitem[Choudhry et~al.(2020)Choudhry, Sharma, Chundury, Kapler, Gray, Ramakrishnan, and Elmqvist]{choudhry2020once}
A.~Choudhry, M.~Sharma, P.~Chundury, T.~Kapler, D.~W. Gray, N.~Ramakrishnan, and N.~Elmqvist.
\newblock Once upon a time in visualization: Understanding the use of textual narratives for causality.
\newblock \emph{IEEE Transactions on Visualization and Computer Graphics}, 27\penalty0 (2):\penalty0 1332--1342, 2020.

\bibitem[Chu et~al.(2014)Chu, Yu, and Wang]{236chu2014optimized}
W.-T. Chu, C.-H. Yu, and H.-H. Wang.
\newblock Optimized comics-based storytelling for temporal image sequences.
\newblock \emph{IEEE Transactions on Multimedia}, 17\penalty0 (2):\penalty0 201--215, 2014.

\bibitem[Cohen(2013)]{000cohen2013statistical}
J.~Cohen.
\newblock \emph{Statistical power analysis for the behavioral sciences}.
\newblock Academic press, 2013.

\bibitem[Cohn(2013)]{000Cohn2013VisualNarrativeStructure}
N.~Cohn.
\newblock Visual narrative structure.
\newblock \emph{Cognitive Science}, 37\penalty0 (3):\penalty0 413--452, 2013.
\newblock \doi{https://doi.org/10.1111/cogs.12016}.
\newblock URL \url{https://onlinelibrary.wiley.com/doi/abs/10.1111/cogs.12016}.

\bibitem[Concannon et~al.(2020)Concannon, Rajan, Shah, Smith, Ursu, and Hook]{356concannon2020brooke}
S.~Concannon, N.~Rajan, P.~Shah, D.~Smith, M.~Ursu, and J.~Hook.
\newblock Brooke leave home: Designing a personalized film to support public engagement with open data.
\newblock In \emph{Proceedings of the 2020 CHI Conference on Human Factors in Computing Systems}, pages 1--14, 2020.

\bibitem[Correa and Silveira(2022)]{correa2022diving}
C.~M. Correa and M.~S. Silveira.
\newblock Diving in the story: exploring tailoring in narrative data visualizations.
\newblock In \emph{Proceedings of the 21st Brazilian Symposium on Human Factors in Computing Systems}, pages 1--7, 2022.

\bibitem[Cruz and Machado(2011)]{206cruz2011generative}
P.~Cruz and P.~Machado.
\newblock Generative storytelling for information visualization.
\newblock \emph{IEEE computer graphics and applications}, 31\penalty0 (2):\penalty0 80--85, 2011.

\bibitem[Cunningham et~al.(2019)Cunningham, Nowina-Krowicki, Walsh, Chung, Wark, and Thomas]{320Cunningham2019ProvenanceNarratives}
A.~Cunningham, M.~Nowina-Krowicki, J.~Walsh, J.~Chung, S.~Wark, and B.~Thomas.
\newblock Provenance narratives and visualisation to support understanding and trust.
\newblock In \emph{El Sawah, S. (ed.) {MODSIM}2019, 23rd International Congress on Modelling and Simulation.} Modelling and Simulation Society of Australia and New Zealand, dec 2019.
\newblock \doi{10.36334/modsim.2019.d1.cunningham}.
\newblock URL \url{https://doi.org/10.36334/modsim.2019.d1.cunningham}.

\bibitem[D.~Jones and Anderson~Crow(2017)]{000Jones2017ScienceStories}
M.~D.~Jones and D.~Anderson~Crow.
\newblock How can we use the ‘science of stories’ to produce persuasive scientific stories?
\newblock \emph{Palgrave Communications}, 3\penalty0 (53):\penalty0 9, 2017.
\newblock \doi{10.1057/s41599-017-0047-7}.
\newblock URL \url{https://doi.org/10.1057/s41599-017-0047-7}.

\bibitem[Dailey et~al.(2022)Dailey, Gilmore, and Rangarajan]{dailey2022visualization}
S.~Dailey, B.~Gilmore, and N.~Rangarajan.
\newblock The visualization of public information: Describing the use of narrative infographics by us municipal governments.
\newblock \emph{Public Policy and Administration}, page 09520767221140954, 2022.

\bibitem[Danner(2020)]{131danner2020story}
P.~Danner.
\newblock Story/telling with data as distributed activity.
\newblock \emph{Technical Communication Quarterly}, 29\penalty0 (2):\penalty0 174--187, 2020.

\bibitem[De~Haan et~al.(2018)De~Haan, Kruikemeier, Lecheler, Smit, and Van~der Nat]{185de2018does}
Y.~De~Haan, S.~Kruikemeier, S.~Lecheler, G.~Smit, and R.~Van~der Nat.
\newblock When does an infographic say more than a thousand words? audience evaluations of news visualizations.
\newblock \emph{Journalism Studies}, 19\penalty0 (9):\penalty0 1293--1312, 2018.

\bibitem[de~Martino et~al.(2013)de~Martino, Rossetti, and Rossi]{225de2013new}
V.~de~Martino, S.~Rossetti, and D.~Rossi.
\newblock New approaches for an effective e-dissemination of statistics: The case of noi italia--100 statistics to understand the country we live in.
\newblock \emph{Statistical Journal of the IAOS}, 29\penalty0 (3):\penalty0 159--166, 2013.

\bibitem[Diakopoulos(2010)]{081diakopoulos2010game}
N.~Diakopoulos.
\newblock Game-y information graphics.
\newblock In \emph{CHI'10 Extended Abstracts on Human Factors in Computing Systems}, pages 3595--3600. 2010.

\bibitem[Diamond et~al.(2021)Diamond, Basu, Cao, and Hussain]{diamond2021canadian}
S.~Diamond, R.~Basu, S.~Cao, and A.~Hussain.
\newblock The canadian cultural diversity dashboard: Data storytelling and visualization for the cultural sector.
\newblock In \emph{International Conference on Human-Computer Interaction}, pages 372--384. Springer, 2021.

\bibitem[Echeverria et~al.(2018)Echeverria, Martinez-Maldonado, Granda, Chiluiza, Conati, and Shum]{061echeverria2018driving}
V.~Echeverria, R.~Martinez-Maldonado, R.~Granda, K.~Chiluiza, C.~Conati, and S.~B. Shum.
\newblock Driving data storytelling from learning design.
\newblock In \emph{Proceedings of the 8th international conference on learning analytics and knowledge}, pages 131--140, 2018.

\bibitem[El~Outa et~al.(2020)El~Outa, Francia, Marcel, Peralta, and Vassiliadis]{el2020towards}
F.~El~Outa, M.~Francia, P.~Marcel, V.~Peralta, and P.~Vassiliadis.
\newblock Towards a conceptual model for data narratives.
\newblock In \emph{International Conference on Conceptual Modeling}, pages 261--270. Springer, 2020.

\bibitem[ElShafie(2018)]{000ElShafi2018}
S.~J. ElShafie.
\newblock {Making Science Meaningful for Broad Audiences through Stories}.
\newblock \emph{Integrative and Comparative Biology}, 58\penalty0 (6):\penalty0 1213--1223, 09 2018.
\newblock ISSN 1540-7063.
\newblock \doi{10.1093/icb/icy103}.
\newblock URL \url{https://doi.org/10.1093/icb/icy103}.

\bibitem[Endert et~al.(2014)Endert, Hossain, Ramakrishnan, North, Fiaux, and Andrews]{165endert2014human}
A.~Endert, M.~S. Hossain, N.~Ramakrishnan, C.~North, P.~Fiaux, and C.~Andrews.
\newblock The human is the loop: new directions for visual analytics.
\newblock \emph{Journal of intelligent information systems}, 43\penalty0 (3):\penalty0 411--435, 2014.

\bibitem[Fernandez-Nieto et~al.(2021)Fernandez-Nieto, Echeverria, Shum, Mangaroska, Kitto, Palominos, Axisa, and Martinez-Maldonado]{fernandez2021storytelling}
G.~M. Fernandez-Nieto, V.~Echeverria, S.~B. Shum, K.~Mangaroska, K.~Kitto, E.~Palominos, C.~Axisa, and R.~Martinez-Maldonado.
\newblock Storytelling with learner data: Guiding student reflection on multimodal team data.
\newblock \emph{IEEE Transactions on Learning Technologies}, 14\penalty0 (5):\penalty0 695--708, 2021.

\bibitem[Fernandez~Nieto et~al.(2022)Fernandez~Nieto, Kitto, Buckingham~Shum, and Mart{\'\i}nez-Maldonado]{fernandez2022beyond}
G.~M. Fernandez~Nieto, K.~Kitto, S.~Buckingham~Shum, and R.~Mart{\'\i}nez-Maldonado.
\newblock Beyond the learning analytics dashboard: Alternative ways to communicate student data insights combining visualisation, narrative and storytelling.
\newblock In \emph{LAK22: 12th International Learning Analytics and Knowledge Conference}, pages 219--229, 2022.

\bibitem[Figueiras(2014)]{234figueiras2014tell}
A.~Figueiras.
\newblock How to tell stories using visualization.
\newblock In \emph{2014 18th International conference on information visualisation}, pages 18--18. IEEE, 2014.

\bibitem[{Figueiras}(2014)]{235figuerias2014narrative}
A.~{Figueiras}.
\newblock Narrative visualization: A case study of how to incorporate narrative elements in existing visualizations.
\newblock In \emph{2014 18th International Conference on Information Visualisation}, pages 46--52, 2014.
\newblock \doi{10.1109/IV.2014.79}.

\bibitem[Figueiras and Vizoso(2022)]{figueiras2022information}
A.~Figueiras and {\'A}.~Vizoso.
\newblock Information visualization: features and challenges in the production of data stories.
\newblock In \emph{Total Journalism: Models, Techniques and Challenges}, pages 83--96. Springer, 2022.

\bibitem[Fish(2020)]{348fish2020storytelling}
C.~Fish.
\newblock Storytelling for making cartographic design decisions for climate change communication in the united states.
\newblock \emph{Cartographica: The International Journal for Geographic Information and Geovisualization}, 55\penalty0 (2):\penalty0 69--84, 2020.

\bibitem[Freixa~Font et~al.(2021)Freixa~Font, P{\'e}rez-Montoro~Guti{\'e}rrez, and Codina]{freixa2021binomial}
P.~Freixa~Font, M.~P{\'e}rez-Montoro~Guti{\'e}rrez, and L.~Codina.
\newblock The binomial of interaction and visualization in digital news media: consolidation, standardization and future challenges.
\newblock \emph{Profesional de la informaci{\'o}n. 2021; 30 (4): e300401}, 2021.

\bibitem[Freytag and MacEwan(1894)]{000freytag1894freytag}
G.~Freytag and E.~MacEwan.
\newblock \emph{Freytag's Technique of the Drama: An Exposition of Dramatic Composition and Art}.
\newblock S. C. Griggs, 1894.
\newblock URL \url{https://books.google.de/books?id=1-MPAAAAYAAJ}.

\bibitem[George-Palilonis and Spillman(2013)]{218george2013storytelling}
J.~George-Palilonis and M.~Spillman.
\newblock Storytelling with interactive graphics: An analysis of editors' attitudes and practices.
\newblock \emph{Visual Communication Quarterly}, 20\penalty0 (1):\penalty0 20--27, 2013.

\bibitem[Gershon and Page(2001)]{184gershon2001storytelling}
N.~Gershon and W.~Page.
\newblock What storytelling can do for information visualization.
\newblock \emph{Communications of the ACM}, 44\penalty0 (8):\penalty0 31--37, 2001.

\bibitem[Graesser et~al.(1991)Graesser, Lang, and Roberts]{graesser1991question}
A.~C. Graesser, K.~L. Lang, and R.~M. Roberts.
\newblock Question answering in the context of stories.
\newblock \emph{Journal of Experimental Psychology: General}, 120\penalty0 (3):\penalty0 254, 1991.

\bibitem[Gratzl et~al.(2016)Gratzl, Lex, Gehlenborg, Cosgrove, and Streit]{080gratzl2016visual}
S.~Gratzl, A.~Lex, N.~Gehlenborg, N.~Cosgrove, and M.~Streit.
\newblock From visual exploration to storytelling and back again.
\newblock In \emph{Computer Graphics Forum}, volume~35, pages 491--500. Wiley Online Library, 2016.

\bibitem[Groshans et~al.(2019)Groshans, Mikhailova, Post, Schlautman, Carbajales-Dale, and Payne]{058groshans2019digital}
G.~Groshans, E.~Mikhailova, C.~Post, M.~Schlautman, P.~Carbajales-Dale, and K.~Payne.
\newblock Digital story map learning for stem disciplines.
\newblock \emph{Education Sciences}, 9\penalty0 (2):\penalty0 75, 2019.

\bibitem[Hasan et~al.(2022)Hasan, Wolff, Knutas, P{\"a}ssil{\"a}, and Kantola]{hasan2022playing}
M.~T. Hasan, A.~Wolff, A.~Knutas, A.~P{\"a}ssil{\"a}, and L.~Kantola.
\newblock Playing games through interactive data comics to explore water quality in a lake: A case study exploring the use of a data-driven storytelling method in co-design.
\newblock In \emph{CHI Conference on Human Factors in Computing Systems Extended Abstracts}, pages 1--7, 2022.

\bibitem[Heer and Robertson(2007)]{heer2007animated}
J.~Heer and G.~Robertson.
\newblock Animated transitions in statistical data graphics.
\newblock \emph{IEEE transactions on visualization and computer graphics}, 13\penalty0 (6):\penalty0 1240--1247, 2007.

\bibitem[Heyer et~al.(2020{\natexlab{a}})Heyer, Raveendranath, and Reda]{000Heyer2020}
J.~Heyer, N.~K. Raveendranath, and K.~Reda.
\newblock Pushing the (visual) narrative: The effects of prior knowledge elicitation in provocative topics.
\newblock In \emph{Proceedings of the 2020 CHI Conference on Human Factors in Computing Systems}, CHI '20, page 1–14, New York, NY, USA, 2020{\natexlab{a}}. Association for Computing Machinery.
\newblock ISBN 9781450367080.
\newblock \doi{10.1145/3313831.3376887}.
\newblock URL \url{https://doi.org/10.1145/3313831.3376887}.

\bibitem[Heyer et~al.(2020{\natexlab{b}})Heyer, Raveendranath, and Reda]{327heyer2020pushing}
J.~Heyer, N.~K. Raveendranath, and K.~Reda.
\newblock Pushing the (visual) narrative: the effects of prior knowledge elicitation in provocative topics.
\newblock In \emph{Proceedings of the 2020 CHI Conference on Human Factors in Computing Systems}, pages 1--14, 2020{\natexlab{b}}.

\bibitem[Hill and Grinnell(2014)]{243hill2014using}
S.~Hill and C.~Grinnell.
\newblock Using digital storytelling with infographics in stem professional writing pedagogy.
\newblock In \emph{2014 IEEE International Professional Communication Conference (IPCC)}, pages 1--7. IEEE, 2014.

\bibitem[Hilviu and Rapp(2015)]{108hilviu2015narrating}
D.~Hilviu and A.~Rapp.
\newblock Narrating the quantified self.
\newblock In \emph{Adjunct Proceedings of the 2015 ACM International Joint Conference on Pervasive and Ubiquitous Computing and Proceedings of the 2015 ACM International Symposium on Wearable Computers}, pages 1051--1056, 2015.

\bibitem[Ho et~al.(2011)Ho, Lundblad, {\AA}str{\"o}m, and Jern]{014ho2011web}
Q.~Ho, P.~Lundblad, T.~{\AA}str{\"o}m, and M.~Jern.
\newblock A web-enabled visualization toolkit for geovisual analytics.
\newblock In \emph{Visualization and Data Analysis 2011}, volume 7868, page 78680R. International Society for Optics and Photonics, 2011.

\bibitem[Holder and Xiong(2022)]{holder2022dispersion}
E.~Holder and C.~Xiong.
\newblock Dispersion vs disparity: Hiding variability can encourage stereotyping when visualizing social outcomes.
\newblock \emph{IEEE Transactions on Visualization and Computer Graphics}, 29\penalty0 (1):\penalty0 624--634, 2022.

\bibitem[Hook(2018)]{076hook2018facts}
J.~Hook.
\newblock Facts, interactivity and videotape: exploring the design space of data in interactive video storytelling.
\newblock In \emph{Proceedings of the 2018 ACM International Conference on Interactive Experiences for TV and Online Video}, pages 43--55, 2018.

\bibitem[Hougaard and Knoche(2019)]{160hougaard2019telling}
B.~I. Hougaard and H.~Knoche.
\newblock Telling the story right: How therapists aid stroke patients interpret personal visualized game performance data.
\newblock In \emph{Proceedings of the 13th EAI International Conference on Pervasive Computing Technologies for Healthcare}, pages 435--443, 2019.

\bibitem[Hullman and Diakopoulos(2011)]{000Hullman2011Framing}
J.~Hullman and N.~Diakopoulos.
\newblock Visualization rhetoric: Framing effects in narrative visualization.
\newblock \emph{IEEE Transactions on Visualization and Computer Graphics}, 17\penalty0 (12):\penalty0 2231--2240, 2011.
\newblock \doi{10.1109/TVCG.2011.255}.

\bibitem[Hullman et~al.(2011)Hullman, Adar, and Shah]{hullman2011impact}
J.~Hullman, E.~Adar, and P.~Shah.
\newblock The impact of social information on visual judgments.
\newblock In \emph{Proceedings of the SIGCHI conference on human factors in computing systems}, pages 1461--1470, 2011.

\bibitem[Hullman et~al.(2013{\natexlab{a}})Hullman, Diakopoulos, and Adar]{hullman2013contextifier}
J.~Hullman, N.~Diakopoulos, and E.~Adar.
\newblock Contextifier: automatic generation of annotated stock visualizations.
\newblock In \emph{Proceedings of the SIGCHI Conference on human factors in computing systems}, pages 2707--2716, 2013{\natexlab{a}}.

\bibitem[Hullman et~al.(2013{\natexlab{b}})Hullman, Drucker, Riche, Lee, Fisher, and Adar]{008hullman2013deeper}
J.~Hullman, S.~Drucker, N.~H. Riche, B.~Lee, D.~Fisher, and E.~Adar.
\newblock A deeper understanding of sequence in narrative visualization.
\newblock \emph{IEEE Transactions on visualization and computer graphics}, 19\penalty0 (12):\penalty0 2406--2415, 2013{\natexlab{b}}.

\bibitem[Isenberg et~al.(2018)Isenberg, Lee, Qu, and Cordeil]{287isenberg2018immersive}
P.~Isenberg, B.~Lee, H.~Qu, and M.~Cordeil.
\newblock Immersive visual data stories.
\newblock In \emph{Immersive Analytics}, pages 165--184. Springer, 2018.

\bibitem[Jacob(2020)]{330jacob2020visualising}
R.~Jacob.
\newblock Visualising global pandemic: A content analysis of infographics on covid--19.
\newblock \emph{Journal of Content, Community \& Communication}, 11\penalty0 (6):\penalty0 116--123, 2020.

\bibitem[Janowski et~al.(2019)Janowski, Ojo, Curry, and Porwol]{305janowski2019mediating}
M.~Janowski, A.~Ojo, E.~Curry, and L.~Porwol.
\newblock Mediating open data consumption-identifying story patterns for linked open statistical data.
\newblock In \emph{Proceedings of the 12th International Conference on Theory and Practice of Electronic Governance}, pages 156--163, 2019.

\bibitem[Jern(2009)]{041jern2009collaborative}
M.~Jern.
\newblock Collaborative web-enabled geoanalytics applied to oecd regional data.
\newblock In \emph{International Conference on Cooperative Design, Visualization and Engineering}, pages 32--43. Springer, 2009.

\bibitem[Jern(2010)]{071jern2010explore}
M.~Jern.
\newblock Explore, collaborate and publish official statistics for measuring regional progress.
\newblock In \emph{Cooperative design, visualization, and engineering}, pages 189--198. Springer, 2010.

\bibitem[Jhala and Young(2009)]{098jhala2009comparing}
A.~Jhala and R.~M. Young.
\newblock Comparing effects of different cinematic visualization strategies on viewer comprehension.
\newblock In \emph{Joint International Conference on Interactive Digital Storytelling}, pages 26--37. Springer, 2009.

\bibitem[Jiang and Kahn(2019)]{52jiang2019data}
S.~Jiang and J.~Kahn.
\newblock Data wrangling practices and process in modeling family migration narratives with big data visualization technologies.
\newblock 2019.

\bibitem[Jiang and Kahn(2020)]{340jiang2020data}
S.~Jiang and J.~Kahn.
\newblock Data wrangling practices and collaborative interactions with aggregated data.
\newblock \emph{International Journal of Computer-Supported Collaborative Learning}, 15\penalty0 (3):\penalty0 257--281, 2020.

\bibitem[Jung et~al.(2017)Jung, Hong, and Nguyen]{256jung2017serendipity}
J.~E. Jung, M.~Hong, and H.~L. Nguyen.
\newblock Serendipity-based storification: from lifelogging to storytelling.
\newblock \emph{Multimedia Tools and Applications}, 76\penalty0 (8):\penalty0 10345--10356, 2017.

\bibitem[Kahneman(2011)]{kahneman2011thinking}
D.~Kahneman.
\newblock \emph{Thinking, fast and slow}.
\newblock Macmillan, 2011.

\bibitem[Kim et~al.(2017{\natexlab{a}})Kim, Bach, Im, Schriber, Gross, and Pfister]{kim2017visualizing}
N.~W. Kim, B.~Bach, H.~Im, S.~Schriber, M.~Gross, and H.~Pfister.
\newblock Visualizing nonlinear narratives with story curves.
\newblock \emph{IEEE Transactions on Visualization and Computer Graphics}, 24\penalty0 (1):\penalty0 595--604, 2017{\natexlab{a}}.

\bibitem[Kim et~al.(2019)Kim, Henry~Riche, Bach, Xu, Brehmer, Hinckley, Pahud, Xia, McGuffin, and Pfister]{053kim2019datatoon}
N.~W. Kim, N.~Henry~Riche, B.~Bach, G.~Xu, M.~Brehmer, K.~Hinckley, M.~Pahud, H.~Xia, M.~J. McGuffin, and H.~Pfister.
\newblock Datatoon: Drawing dynamic network comics with pen+ touch interaction.
\newblock In \emph{Proceedings of the 2019 CHI Conference on Human Factors in Computing Systems}, pages 1--12, 2019.

\bibitem[Kim et~al.(2017{\natexlab{b}})Kim, Wongsuphasawat, Hullman, and Heer]{kim2017graphscape}
Y.~Kim, K.~Wongsuphasawat, J.~Hullman, and J.~Heer.
\newblock Graphscape: A model for automated reasoning about visualization similarity and sequencing.
\newblock In \emph{Proceedings of the 2017 CHI Conference on Human Factors in Computing Systems}, pages 2628--2638, 2017{\natexlab{b}}.

\bibitem[Kong et~al.(2019)Kong, Zhu, Liu, and Karahalios]{307kong2019understanding}
H.-K. Kong, W.~Zhu, Z.~Liu, and K.~Karahalios.
\newblock Understanding visual cues in visualizations accompanied by audio narrations.
\newblock In \emph{Proceedings of the 2019 CHI Conference on Human Factors in Computing Systems}, pages 1--13, 2019.

\bibitem[Kosara and Mackinlay(2013)]{149kosara2013storytelling}
R.~Kosara and J.~Mackinlay.
\newblock Storytelling: The next step for visualization.
\newblock \emph{Computer}, 46\penalty0 (5):\penalty0 44--50, 2013.

\bibitem[Lan et~al.(2022)Lan, O’Brien, Cheshire, Singleton, and Longley]{lan2022data}
T.~Lan, O.~O’Brien, J.~Cheshire, A.~Singleton, and P.~Longley.
\newblock From data to narratives: Scrutinising the spatial dimensions of social and cultural phenomena through lenses of interactive web mapping.
\newblock \emph{Journal of Geovisualization and Spatial Analysis}, 6\penalty0 (2):\penalty0 22, 2022.

\bibitem[Lan et~al.(2021)Lan, Shi, Wu, Jiao, and Cao]{lan2021kineticharts}
X.~Lan, Y.~Shi, Y.~Wu, X.~Jiao, and N.~Cao.
\newblock Kineticharts: Augmenting affective expressiveness of charts in data stories with animation design.
\newblock \emph{IEEE Transactions on Visualization and Computer Graphics}, 28\penalty0 (1):\penalty0 933--943, 2021.

\bibitem[Latif et~al.(2021)Latif, Zhou, Kim, Beck, and Kim]{latif2021kori}
S.~Latif, Z.~Zhou, Y.~Kim, F.~Beck, and N.~W. Kim.
\newblock Kori: Interactive synthesis of text and charts in data documents.
\newblock \emph{IEEE Transactions on Visualization and Computer Graphics}, 28\penalty0 (1):\penalty0 184--194, 2021.

\bibitem[Lee et~al.(2015)Lee, Riche, Isenberg, and Carpendale]{245lee2015more}
B.~Lee, N.~H. Riche, P.~Isenberg, and S.~Carpendale.
\newblock More than telling a story: Transforming data into visually shared stories.
\newblock \emph{IEEE computer graphics and applications}, 35\penalty0 (5):\penalty0 84--90, 2015.

\bibitem[Liao et~al.(2014)Liao, Hsu, and Ma]{232liao2014storytelling}
I.~Liao, W.-H. Hsu, and K.-L. Ma.
\newblock Storytelling via navigation: A novel approach to animation for scientific visualization.
\newblock In \emph{International Symposium on Smart Graphics}, pages 1--14. Springer, 2014.

\bibitem[Lidskog et~al.(2020)Lidskog, Berg, Gustafsson, and L{\"o}fmarck]{341lidskog2020cold}
R.~Lidskog, M.~Berg, K.~M. Gustafsson, and E.~L{\"o}fmarck.
\newblock Cold science meets hot weather: Environmental threats, emotional messages and scientific storytelling.
\newblock \emph{Media and Communication}, 8\penalty0 (1):\penalty0 118--128, 2020.

\bibitem[Liem et~al.(2020)Liem, Perin, and Wood]{326liem2020structure}
J.~Liem, C.~Perin, and J.~Wood.
\newblock Structure and empathy in visual data storytelling: Evaluating their influence on attitude.
\newblock In \emph{Computer Graphics Forum}, volume~39, pages 277--289. Wiley Online Library, 2020.

\bibitem[Liest{\o}l(2018)]{VRliestol2018story}
G.~Liest{\o}l.
\newblock Story and storage--narrative theory as a tool for creativity in augmented reality storytelling.
\newblock \emph{Virtual Creativity}, 8\penalty0 (1):\penalty0 75--89, 2018.

\bibitem[Lim et~al.(2022)Lim, Peralta, Rubel, Jiang, Kahn, and Herbel-Eisenmann]{lim2022keeping}
V.~Y. Lim, L.~M.~M. Peralta, L.~H. Rubel, S.~Jiang, J.~B. Kahn, and B.~Herbel-Eisenmann.
\newblock Keeping pace with innovations in data visualizations: A commentary for mathematics education in times of crisis.
\newblock \emph{ZDM--Mathematics Education}, pages 1--10, 2022.

\bibitem[Link et~al.(2021)Link, Henke, and M{\"o}hring]{link2021credibility}
E.~Link, J.~Henke, and W.~M{\"o}hring.
\newblock Credibility and enjoyment through data? effects of statistical information and data visualizations on message credibility and reading experience.
\newblock \emph{Journalism studies}, 22\penalty0 (5):\penalty0 575--594, 2021.

\bibitem[Lopezosa et~al.(2022)Lopezosa, P{\'e}rez-Montoro, and Guallar]{lopezosa2022data}
C.~Lopezosa, M.~P{\'e}rez-Montoro, and J.~Guallar.
\newblock Data visualization in the news media: Trends and challenges.
\newblock \emph{Technology, Business, Innovation, and Entrepreneurship in Industry 4.0}, pages 315--334, 2022.

\bibitem[Losev et~al.(2022)Losev, Raynor, Carpendale, and Tory]{losev2022embracing}
T.~Losev, J.~Raynor, S.~Carpendale, and M.~Tory.
\newblock Embracing disciplinary diversity in visualization.
\newblock \emph{IEEE Computer Graphics and Applications}, 42\penalty0 (6):\penalty0 64--71, 2022.

\bibitem[Lu et~al.(2020)Lu, Wang, Ye, Gu, Ding, Xu, and Chen]{089lu2020illustrating}
J.~Lu, J.~Wang, H.~Ye, Y.~Gu, Z.~Ding, M.~Xu, and W.~Chen.
\newblock Illustrating changes in time-series data with data video.
\newblock \emph{IEEE computer graphics and applications}, 40\penalty0 (2):\penalty0 18--31, 2020.

\bibitem[Lugmayr et~al.(2017)Lugmayr, Sutinen, Suhonen, Sedano, Hlavacs, and Montero]{125lugmayr2017serious}
A.~Lugmayr, E.~Sutinen, J.~Suhonen, C.~I. Sedano, H.~Hlavacs, and C.~S. Montero.
\newblock Serious storytelling--a first definition and review.
\newblock \emph{Multimedia tools and applications}, 76\penalty0 (14):\penalty0 15707--15733, 2017.

\bibitem[Lugmayr et~al.(2018)Lugmayr, Lim, Hollick, Khuu, and Chan]{303lugmayr2018financial}
A.~Lugmayr, Y.~J. Lim, J.~Hollick, J.~Khuu, and F.~Chan.
\newblock Financial data visualization in 3d on immersive virtual reality displays.
\newblock In \emph{International Workshop on Enterprise Applications, Markets and Services in the Finance Industry}, pages 118--130. Springer, 2018.

\bibitem[Lugrin et~al.(2010)Lugrin, Cavazza, Pizzi, Vogt, and Andr{\'e}]{207lugrin2010exploring}
J.-L. Lugrin, M.~Cavazza, D.~Pizzi, T.~Vogt, and E.~Andr{\'e}.
\newblock Exploring the usability of immersive interactive storytelling.
\newblock In \emph{Proceedings of the 17th ACM symposium on virtual reality software and technology}, pages 103--110, 2010.

\bibitem[Lundblad and Jern(2013)]{224lundblad2013geovisual}
P.~Lundblad and M.~Jern.
\newblock Geovisual analytics and storytelling using html5.
\newblock In \emph{2013 17th International Conference on Information Visualisation}, pages 263--271. IEEE, 2013.

\bibitem[Lunterova et~al.(2019)Lunterova, Spetko, and Palamas]{70lunterova2019explorative}
A.~Lunterova, O.~Spetko, and G.~Palamas.
\newblock Explorative visualization of food data to raise awareness of nutritional value.
\newblock In \emph{International Conference on Human-Computer Interaction}, pages 180--191. Springer, 2019.

\bibitem[Lyu et~al.(2020)Lyu, Cheng, and Lin]{331lyu2020visual}
Y.~Lyu, T.~F. Cheng, and R.~Lin.
\newblock Visual data storytelling: A case study of turning big data into chinese painting.
\newblock In \emph{International Conference on Human-Computer Interaction}, pages 526--535. Springer, 2020.

\bibitem[M.~Amerian(2015)]{000Amerian2015KeyConceptsNarratology}
L.~J. M.~Amerian.
\newblock Key concepts and basic notes on narratology and narrative.
\newblock \emph{Scientific Journal of Review}, 4\penalty0 (10):\penalty0 182--192, 2015.

\bibitem[Ma et~al.(2011)Ma, Liao, Frazier, Hauser, and Kostis]{121ma2011scientific}
K.-L. Ma, I.~Liao, J.~Frazier, H.~Hauser, and H.-N. Kostis.
\newblock Scientific storytelling using visualization.
\newblock \emph{IEEE Computer Graphics and Applications}, 32\penalty0 (1):\penalty0 12--19, 2011.

\bibitem[Madni*(2015)]{VRmadni2015expanding}
A.~M. Madni*.
\newblock Expanding stakeholder participation in upfront system engineering through storytelling in virtual worlds.
\newblock \emph{Systems Engineering}, 18\penalty0 (1):\penalty0 16--27, 2015.

\bibitem[Mahyar et~al.(2015)Mahyar, Kim, and Kwon]{mahyar2015towards}
N.~Mahyar, S.-H. Kim, and B.~C. Kwon.
\newblock Towards a taxonomy for evaluating user engagement in information visualization.
\newblock In \emph{Workshop on Personal Visualization: Exploring Everyday Life}, volume~3, page~2, 2015.

\bibitem[Majooni et~al.(2018)Majooni, Masood, and Akhavan]{289majooni2018eye}
A.~Majooni, M.~Masood, and A.~Akhavan.
\newblock An eye-tracking study on the effect of infographic structures on viewer’s comprehension and cognitive load.
\newblock \emph{Information Visualization}, 17\penalty0 (3):\penalty0 257--266, 2018.

\bibitem[Mantri et~al.(2022)Mantri, Subramonyam, Michal, and Xiong]{mantri2022viewers}
P.~Mantri, H.~Subramonyam, A.~L. Michal, and C.~Xiong.
\newblock How do viewers synthesize conflicting information from data visualizations?
\newblock \emph{IEEE transactions on visualization and computer graphics}, 29\penalty0 (1):\penalty0 1005--1015, 2022.

\bibitem[{Marjanovic}(2016)]{258MarjanovicEmpoweringBusiness}
O.~{Marjanovic}.
\newblock Empowering business users to explore visual data through boundary objects and storytelling.
\newblock In \emph{2016 49th Hawaii International Conference on System Sciences (HICSS)}, pages 5032--5041, 2016.
\newblock \doi{10.1109/HICSS.2016.624}.

\bibitem[Marques et~al.(2022)Marques, Branco, and Costa]{marques2022narrative}
A.~B. Marques, V.~Branco, and R.~Costa.
\newblock Narrative visualization with augmented reality.
\newblock \emph{Multimodal Technologies and Interaction}, 6\penalty0 (12):\penalty0 105, 2022.

\bibitem[Martinez-Maldonado et~al.(2020)Martinez-Maldonado, Echeverria, Fernandez~Nieto, and Buckingham~Shum]{352martinez2020data}
R.~Martinez-Maldonado, V.~Echeverria, G.~Fernandez~Nieto, and S.~Buckingham~Shum.
\newblock From data to insights: a layered storytelling approach for multimodal learning analytics.
\newblock In \emph{Proceedings of the 2020 chi conference on human factors in computing systems}, pages 1--15, 2020.

\bibitem[Mason(2010)]{000mason2010sample}
M.~Mason.
\newblock Sample size and saturation in phd studies using qualitative interviews.
\newblock In \emph{Forum qualitative Sozialforschung/Forum: qualitative social research}, volume~11, 2010.

\bibitem[Mayr and Windhager(2018)]{277mayr2018once}
E.~Mayr and F.~Windhager.
\newblock Once upon a spacetime: Visual storytelling in cognitive and geotemporal information spaces.
\newblock \emph{ISPRS International Journal of Geo-Information}, 7\penalty0 (3):\penalty0 96, 2018.

\bibitem[Metoyer et~al.(2020)Metoyer, Chuanromanee, Girgis, Zhi, and Kinyon]{318metoyer2020supporting}
R.~A. Metoyer, T.~S. Chuanromanee, G.~M. Girgis, Q.~Zhi, and E.~C. Kinyon.
\newblock Supporting storytelling with evidence in holistic review processes: A participatory design approach.
\newblock \emph{Proceedings of the ACM on Human-Computer Interaction}, 4\penalty0 (CSCW1):\penalty0 1--24, 2020.

\bibitem[Meuschke et~al.(2022)Meuschke, Garrison, Smit, Bach, Mittenentzwei, Wei{\ss}, Bruckner, Lawonn, and Preim]{meuschke2022narrative}
M.~Meuschke, L.~A. Garrison, N.~N. Smit, B.~Bach, S.~Mittenentzwei, V.~Wei{\ss}, S.~Bruckner, K.~Lawonn, and B.~Preim.
\newblock Narrative medical visualization to communicate disease data.
\newblock \emph{Computers \& Graphics}, 107:\penalty0 144--157, 2022.

\bibitem[Michel and Ladd(2015)]{244michel2015snow}
J.~P. Michel and M.~Ladd.
\newblock “snow fall”-ing special collections and archives.
\newblock \emph{Journal of Web Librarianship}, 9\penalty0 (2-3):\penalty0 121--131, 2015.

\bibitem[Minelli et~al.(2014)Minelli, Baracchi, Mocci, and Lanza]{179minelli2014visual}
R.~Minelli, L.~Baracchi, A.~Mocci, and M.~Lanza.
\newblock Visual storytelling of development sessions.
\newblock In \emph{2014 IEEE International Conference on Software Maintenance and Evolution}, pages 416--420. IEEE, 2014.

\bibitem[M{\"o}rth et~al.(2022)M{\"o}rth, Bruckner, and Smit]{morth2022scrollyvis}
E.~M{\"o}rth, S.~Bruckner, and N.~N. Smit.
\newblock Scrollyvis: Interactive visual authoring of guided dynamic narratives for scientific scrollytelling.
\newblock \emph{IEEE Transactions on Visualization and Computer Graphics}, 2022.

\bibitem[Murray and Sools(2014)]{000Murray2014Narrativeresearch}
M.~Murray and A.~Sools.
\newblock \emph{Narrative Research in Clinical and Health Psychology}, pages~--.
\newblock Palgrave Macmillan Ltd., United Kingdom, 2014.

\bibitem[Nardi(2016)]{246nardi2016form}
E.~Nardi.
\newblock Where form and substance meet: Using the narrative approach of re-storying to generate research findings and community rapprochement in (university) mathematics education.
\newblock \emph{Educational Studies in Mathematics}, 92\penalty0 (3):\penalty0 361--377, 2016.

\bibitem[Nielsen(2000)]{000nielsen2000you}
J.~Nielsen.
\newblock Why you only need to test with 5 users, 2000.

\bibitem[Obie et~al.(2019{\natexlab{a}})Obie, Chua, Avazpour, Abdelrazek, Grundy, and Bednarz]{009obie2019framework}
H.~O. Obie, C.~Chua, I.~Avazpour, M.~Abdelrazek, J.~Grundy, and T.~Bednarz.
\newblock A framework for authoring logically ordered visual data stories.
\newblock In \emph{2019 IEEE Symposium on Visual Languages and Human-Centric Computing (VL/HCC)}, pages 257--259. IEEE, 2019{\natexlab{a}}.

\bibitem[Obie et~al.(2019{\natexlab{b}})Obie, Chua, Avazpour, Abdelrazek, Grundy, and Bednarz]{302obie2019study}
H.~O. Obie, C.~Chua, I.~Avazpour, M.~Abdelrazek, J.~Grundy, and T.~Bednarz.
\newblock A study of the effects of narration on comprehension and memorability of visualisations.
\newblock \emph{Journal of Computer Languages}, 52:\penalty0 113--124, 2019{\natexlab{b}}.

\bibitem[Obie et~al.(2020{\natexlab{a}})Obie, Chua, Avazpour, Abdelrazek, Grundy, and Bednarz]{030obie2020authoring}
H.~O. Obie, C.~Chua, I.~Avazpour, M.~Abdelrazek, J.~Grundy, and T.~Bednarz.
\newblock Authoring logically sequenced visual data stories with gravity.
\newblock \emph{Journal of Computer Languages}, 58:\penalty0 100961, 2020{\natexlab{a}}.

\bibitem[Obie et~al.(2020{\natexlab{b}})Obie, Chua, Avazpour, Abdelrazek, Grundy, and Bednarz]{obie2020effect}
H.~O. Obie, C.~Chua, I.~Avazpour, M.~Abdelrazek, J.~Grundy, and T.~Bednarz.
\newblock The effect of narration on user comprehension and recall of information visualisations.
\newblock In \emph{2020 IEEE Symposium on Visual Languages and Human-Centric Computing (VL/HCC)}, pages 1--4. IEEE, 2020{\natexlab{b}}.

\bibitem[Obie et~al.(2022)Obie, Ho, Avazpour, Grundy, Abdelrazek, Bednarz, and Chua]{obie2022gravity++}
H.~O. Obie, D.~T.~C. Ho, I.~Avazpour, J.~Grundy, M.~Abdelrazek, T.~Bednarz, and C.~Chua.
\newblock Gravity++: A graph-based framework for constructing interactive visualization narratives.
\newblock \emph{Journal of Computer Languages}, 71:\penalty0 101125, 2022.

\bibitem[O'Brien and Toms(2008)]{o2008user}
H.~L. O'Brien and E.~G. Toms.
\newblock What is user engagement? a conceptual framework for defining user engagement with technology.
\newblock \emph{Journal of the American society for Information Science and Technology}, 59\penalty0 (6):\penalty0 938--955, 2008.

\bibitem[Ojo and Heravi(2018)]{116ojo2018patterns}
A.~Ojo and B.~Heravi.
\newblock Patterns in award winning data storytelling: Story types, enabling tools and competences.
\newblock \emph{Digital journalism}, 6\penalty0 (6):\penalty0 693--718, 2018.

\bibitem[Otten et~al.(2015)Otten, Cheng, and Drewnowski]{otten2015infographics}
J.~J. Otten, K.~Cheng, and A.~Drewnowski.
\newblock Infographics and public policy: using data visualization to convey complex information.
\newblock \emph{Health Affairs}, 34\penalty0 (11):\penalty0 1901--1907, 2015.

\bibitem[Page et~al.(2021)Page, McKenzie, Bossuyt, Boutron, Hoffmann, Mulrow, Shamseer, Tetzlaff, Akl, Brennan, et~al.]{page2021prisma}
M.~J. Page, J.~E. McKenzie, P.~M. Bossuyt, I.~Boutron, T.~C. Hoffmann, C.~D. Mulrow, L.~Shamseer, J.~M. Tetzlaff, E.~A. Akl, S.~E. Brennan, et~al.
\newblock The prisma 2020 statement: an updated guideline for reporting systematic reviews.
\newblock \emph{Bmj}, 372, 2021.

\bibitem[Park et~al.(2022)Park, Suhail, Zheng, Dunne, Ragan, and Elmqvist]{park2022storyfacets}
D.~Park, M.~Suhail, M.~Zheng, C.~Dunne, E.~Ragan, and N.~Elmqvist.
\newblock Storyfacets: A design study on storytelling with visualizations for collaborative data analysis.
\newblock \emph{Information Visualization}, 21\penalty0 (1):\penalty0 3--16, 2022.

\bibitem[Peng(2017)]{146peng2017storytelling}
Q.~Peng.
\newblock Storytelling tools in support of user experience design.
\newblock In \emph{Proceedings of the 2017 CHI Conference Extended Abstracts on Human Factors in Computing Systems}, pages 316--319, 2017.

\bibitem[P{\'e}rez-Montoro(2018)]{095perez2018interaction}
M.~P{\'e}rez-Montoro.
\newblock \emph{Interaction in Digital News Media: From Principles to Practice}.
\newblock Springer, 2018.

\bibitem[Ping and Chen(2018)]{105ping2018litstoryteller}
Q.~Ping and C.~Chen.
\newblock Litstoryteller+: an interactive system for multi-level scientific paper visual storytelling with a supportive text mining toolbox.
\newblock \emph{Scientometrics}, 116\penalty0 (3):\penalty0 1887--1944, 2018.

\bibitem[Ploehn et~al.(2020)Ploehn, Steenson, and Byrne]{ploehn2020tsuga}
C.~Ploehn, M.~Steenson, and D.~Byrne.
\newblock Tsuga convictio: Visualizing for the ecological, feminine, and embodied.
\newblock In \emph{2020 IEEE VIS Arts Program (VISAP)}, pages 9--18. IEEE, 2020.

\bibitem[Polan{\'\i}a and de~Francisco~Vela(2022)]{polania2022designing}
A.~Polan{\'\i}a and S.~de~Francisco~Vela.
\newblock Designing interfaces for legal artifacts. using storytelling to enhance digital legal experiences.
\newblock In \emph{Advances in Design and Digital Communication III: Proceedings of the 6th International Conference on Design and Digital Communication, Digicom 2022, November 3--5, 2022, Barcelos, Portugal}, pages 137--147. Springer, 2022.

\bibitem[Qiang and Bingjie(2016)]{133Qiang2016StoryCake}
L.~Qiang and C.~Bingjie.
\newblock Storycake: A hierarchical plot visualization method for storytelling in polar coordinates.
\newblock In \emph{2016 International Conference on Cyberworlds (CW)}, pages 211--218, 2016.
\newblock \doi{10.1109/CW.2016.43}.

\bibitem[Ren et~al.(2017)Ren, Brehmer, Lee, H{\"o}llerer, and Choe]{39ren2017chartaccent}
D.~Ren, M.~Brehmer, B.~Lee, T.~H{\"o}llerer, and E.~K. Choe.
\newblock Chartaccent: Annotation for data-driven storytelling.
\newblock In \emph{2017 IEEE Pacific Visualization Symposium (PacificVis)}, pages 230--239. Ieee, 2017.

\bibitem[Ren et~al.(2018)Ren, Lee, and H\"{o}llerer]{187RenXRcreator}
D.~Ren, B.~Lee, and T.~H\"{o}llerer.
\newblock Xrcreator: Interactive construction of immersive data-driven stories.
\newblock In \emph{Proceedings of the 24th ACM Symposium on Virtual Reality Software and Technology}, VRST '18, New York, NY, USA, 2018. Association for Computing Machinery.
\newblock ISBN 9781450360869.
\newblock \doi{10.1145/3281505.3283400}.
\newblock URL \url{https://doi.org/10.1145/3281505.3283400}.

\bibitem[Rickhaus(2022)]{rickhaus2022visual}
M.~Rickhaus.
\newblock Visual anatomy of an article--lessons learned and taught in five figures.
\newblock \emph{Chimia}, 76\penalty0 (9):\penalty0 748--748, 2022.

\bibitem[Riedl and Young(2006)]{190riedl2006linear}
M.~O. Riedl and R.~M. Young.
\newblock From linear story generation to branching story graphs.
\newblock \emph{IEEE Computer Graphics and Applications}, 26\penalty0 (3):\penalty0 23--31, 2006.

\bibitem[Rodrigues et~al.(2019)Rodrigues, Figueiras, and Alexandre]{114rodrigues2019once}
S.~Rodrigues, A.~Figueiras, and I.~Alexandre.
\newblock Once upon a time in a land far away: Guidelines for spatio-temporal narrative visualization.
\newblock In \emph{2019 23rd International Conference Information Visualisation (IV)}, pages 44--49. IEEE, 2019.

\bibitem[Rodr\'{\i}guez et~al.(2015)Rodr\'{\i}guez, Nunes, and Devezas]{157RodriguezTellingStories}
M.~T. Rodr\'{\i}guez, S.~Nunes, and T.~Devezas.
\newblock Telling stories with data visualization.
\newblock In \emph{Proceedings of the 2015 Workshop on Narrative \& Hypertext}, NHT '15, page 7–11, New York, NY, USA, 2015. Association for Computing Machinery.
\newblock ISBN 9781450337977.
\newblock \doi{10.1145/2804565.2804567}.
\newblock URL \url{https://doi.org/10.1145/2804565.2804567}.

\bibitem[Roels et~al.(2016)Roels, Baeten, and Signer]{271roels2016interactive}
R.~Roels, Y.~Baeten, and B.~Signer.
\newblock Interactive and narrative data visualisation for presentation-based knowledge transfer.
\newblock In \emph{International Conference on Computer Supported Education}, pages 237--258. Springer, 2016.

\bibitem[Rohrmann(1992)]{rohrmann1992evaluation}
B.~Rohrmann.
\newblock The evaluation of risk communication effectiveness.
\newblock \emph{Acta psychologica}, 81\penalty0 (2):\penalty0 169--192, 1992.

\bibitem[Rosling(2007)]{roslingTed}
H.~Rosling.
\newblock The best stats you've ever seen {TED Talk}.
\newblock \url{https://www.youtube.com/watch?v=hVimVzgtD6w}, January 2007.

\bibitem[Roth(2020)]{328roth2020cartographic}
R.~E. Roth.
\newblock Cartographic design as visual storytelling: Synthesis and review of map-based narratives, genres, and tropes.
\newblock \emph{The Cartographic Journal}, pages 1--32, 2020.

\bibitem[Rubio~Tamayo et~al.(2018)Rubio~Tamayo, Barro~Hern{\'a}ndez, G{\'o}mez~G{\'o}mez, et~al.]{VRrubio2018digital}
J.~L. Rubio~Tamayo, M.~Barro~Hern{\'a}ndez, H.~G{\'o}mez~G{\'o}mez, et~al.
\newblock Digital data visualization with interactive and virtual reality tools. review of current state of the art and proposal of a model.
\newblock 2018.

\bibitem[Saini et~al.(2019)Saini, Mathur, Thukral, Singhal, and Parnami]{17saini2019aesop}
A.~Saini, K.~Mathur, A.~Thukral, N.~Singhal, and A.~Parnami.
\newblock Aesop: Authoring engaging digital storytelling experiences.
\newblock In \emph{The Adjunct Publication of the 32nd Annual ACM Symposium on User Interface Software and Technology}, pages 56--59, 2019.

\bibitem[Sallam et~al.(2022)Sallam, Sakamoto, Leboe-McGowan, Latulipe, and Irani]{sallam2022towards}
S.~Sallam, Y.~Sakamoto, J.~Leboe-McGowan, C.~Latulipe, and P.~Irani.
\newblock Towards design guidelines for effective health-related data videos: an empirical investigation of affect, personality, and video content.
\newblock In \emph{Proceedings of the 2022 CHI Conference on Human Factors in Computing Systems}, pages 1--22, 2022.

\bibitem[Sanei et~al.()Sanei, Wang, Zhu, and Jiang]{saneiremixing}
H.~Sanei, C.~Wang, L.~Zhu, and S.~Jiang.
\newblock Remixing as a key practice for coding and data storytelling.

\bibitem[Satyanarayan and Heer(2014)]{31satyanarayan2014authoring}
A.~Satyanarayan and J.~Heer.
\newblock Authoring narrative visualizations with ellipsis.
\newblock In \emph{Computer Graphics Forum}, volume~33, pages 361--370. Wiley Online Library, 2014.

\bibitem[Schr{\"o}der et~al.(2022{\natexlab{a}})Schr{\"o}der, Belavadi, Ziefle, and Calero~Valdez]{schroder2022pension}
K.~Schr{\"o}der, P.~Belavadi, M.~Ziefle, and A.~Calero~Valdez.
\newblock The pension story-data-driven storytelling with pension data.
\newblock In \emph{International Conference on Human-Computer Interaction}, pages 404--415. Springer, 2022{\natexlab{a}}.
\newblock \doi{10.1007/978-3-031-06018-2\_28}.

\bibitem[Schr{\"o}der et~al.(2022{\natexlab{b}})Schr{\"o}der, Kohl, de~Jongh, Putzu, Ziefle, and Calero~Valdez]{schroder2022rethinking}
K.~Schr{\"o}der, S.~Kohl, F.~de~Jongh, M.~Putzu, M.~Ziefle, and A.~Calero~Valdez.
\newblock Rethinking pension communication--the role of metaphors in information visualization.
\newblock In \emph{International Conference on Human-Computer Interaction}, pages 416--429. Springer, 2022{\natexlab{b}}.
\newblock \doi{10.1007/978-3-031-06018-2\_29}.

\bibitem[Schroeder et~al.(2023)Schroeder, Eberhardt, Eberhardt, and Henkel]{schroeder2023show}
K.~Schroeder, I.~Eberhardt, W.~Eberhardt, and A.~Henkel.
\newblock Show me my future: Data-driven storytelling and pension communication.
\newblock \emph{Netspar Design Paper}, 2023.

\bibitem[Schumann et~al.(2013)Schumann, Buttler, and Lukosch]{019schumann2013approach}
J.~Schumann, T.~Buttler, and S.~Lukosch.
\newblock An approach for asynchronous awareness support in collaborative non-linear storytelling.
\newblock \emph{Computer Supported Cooperative Work (CSCW)}, 22\penalty0 (2):\penalty0 271--308, 2013.

\bibitem[Segel and Heer(2010)]{203segel2010narrative}
E.~Segel and J.~Heer.
\newblock Narrative visualization: Telling stories with data.
\newblock \emph{IEEE Transactions on Visualization and Computer Graphics}, 16\penalty0 (6):\penalty0 1139--1148, 2010.

\bibitem[Seyser and Zeiller(2018)]{295seyser2018scrollytelling}
D.~Seyser and M.~Zeiller.
\newblock Scrollytelling--an analysis of visual storytelling in online journalism.
\newblock In \emph{2018 22nd International Conference Information Visualisation (IV)}, pages 401--406. IEEE, 2018.

\bibitem[Shan et~al.(2022)Shan, Wang, and Li]{shan2022research}
X.~Shan, D.~Wang, and J.~Li.
\newblock Research on data storytelling strategies for cultural heritage transmission and dissemination.
\newblock In \emph{HCI International 2022--Late Breaking Posters: 24th International Conference on Human-Computer Interaction, HCII 2022, Virtual Event, June 26--July 1, 2022, Proceedings, Part I}, pages 344--353. Springer, 2022.

\bibitem[Shi et~al.(2020)Shi, Xu, Sun, Shi, and Cao]{shi2020calliope}
D.~Shi, X.~Xu, F.~Sun, Y.~Shi, and N.~Cao.
\newblock Calliope: Automatic visual data story generation from a spreadsheet.
\newblock \emph{IEEE Transactions on Visualization and Computer Graphics}, 27\penalty0 (2):\penalty0 453--463, 2020.

\bibitem[Shi et~al.(2021)Shi, Sun, Xu, Lan, Gotz, and Cao]{shi2021autoclips}
D.~Shi, F.~Sun, X.~Xu, X.~Lan, D.~Gotz, and N.~Cao.
\newblock Autoclips: An automatic approach to video generation from data facts.
\newblock In \emph{Computer Graphics Forum}, volume~40, pages 495--505. Wiley Online Library, 2021.

\bibitem[Shi et~al.(2022)Shi, Gao, Jiao, and Cao]{shi2022breaking}
Y.~Shi, T.~Gao, X.~Jiao, and N.~Cao.
\newblock Breaking the fourth wall of data stories through interaction.
\newblock \emph{IEEE Transactions on Visualization and Computer Graphics}, 29\penalty0 (1):\penalty0 972--982, 2022.

\bibitem[Shin et~al.(2022)Shin, Kim, Han, Xie, Whitelaw, Kwon, Ko, and Elmqvist]{shin2022roslingifier}
M.~Shin, J.~Kim, Y.~Han, L.~Xie, M.~Whitelaw, B.~C. Kwon, S.~Ko, and N.~Elmqvist.
\newblock Roslingifier: Semi-automated storytelling for animated scatterplots.
\newblock \emph{IEEE Transactions on Visualization and Computer Graphics}, 2022.

\bibitem[Shleifer(2012)]{shleifer2012psychologists}
A.~Shleifer.
\newblock Psychologists at the gate: a review of daniel kahneman's thinking, fast and slow.
\newblock \emph{Journal of Economic Literature}, 50\penalty0 (4):\penalty0 1080--91, 2012.

\bibitem[Shu et~al.(2021)Shu, Wu, Wu, Liang, Cui, Wu, and Qu]{346shu2021dancingwords}
X.~Shu, J.~Wu, X.~Wu, H.~Liang, W.~Cui, Y.~Wu, and H.~Qu.
\newblock Dancingwords: exploring animated word clouds to tell stories.
\newblock \emph{Journal of Visualization}, 24\penalty0 (1):\penalty0 85--100, 2021.

\bibitem[Sim{\~o}es et~al.(2018)Sim{\~o}es, Antunes, and Carri{\c{c}}o]{065simoes2018eliciting}
D.~Sim{\~o}es, P.~Antunes, and L.~Carri{\c{c}}o.
\newblock Eliciting and modeling business process stories.
\newblock \emph{Business \& Information Systems Engineering}, 60\penalty0 (2):\penalty0 115--132, 2018.

\bibitem[Smith and Moore(2020)]{353smith2020storytelling}
T.~L. Smith and E.~B. Moore.
\newblock Storytelling to sensemaking: A systematic framework for designing auditory description display for interactives.
\newblock In \emph{Proceedings of the 2020 CHI Conference on Human Factors in Computing Systems}, pages 1--12, 2020.

\bibitem[So et~al.(2020)So, Bogucka, {\v{S}}{\'c}epanovi{\'c}, Joglekar, Zhou, and Quercia]{so2020humane}
W.~So, E.~P. Bogucka, S.~{\v{S}}{\'c}epanovi{\'c}, S.~Joglekar, K.~Zhou, and D.~Quercia.
\newblock Humane visual ai: Telling the stories behind a medical condition.
\newblock \emph{IEEE Transactions on Visualization and Computer Graphics}, 27\penalty0 (2):\penalty0 678--688, 2020.

\bibitem[Soler-Dom{\'\i}nguez et~al.(2019)Soler-Dom{\'\i}nguez, Contero, and Alca{\~n}iz]{186VRsoler2019workflow}
J.~L. Soler-Dom{\'\i}nguez, M.~Contero, and M.~Alca{\~n}iz.
\newblock Workflow and tools to track and visualize behavioural data from a virtual reality environment using a lightweight gis.
\newblock \emph{SoftwareX}, 10:\penalty0 100269, 2019.

\bibitem[Stalph and Heravi(2021)]{stalph2021exploring}
F.~Stalph and B.~Heravi.
\newblock Exploring data visualisations: An analytical framework based on dimensional components of data artefacts in journalism.
\newblock \emph{Digital Journalism}, pages 1--23, 2021.

\bibitem[Steinert et~al.(2022)Steinert, Gro{\ss}e, Hirth, and Kr{\"o}mker]{steinert2022mobility}
T.~Steinert, U.~Gro{\ss}e, M.~Hirth, and H.~Kr{\"o}mker.
\newblock Mobility data stories for a better understanding of mobility data.
\newblock In \emph{HCI International 2022--Late Breaking Papers: HCI for Today's Community and Economy: 24th International Conference on Human-Computer Interaction, HCII 2022, Virtual Event, June 26--July 1, 2022, Proceedings}, pages 533--546. Springer, 2022.

\bibitem[Sun et~al.(2022)Sun, Cai, Cui, Wu, Shi, and Cao]{sun2022erato}
M.~Sun, L.~Cai, W.~Cui, Y.~Wu, Y.~Shi, and N.~Cao.
\newblock Erato: Cooperative data story editing via fact interpolation.
\newblock \emph{IEEE Transactions on Visualization and Computer Graphics}, 29\penalty0 (1):\penalty0 983--993, 2022.

\bibitem[Tang et~al.(2020)Tang, Tang, Hong, Yu, Ren, and Wu]{317tang2020design}
T.~Tang, J.~Tang, J.~Hong, L.~Yu, P.~Ren, and Y.~Wu.
\newblock Design guidelines for augmenting short-form videos using animated data visualizations.
\newblock \emph{Journal of Visualization}, 23\penalty0 (4):\penalty0 707--720, 2020.

\bibitem[Th{\"o}ny et~al.(2018)Th{\"o}ny, Schn{\"u}rer, Sieber, Hurni, and Pajarola]{144thony2018storytelling}
M.~Th{\"o}ny, R.~Schn{\"u}rer, R.~Sieber, L.~Hurni, and R.~Pajarola.
\newblock Storytelling in interactive 3d geographic visualization systems.
\newblock \emph{ISPRS International Journal of Geo-Information}, 7\penalty0 (3):\penalty0 123, 2018.

\bibitem[Tong et~al.(2018)Tong, Roberts, Borgo, Walton, Laramee, Wegba, Lu, Wang, Qu, Luo, et~al.]{139tong2018storytelling}
C.~Tong, R.~Roberts, R.~Borgo, S.~Walton, R.~S. Laramee, K.~Wegba, A.~Lu, Y.~Wang, H.~Qu, Q.~Luo, et~al.
\newblock Storytelling and visualization: An extended survey.
\newblock \emph{Information}, 9\penalty0 (3):\penalty0 65, 2018.

\bibitem[Tversky and Kahneman(1974)]{tversky1974judgment}
A.~Tversky and D.~Kahneman.
\newblock Judgment under uncertainty: Heuristics and biases.
\newblock \emph{science}, 185\penalty0 (4157):\penalty0 1124--1131, 1974.

\bibitem[Tversky et~al.(2002)Tversky, Morrison, and Betrancourt]{tversky2002animation}
B.~Tversky, J.~B. Morrison, and M.~Betrancourt.
\newblock Animation: can it facilitate?
\newblock \emph{International journal of human-computer studies}, 57\penalty0 (4):\penalty0 247--262, 2002.

\bibitem[Tyagi et~al.(2022)Tyagi, Estro, Kuenning, Zadok, and Mueller]{tyagi2022pc}
A.~Tyagi, T.~Estro, G.~Kuenning, E.~Zadok, and K.~Mueller.
\newblock Pc-expo: A metrics-based interactive axes reordering method for parallel coordinate displays.
\newblock \emph{IEEE Transactions on Visualization and Computer Graphics}, 29\penalty0 (1):\penalty0 712--722, 2022.

\bibitem[Van Den~Bosch et~al.(2022)Van Den~Bosch, Peeters, and Claes]{van2022more}
C.~Van Den~Bosch, N.~Peeters, and S.~Claes.
\newblock More weather tomorrow. engaging families with data through a personalised weather forecast.
\newblock In \emph{ACM International Conference on Interactive Media Experiences}, pages 1--10, 2022.

\bibitem[Van~Ho et~al.(2012)Van~Ho, Lundblad, {\AA}str{\"o}m, and Jern]{015van2012web}
Q.~Van~Ho, P.~Lundblad, T.~{\AA}str{\"o}m, and M.~Jern.
\newblock A web-enabled visualization toolkit for geovisual analytics.
\newblock \emph{Information Visualization}, 11\penalty0 (1):\penalty0 22--42, 2012.

\bibitem[Virtue et~al.(2008)Virtue, Parrish, and Jung-Beeman]{000Virtue2008Inferences}
S.~Virtue, T.~Parrish, and M.~Jung-Beeman.
\newblock {Inferences during Story Comprehension: Cortical Recruitment Affected by Predictability of Events and Working Memory Capacity}.
\newblock \emph{Journal of Cognitive Neuroscience}, 20\penalty0 (12):\penalty0 2274--2284, 12 2008.
\newblock ISSN 0898-929X.
\newblock \doi{10.1162/jocn.2008.20160}.
\newblock URL \url{https://doi.org/10.1162/jocn.2008.20160}.

\bibitem[Walsh et~al.(2019)Walsh, Cunningham, Wark, Nowina-Krowicki, and Thomas]{321Walsh2019Modelling}
J.~Walsh, A.~Cunningham, S.~Wark, M.~Nowina-Krowicki, and B.~H. Thomas.
\newblock Narrative visualisation of simulations: Finding the stories within the data.
\newblock In \emph{El Sawah, S. (ed.) {MODSIM}2019, 23rd International Congress on Modelling and Simulation.} Modelling and Simulation Society of Australia and New Zealand, dec 2019.
\newblock \doi{10.36334/modsim.2019.B3.walsh}.
\newblock URL \url{https://doi.org/10.36334/modsim.2019.B3.walsh}.

\bibitem[Wang et~al.(2016)Wang, Liu, Qu, Luo, and Ma]{264wang2016guided}
Y.~Wang, D.~Liu, H.~Qu, Q.~Luo, and X.~Ma.
\newblock A guided tour of literature review: Facilitating academic paper reading with narrative visualization.
\newblock In \emph{Proceedings of the 9th International Symposium on Visual Information Communication and Interaction}, pages 17--24, 2016.

\bibitem[Wang et~al.(2019{\natexlab{a}})Wang, Dingwall, and Bach]{154wang2019teaching}
Z.~Wang, H.~Dingwall, and B.~Bach.
\newblock Teaching data visualization and storytelling with data comic workshops.
\newblock In \emph{Extended Abstracts of the 2019 CHI Conference on Human Factors in Computing Systems}, pages 1--9, 2019{\natexlab{a}}.

\bibitem[Wang et~al.(2019{\natexlab{b}})Wang, Wang, Farinella, Murray-Rust, Henry~Riche, and Bach]{43wang2019comparing}
Z.~Wang, S.~Wang, M.~Farinella, D.~Murray-Rust, N.~Henry~Riche, and B.~Bach.
\newblock Comparing effectiveness and engagement of data comics and infographics.
\newblock In \emph{Proceedings of the 2019 CHI Conference on Human Factors in Computing Systems}, pages 1--12, 2019{\natexlab{b}}.

\bibitem[Wang et~al.(2021)Wang, Romat, Chevalier, Riche, Murray-Rust, and Bach]{wang2021interactive}
Z.~Wang, H.~Romat, F.~Chevalier, N.~H. Riche, D.~Murray-Rust, and B.~Bach.
\newblock Interactive data comics.
\newblock \emph{IEEE Transactions on Visualization and Computer Graphics}, 28\penalty0 (1):\penalty0 944--954, 2021.

\bibitem[Watson et~al.(2019)Watson, Sohn, Schriber, Gross, Muniz, and Kapadia]{136watson2019storyprint}
K.~Watson, S.~S. Sohn, S.~Schriber, M.~Gross, C.~M. Muniz, and M.~Kapadia.
\newblock Storyprint: an interactive visualization of stories.
\newblock In \emph{Proceedings of the 24th International Conference on Intelligent User Interfaces}, pages 303--311, 2019.

\bibitem[Weber et~al.(2018)Weber, Engebretsen, and Kennedy]{050weber2018data}
W.~Weber, M.~Engebretsen, and H.~Kennedy.
\newblock Data stories: Rethinking journalistic storytelling in the context of data journalism.
\newblock \emph{Studies in Communication Sciences}, 2018\penalty0 (1):\penalty0 191--206, 2018.

\bibitem[Weng et~al.(2018)Weng, Chen, Deng, Wu, Chen, and Wu]{weng2018srvis}
D.~Weng, R.~Chen, Z.~Deng, F.~Wu, J.~Chen, and Y.~Wu.
\newblock Srvis: Towards better spatial integration in ranking visualization.
\newblock \emph{IEEE transactions on visualization and computer graphics}, 25\penalty0 (1):\penalty0 459--469, 2018.

\bibitem[Willemsen and Kiss(2020)]{358Willemnsen2020Storyworlds}
S.~Willemsen and M.~Kiss.
\newblock Keeping track of time: The role of spatial and embodied cognition in the comprehension of nonlinear storyworlds.
\newblock \emph{Style}, 54\penalty0 (2):\penalty0 172--198, 2020.
\newblock ISSN 00394238, 23746629.
\newblock URL \url{https://www.jstor.org/stable/10.5325/style.54.2.0172}.

\bibitem[Wong and Lee(2011)]{214Wong2011Collaborative}
Y.-L. Wong and C.-S. Lee.
\newblock Creative storytelling enhanced through social media and intelligent recommendation.
\newblock In \emph{Proceedings of the 8th ACM Conference on Creativity and Cognition}, C\&C '11, page 399–400, New York, NY, USA, 2011. Association for Computing Machinery.
\newblock ISBN 9781450308205.
\newblock \doi{10.1145/2069618.2069715}.
\newblock URL \url{https://doi.org/10.1145/2069618.2069715}.

\bibitem[Xu and Ragan(2019)]{VRxu2019effects}
Q.~Xu and E.~D. Ragan.
\newblock Effects of character guide in immersive virtual reality stories.
\newblock In \emph{International Conference on Human-Computer Interaction}, pages 375--391. Springer, 2019.

\bibitem[Xu et~al.(2022)Xu, Yang, Yip, Fan, Wei, and Qu]{xu2022wow}
X.~Xu, L.~Yang, D.~Yip, M.~Fan, Z.~Wei, and H.~Qu.
\newblock From ‘wow’to ‘why’: Guidelines for creating the opening of a data video with cinematic styles.
\newblock In \emph{Proceedings of the 2022 CHI Conference on Human Factors in Computing Systems}, pages 1--20, 2022.

\bibitem[Yang et~al.(2021)Yang, Xu, Lan, Liu, Guo, Shi, Qu, and Cao]{yang2021design}
L.~Yang, X.~Xu, X.~Lan, Z.~Liu, S.~Guo, Y.~Shi, H.~Qu, and N.~Cao.
\newblock A design space for applying the freytag's pyramid structure to data stories.
\newblock \emph{IEEE Transactions on Visualization and Computer Graphics}, 28\penalty0 (1):\penalty0 922--932, 2021.

\bibitem[Ya’acob et~al.(2021)Ya’acob, Yusof, Ten, and Zainuddin]{ya2021analytical}
S.~Ya’acob, S.~M. Yusof, D.~W.~H. Ten, and N.~M. Zainuddin.
\newblock An analytical reasoning framework for visual analytics representation.
\newblock In \emph{International Visual Informatics Conference}, pages 41--52. Springer, 2021.

\bibitem[Yousuf and Conlan(2014)]{372yousuf2014constructing}
B.~Yousuf and O.~Conlan.
\newblock Constructing narrative visualizations as a means of increasing learner engagement.
\newblock In \emph{HT (Doctoral Consortium/Late-breaking Results/Workshops)}, 2014.

\bibitem[Yu et~al.(2016)Yu, Harrison, and Lu]{63yu2016effectiveness}
L.~Yu, L.~Harrison, and A.~Lu.
\newblock Effectiveness of feature-driven storytelling in 3d time-varying data visualization.
\newblock \emph{Journal of Imaging Science and Technology}, 60\penalty0 (6):\penalty0 60408--1, 2016.

\bibitem[Zanda et~al.(2019)Zanda, Lutz, Heymann, and Bleisch]{313zanda2019technological}
A.~Zanda, J.~Lutz, A.~Heymann, and S.~Bleisch.
\newblock Technological infrastructure supporting the story network principle of the atlas of the ageing society.
\newblock \emph{Geografie}, 2019.

\bibitem[Zdanovic et~al.(2022)Zdanovic, Lembcke, and Bogers]{zdanovic2022influence}
D.~Zdanovic, T.~J. Lembcke, and T.~Bogers.
\newblock The influence of data storytelling on the ability to recall information.
\newblock In \emph{Proceedings of the 2022 Conference on Human Information Interaction and Retrieval}, pages 67--77, 2022.

\bibitem[Zeng et~al.(2020)Zeng, Dong, Chen, and Cheng]{174zeng2020vistory}
W.~Zeng, A.~Dong, X.~Chen, and Z.-l. Cheng.
\newblock Vistory: interactive storyboard for exploring visual information in scientific publications.
\newblock \emph{Journal of visualization}, pages 1--16, 2020.

\bibitem[Zhang(2018)]{180zhang2018visualization}
X.~Zhang.
\newblock Visualization, technologies, or the public? exploring the articulation of data-driven journalism in the twittersphere.
\newblock \emph{Digital Journalism}, 6\penalty0 (6):\penalty0 737--758, 2018.

\bibitem[Zhang et~al.(2022)Zhang, Reynolds, Lugmayr, Damjanov, and Hassan]{zhang2022visual}
Y.~Zhang, M.~Reynolds, A.~Lugmayr, K.~Damjanov, and G.~M. Hassan.
\newblock A visual data storytelling framework.
\newblock In \emph{Informatics}, volume~9, page~73. MDPI, 2022.

\bibitem[Zhao et~al.(2021)Zhao, Shi, Liu, Zhao, Zhou, Zhang, Chen, Zhao, Zhu, and Chen]{zhao2021evaluating}
Y.~Zhao, J.~Shi, J.~Liu, J.~Zhao, F.~Zhou, W.~Zhang, K.~Chen, X.~Zhao, C.~Zhu, and W.~Chen.
\newblock Evaluating effects of background stories on graph perception.
\newblock \emph{IEEE Transactions on Visualization and Computer Graphics}, 2021.

\bibitem[Zhao et~al.(2015)Zhao, Marr, and Elmqvist]{000zhao2015data}
Z.~Zhao, R.~Marr, and N.~Elmqvist.
\newblock Data comics: Sequential art for data-driven storytelling.
\newblock \emph{tech. report}, 2015.

\bibitem[Zhao et~al.(2019)Zhao, Marr, Shaffer, and Elmqvist]{170zhao2019understanding}
Z.~Zhao, R.~Marr, J.~Shaffer, and N.~Elmqvist.
\newblock Understanding partitioning and sequence in data-driven storytelling.
\newblock In \emph{International Conference on Information}, pages 327--338. Springer, 2019.

\bibitem[Zhi et~al.(2019{\natexlab{a}})Zhi, Lin, Talkad~Sukumar, and Metoyer]{82zhi2019gameviews}
Q.~Zhi, S.~Lin, P.~Talkad~Sukumar, and R.~Metoyer.
\newblock Gameviews: Understanding and supporting data-driven sports storytelling.
\newblock In \emph{Proceedings of the 2019 CHI Conference on Human Factors in Computing Systems}, pages 1--13, 2019{\natexlab{a}}.

\bibitem[Zhi et~al.(2019{\natexlab{b}})Zhi, Ottley, and Metoyer]{103zhi2019linking}
Q.~Zhi, A.~Ottley, and R.~Metoyer.
\newblock Linking and layout: Exploring the integration of text and visualization in storytelling.
\newblock In \emph{Computer Graphics Forum}, volume~38, pages 675--685. Wiley Online Library, 2019{\natexlab{b}}.

\bibitem[Zhu et~al.(2020)Zhu, Sun, Jiang, Zha, and Liang]{336zhu2020survey}
S.~Zhu, G.~Sun, Q.~Jiang, M.~Zha, and R.~Liang.
\newblock A survey on automatic infographics and visualization recommendations.
\newblock \emph{Visual Informatics}, 4\penalty0 (3):\penalty0 24--40, 2020.

\end{thebibliography}

\end{document}